%% file: main.tex
\documentclass[sigconf,nonacm]{acmart}

\usepackage{caption}
\usepackage{subcaption}
\usepackage[ruled, vlined, linesnumbered]{algorithm2e}
\usepackage{tikz}
\usepackage{pgfplots}
\usepackage{multirow}

\newcommand{\vc}{{\mathbf{c}}}
\newcommand{\vd}{{\mathbf{d}}}
\newcommand{\ve}{{\mathbf{e}}}
\newcommand{\vf}{{\mathbf{f}}}

\newcommand{\vp}{{\mathbf{p}}}
\newcommand{\vq}{{\mathbf{q}}}

\newcommand{\vu}{{\mathbf{u}}}

\newcommand{\valpha}{{\boldsymbol{\alpha}}}
\newcommand{\vlambda}{{\boldsymbol{\lambda}}}

\newcommand{\mB}{{\mathbf{B}}}
\newcommand{\mC}{{\mathbf{C}}}

\newcommand{\mF}{{\mathbf{F}}}
\newcommand{\mG}{{\mathbf{G}}}

\newcommand{\mK}{{\mathbf{K}}}
\newcommand{\mL}{{\mathbf{L}}}

\newcommand{\mO}{{\mathbf{O}}}

\newcommand{\mR}{{\mathbf{R}}}

\newcommand{\mY}{{\mathbf{Y}}}

\DeclareMathOperator{\Kernelof}{Ker}

\begin{document}

\title{Utilizing Sparsity in the GPU-accelerated Assembly of Schur Complement Matrices in Domain Decomposition Methods}

\author{Jakub Homola}
\affiliation{%
  \department{IT4Innovations}
  \institution{VSB -- Technical University of Ostrava}
  \city{Ostrava}
  \country{Czech Republic}
}
\email{jakub.homola@vsb.cz}

\author{Ond\v{r}ej Meca}
\affiliation{%
  \department{IT4Innovations}
  \institution{VSB -- Technical University of Ostrava}
  \city{Ostrava}
  \country{Czech Republic}
}
\email{ondrej.meca@vsb.cz}

\author{Lubom\'{i}r \v{R}\'{i}ha}
\affiliation{%
  \department{IT4Innovations}
  \institution{VSB -- Technical University of Ostrava}
  \city{Ostrava}
  \country{Czech Republic}
}
\email{lubomir.riha@vsb.cz}

\author{Tom\'{a}\v{s} Brzobohat\'{y}}
\affiliation{%
  \department{IT4Innovations}
  \institution{VSB -- Technical University of Ostrava}
  \city{Ostrava}
  \country{Czech Republic}
}
\email{tomas.brzobohaty@vsb.cz}

\begin{abstract}

Schur complement matrices emerge in many domain decomposition methods that can solve complex engineering problems
using supercomputers.
Today, as most of the high-performance clusters' performance lies in GPUs, these methods should also be accelerated.

Typically, the offloaded components are the explicitly assembled dense Schur complement matrices used later in the iterative solver for multiplication with a vector.
As the explicit assembly is expensive, it represents a significant overhead associated with this approach to acceleration.
It has already been shown that the overhead can be minimized by assembling the Schur complements directly on the GPU.

This paper shows that the GPU assembly can be further improved by wisely utilizing the sparsity 
of the input matrices. 
In the context of FETI methods,
we achieved a speedup of 5.1 in the GPU section of the code
and 3.3 for the whole assembly, making the acceleration beneficial from as few as 10 iterations.

\end{abstract}


\maketitle

\section{Introduction}

Numerical methods are commonly used to simulate complex engineering problems in many scientific and industrial tools. Based on partial differential equations (PDEs), they transform a discretized model into a system of linear or nonlinear equations.
With the increase in the size of the problem, the system becomes impractical or even impossible to solve on a single machine.
Using domain decomposition methods (DDM), we can utilize the capacity of supercomputers to solve the system faster or in more detail.

The main idea of DDM, the decomposition of a problem into smaller parts (subdomains), works in conjunction with the architecture of high-performance clusters, as subdomains can be processed by different compute nodes concurrently.
It has been proven that representations of DDM, such as Finite Element Tearing and Interconnecting (FETI)~\cite{FETI} or Balancing Domain Decomposition by Constraints (BDDC)~\cite{BDDC},
can efficiently utilize supercomputers with hundreds of computational nodes~\cite{espreso-pasc,BDDC_SCALE}.

Nowadays, most high-performance systems are based on GPUs. They offer higher FLOP/s performance, memory bandwidth, and energy efficiency than general-purpose CPUs.
Hence, to fully utilize the power of such systems with DDM, a substantial amount of work must be offloaded to GPU accelerators.
This is generally challenging, especially for numerical methods that generate sparse systems. 

The DDM typically uses an iterative solver to solve the given problem. In the iterative solver, most computation is often related to some form of Schur complement matrix (SC). In general, SC matrices can be processed \textit{implicitly} or \textit{explicitly}. The implicit approach involves a sparse matrix factorization for each subdomain in the preprocessing phase and triangular solutions using the obtained factors in each iteration. The explicit approach assembles a dense SC in the preprocessing phase. Then, in each iteration, the SC matrices are used for multiplication with a vector.

The dense matrix-vector multiplication is generally faster.
Hence, the explicit approach can substantially speed up the iterations, especially if SC matrices are offloaded to GPU, which provides higher memory throughput than CPU.
On the other hand, the explicit assembly is generally an expensive operation.
Hence, the explicit approach is beneficial only if the additional time spent in preprocessing is amortized by the time saved during iterations.
This \emph{amortization point} can be expressed by the number of iterations. Naturally, the ultimate goal is to get the amortization point as low as possible, i.e., to explicitly assemble the SC as fast as possible. 

\begin{figure}
    \centering
    \includegraphics[width=0.95\linewidth]{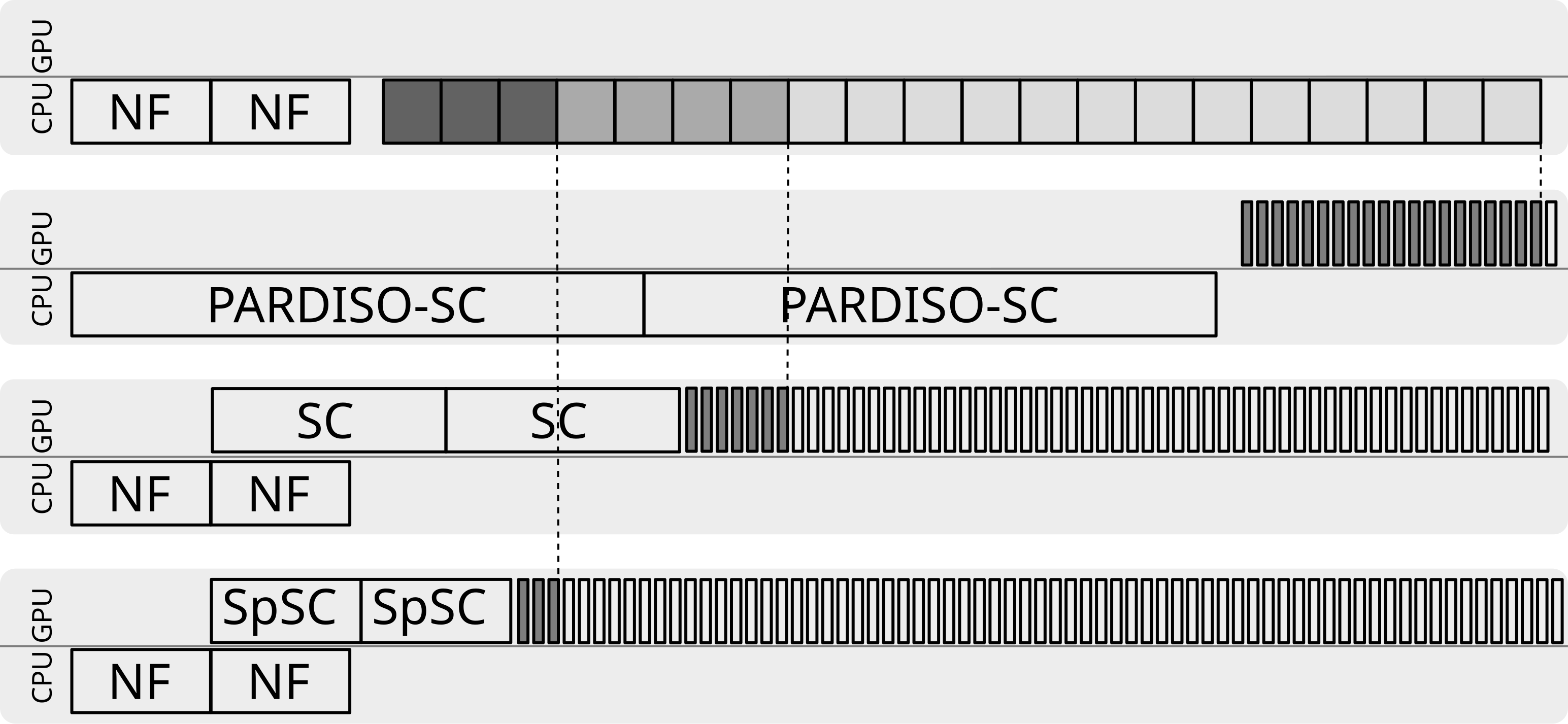}
    \caption{Solver approaches with outlined amortization points.
    From the top:
    (1) implicit approach with Numerical Factorization (NF) in preprocessing only. (2) Explicit SC computed with the PARDISO library.
    (3) Explicit SC on the GPU.
    (4) Explicit SC on the GPU utilizing sparsity.}
    \label{fig:overview}
\end{figure}

Differences between several approaches for assembling SC are depicted in Figure~\ref{fig:overview}, where large boxes denote the preprocessing for subdomains. Then, thin boxes denote iterations of the solver. Dashed vertical lines sketch the amortization points; each optimization moves the amortization point to the left, i.e., acceleration starts to be beneficial earlier.
The early implementations assemble the SC on the CPU and copy them to the accelerator to be used by the iterative solver~\cite{ESPRESO-SC, BDDS_ACC, FETI_PHI} (second approach in Figure~\ref{fig:overview}). Despite using an augmented incomplete factorization (a method from the PARDISO library~\cite{pardiso} that can utilize the sparsity of input matrices), the preprocessing was approximately 10-20 times slower, and the amortization point was in hundreds of iterations. Making the acceleration beneficial only for, e.g., ill-conditioned problems with thousands of iterations needed.

In the most recent research~\cite{PDSEC}, it has been shown that with appropriate settings and wisely chosen kernels, one can use the GPU also for assembling the SC matrices, reducing the amortization point by an order of magnitude to a range between 10 and 100 iterations, making the approach attractive for a broader set of problems (third approach in Figure~\ref{fig:overview}).

The explicit assembly on GPU is dominated by two kernels: TRSM for solving a triangular system with multiple right-hand sides and SYRK for multiplying a general matrix with its transposition.
Both kernels can be optimized by better utilizing the sparsity of the input matrices (fourth approach in Figure~\ref{fig:overview}).
Several approaches are described in this paper.
The optimized kernels are based on splitting the input sparse matrices into blocks, identifying zero values, and omitting them in the computation.
With the appropriate configuration, we achieved a speed-up of up to 5.1 for the GPU part of the code and up to 3.1 for a whole assembly of the SC in the context of a FETI solver on the Karolina GPU node with 8x Nvidia A100 GPUs and 2x 64-core AMD EPYC 7763 CPUs.
The observed speed-up increases with subdomain size, which is also preferable from the iterative solver point of view.
With our newly implemented approach, the amortization point stays almost constant at around 10 iterations for matrix sizes ranging from 1,000 to 70,000 unknowns, making the acceleration beneficial early and easily predictable.

Despite the approach being described in the context of the FETI method on a system generated by the Finite Element Method (FEM), it is general enough to be easily adopted by other methods that assemble SC from sparse matrices (e.g., BDDC).
Thus, our approach is a big step forward in efficiently utilizing modern GPU-based clusters with DDM methods for solving sparse linear systems.

\section{Finite Element Tearing and Interconnecting}
\label{sec:feti}

This section describes the basic concept of the FETI method~\cite{FETI}.
It decomposes the problem into non-overlapping subdomains.
Then, an iterative method is used to compute the solution of the dual problem on the subdomain interfaces.
In FETI, Schur complements emerge in the so-called \textit{dual operator} that maps dual variables to subdomains.
The following subsections describe it in more detail, focusing on the dual operator, the sparsity of its matrices, and parallel implementation.
These aspects help to understand the optimizations described in Section~\ref{sec:algorithm}.

\subsection{Theory}
We start with a linear system generated by FEM
\begin{equation}
    \label{eq:femsystem}
    \tilde{\mK} \tilde{\vu} = \tilde{\vf},
\end{equation}
where $\tilde{\mK}$ is the sparse Symmetric Positive Definite (SPD) system matrix, $\tilde{\vf}$ is the right-hand side vector, and $\tilde{\vu}$ is the solution.
By rearranging the order of unknowns and adding several carefully crafted equations and variables, we can transform the system into an equivalent system with a block structure,
\begin{equation}
    \label{eq:fetisystemblocked}
    \begin{bmatrix}
        \mK_1 & & & & \mB_1^\top \\
        & \mK_2 & & & \mB_2^\top \\
        & & \ddots & & \vdots \\
        & & & \mK_N & \mB_N^\top \\
        \mB_1 & \mB_2 & \dots & \mB_N & \mO
    \end{bmatrix}
    \begin{bmatrix}
        \vu_1 \\
        \vu_2 \\
        \vdots \\
        \vu_N \\
        \vlambda
    \end{bmatrix}
    =
    \begin{bmatrix}
        \vf_1 \\
        \vf_2 \\
        \vdots \\
        \vf_N \\
        \vc
    \end{bmatrix}
\end{equation}
or, shortly,
\begin{equation}
    \label{eq:fetisystem}
    \begin{bmatrix}
        \mK & \mB^\top \\
        \mB & \mO
    \end{bmatrix}
    \begin{bmatrix}
        \vu \\
        \vlambda
    \end{bmatrix}
    =
    \begin{bmatrix}
        \vf \\
        \vc
    \end{bmatrix}
\end{equation}
where each block represents one of the $N$ subdomains into which we decomposed the spatial domain.
$\mK$ is the block diagonal system matrix comprising of sparse symmetric positive semi-definite (SPSD) subdomain matrices $\mK_i$, $\vlambda$ is the Lagrange multiplier vector (we call all vectors with size equal to $\vlambda$ \textit{dual vectors}), and the matrix $\mB$ is a sparse matrix representing the subdomain gluing.

We denote $\mR_i$ to be the matrix containing the basis vectors of $\Kernelof \mK_i$ in its columns, and we create $\mR$ by placing $\mR_i$ on the main block diagonal.
From the solvability of the first equation in \eqref{eq:fetisystem}, we eventually get
\begin{equation}
    \label{eq:solvability}
    - \mR^\top \mB^\top \vlambda = - \mR^\top \vf.
\end{equation}
Given the Lagrange multipliers $\vlambda$, we can express $\vu$ using
\begin{equation}
    \label{eq:solutionUeval}
    \vu = \mK^+ (\vf - \mB^\top \vlambda) + \mR \valpha,
\end{equation}
where $\mK^+$ is a generalized inverse of $\mK$ satisfying $\mK \mK^+ \mK = \mK$.
Using \eqref{eq:solutionUeval} in the second equation in \eqref{eq:fetisystem}, we eventually get
\begin{equation}
    \label{eq:alphaconstraint}
    \mB \mK^+ \mB^\top \vlambda - \mB \mR \valpha = \mB \mK^+ \vf - \vc.
\end{equation}
We define $\mF = \mB \mK^+ \mB^\top$ (the \textit{dual operator}), $\mG = \mB \mR$, $\vd = \mB \mK^+ \vf - \vc$, $\ve = \mR^\top \vf$ and combine \eqref{eq:solvability} with \eqref{eq:alphaconstraint}, and get
\begin{equation}
    \label{eq:tfetidualproblem}
    \begin{bmatrix}
        \mF & -\mG \\
        -\mG^\top & \mO
    \end{bmatrix}
    \begin{bmatrix}
        \vlambda \\
        \valpha
    \end{bmatrix}
    =
    \begin{bmatrix}
        \vd \\
        -\ve
    \end{bmatrix}.
\end{equation}

The system \eqref{eq:tfetidualproblem} can be solved, e.g., by the preconditioned conjugate projected gradient method (PCPG)~\cite{PCPG}. In each iteration, the operator $\mF$ is applied. From this, we obtain the vectors $\vlambda$ and $\valpha$, which we use to evaluate the solution $\vu$.

The matrix $\mF$, defined previously as
\begin{equation}
    \mF = \mB \mK^+ \mB^\top,
\end{equation}
is sometimes called the \textbf{dual operator}, \textbf{dual Schur complement matrix}, or the \textbf{FETI operator}. 
Each row and column corresponds to a single Lagrange multiplier (gluing connection between subdomain surfaces). 
Due to the block structure of the matrices, we can define a \textit{local dual operator} for each subdomain as
\begin{equation}
    \label{eq:localdualoperator}
    \tilde{\mF}_i = \tilde{\mB}_i \mK_i^+ \tilde{\mB}_i^\top.
\end{equation}
It is a (possibly non-contiguous) submatrix of $\mF$ defined by only the Lagrange multipliers connected to the $i$-th subdomain. The individual $\tilde{\mF}_i$ can be combined additively to form $\mF$.
Therefore, applying the dual operator $\mF$ primarily consists of the (possibly concurrent) application of the local dual operator $\tilde{\mF}_i$ for each subdomain.

\begin{figure}
 \centering
 \includegraphics[width=\columnwidth]{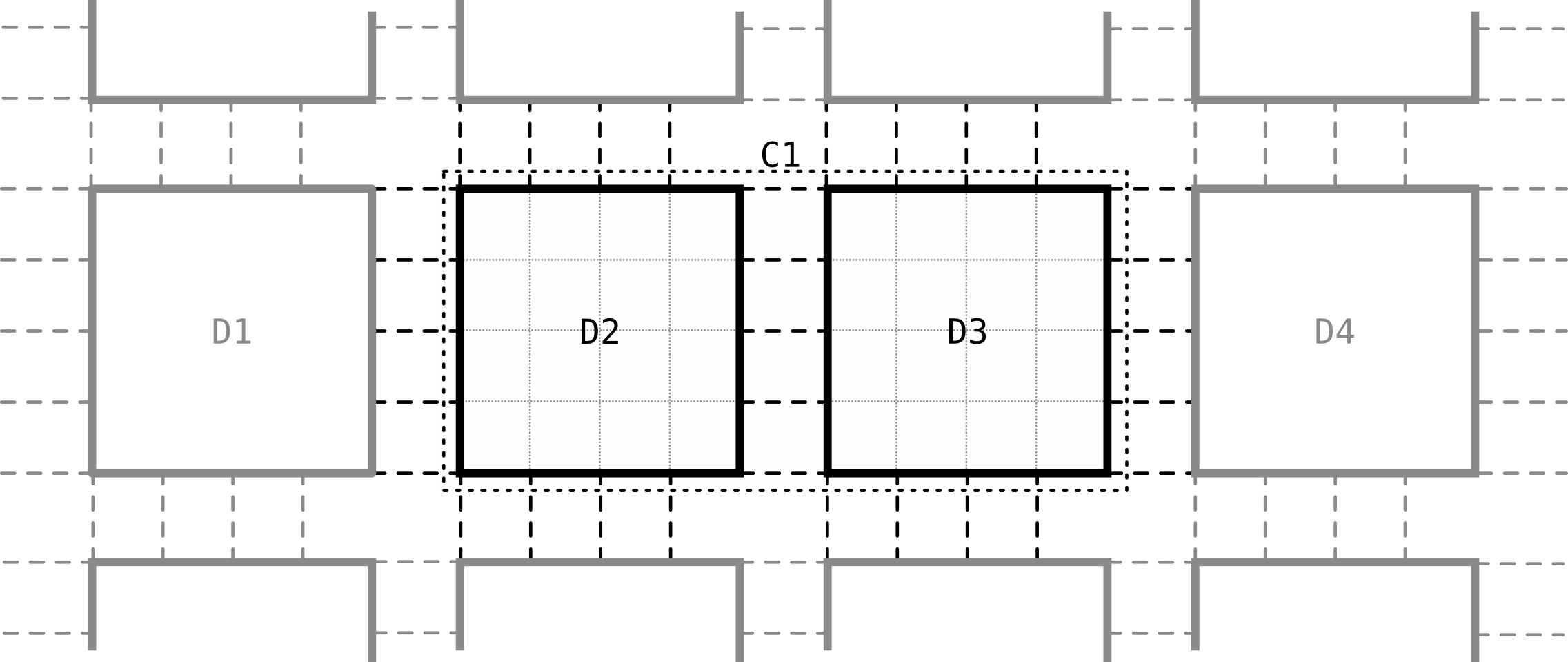}
 \Description[Domain Decomposition]{Example of domain decomposition and clusterization of subdomains.}
 \caption{Example of domain decomposition and clusterization of subdomains. Clusterization allows a hybrid parallelization for both shared and distributed memory systems.}
 \label{fig:domains}
\end{figure}

\subsection{Implementation}

The fundamental principle of a FETI solver is to divide the workload and distribute it among available hardware resources.
With current architectures, a good distribution usually considers both shared and distributed memory.
One of the possible decompositions of a domain into subdomains is shown in Fig.~\ref{fig:domains}.
The figure shows several subdomains and their gluings to neighboring subdomains (the Lagrange multipliers), which are drawn with dashed lines. 
Several subdomains are grouped into a \textit{cluster} (in the figure, subdomains D2 and D3 form the cluster C1, as highlighted by the dotted line).
Each process (distributed memory) then handles a single cluster.
In each cluster, threads handle subdomains (sharing the memory).
Typically, the number of subdomains per cluster is an integer multiple of the number of threads to utilize the available cores equally and achieve optimal performance of the FETI solver~\cite{espreso-pasc}.

The generalized inverse $\mK_i^+$ in \eqref{eq:localdualoperator} can be expressed as $\mK_i^+ = \mK_{i,reg}^{-1} = (\mL_i \mL_i^\top)^{-1} = \mL_i^{-\top} \mL_i^{-1}$, where $\mK_{i,reg}$ is the regularized subdomain matrix that can be obtained, e.g., through analytic regularization~\cite{brzyngeninverse}.
Hence, the local dual operator can be written as
\begin{equation}
    \label{eq:localdualoperatorwithU}
    \tilde{\mF}_i = \tilde{\mB}_i \mL_i^{-\top} \mL_i^{-1} \tilde{\mB}_i^\top.
\end{equation}

It can be applied \textit{implicitly}, where individual matrices are applied from right to left,
\begin{equation}
\label{eq:dualop_apply_impl}
    \tilde{\vq}_i
    =
    \tilde{\mF}_i \tilde{\vp}_i
    =
    \tilde{\mB}_i (\mL_i^{-\top} (\mL_i^{-1} (\tilde{\mB}_i^\top \tilde{\vp}_i))),
\end{equation}
or \textit{explicitly}, where all matrices are multiplied at first and then applied at once,
\begin{equation}
\label{eq:dualop_apply_expl}
    \tilde{\vq}_i
    =
    \tilde{\mF}_i \tilde{\vp}_i
    =
    (\tilde{\mB}_i \mL_i^{-\top} \mL_i^{-1} \tilde{\mB}_i^\top) \tilde{\vp}_i
\end{equation}

The implicit application consists of performing a sparse matrix-vector multiplication (SPMV), 
followed by two triangular solves (TRSV) and sparse matrix-vector multiplication.
The explicit application is a fast dense matrix-vector multiplication (GEMV), but the dense matrix $\tilde{\mF}_i$ needs to be assembled first, which can be done, e.g., with two triangular solves with a right-hand-side matrix (TRSM) and a single sparse-dense matrix multiplication (SPMM).
Comparing both approaches, \textbf{applying $\tilde{\mF}_i$ explicitly is typically faster, but the explicit assembly is a highly expensive operation, which makes the explicit approach beneficial only after a sufficient number of iterations is performed}. The amortization point (the number of iterations after which the explicit approach is faster) depends on the additional time spent in the explicit assembly and the time saved by the faster application.

The explicit assembly can also be done using a more sophisticated algorithm. The matrix $\tilde{\mF}_i$ can be expressed as the negative of a Schur complement of the matrix
\begin{equation}
\label{eq:matrix_for_sc}
    \begin{bmatrix}
        \mK_{i,reg} & \tilde{\mB}_i^\top \\
        \tilde{\mB}_i & \mO
    \end{bmatrix}.
\end{equation}
With this, we can use an augmented incomplete factorization designed and optimized for evaluating the SC of sparse matrices~\cite{pardiso}.

In both implicit and explicit approaches, the factorization $\mK_{i,reg} = \mL_i \mL_i^\top$ must be performed. There are many libraries available for this task~\cite{ssolvers}.
To handle the sparsity pattern efficiently, they typically divide the factorization into two stages~\cite{Csparse} -- symbolic and numerical. During the symbolic factorization, they search for a permutation of the matrix such that the resulting factor $\mL_i$ has the number of non-zeros as low as possible (minimizing fill-in), and they also create the non-zero pattern of the factor. In the numerical factorization, the factor is computed and filled with values.
Dividing the factorization into two steps allows skipping symbolic factorization in multi-step simulations when matrices have a fixed sparsity pattern, decreasing the factorization time.

Utilizing the two-step factorization, we can define three stages in a FETI solver:
\begin{description}
\item [initialization:] symbolic factorization and preparation of other persistent structures;
\item [preprocessing:] numerical factorization and explicit assembly of $\tilde{\mF}_i$ in the case of explicit approach;
\item [solution:] application of the dual operator to a dual vector in each iteration of an iterative solver.
\end{description}
The \textit{initialization} phase is usually called only once at the beginning of the computation; therefore, it has minimal impact on the overall performance of long-running simulations.
W.r.t. the explicit dual operator, the \textit{solution} phase contains only the multiplication of the dense $\tilde{\mF}_i$ matrices with vectors, which has been thoroughly optimized in previous studies~\cite{FETI_radim, PDSEC}.
Hence, this paper focuses only on the \textit{preprocessing} phase that is called when values in $\mK$ are changed, i.e., repeatedly.
In this phase, each process calls the numerical factorization in a loop for each subdomain in the process's cluster. In the explicit approach, the loop also contains the explicit assembly of the SC (the dual operator) that can be done on the GPU.
The following section describes optimizing this assembly by better utilizing the sparsity of the matrices $\mL_i$ and $\tilde{\mB}_i$.

\section{Utilizing the Sparsity in Computation of Schur Complement}
\label{sec:algorithm}

This section describes the main ideas for utilizing sparsity of input matrices in the explicit assembly of symmetric Schur complement matrices.
Although the approach is described in the context of FETI algorithms, it is general enough to be applied in other methods that compute the SC in the form $\mF = \mB \mK^{-1} \mB^\top$, where $\mB$ has a small number of connections (non-zeros) to a possibly sparse matrix $\mK$ (e.g., $\mB$ corresponds to subdomain surfaces).

The input for the algorithm is the matrix $\tilde{\mB}_i^\top$ together with the factor $\mL_i$ (lower triangular matrix) of $\mK_{i,reg}$.
The sparsity pattern of both matrices is determined by the fill-in reducing permutation computed during the symbolic factorization.
Our optimization heavily depends on permuting the matrix $\tilde{\mB}_i^\top$ to create areas full of zeros.
Permuting its rows, which also forces a permutation of $\mK_{i,reg}$, would interfere with the fill-reducing permutation and be counterproductive. Hence, we only permute its columns.

\begin{figure}[ht]
    \begin{subfigure}[b]{0.49\columnwidth}
        \centering
        \includegraphics[width=.8\textwidth]{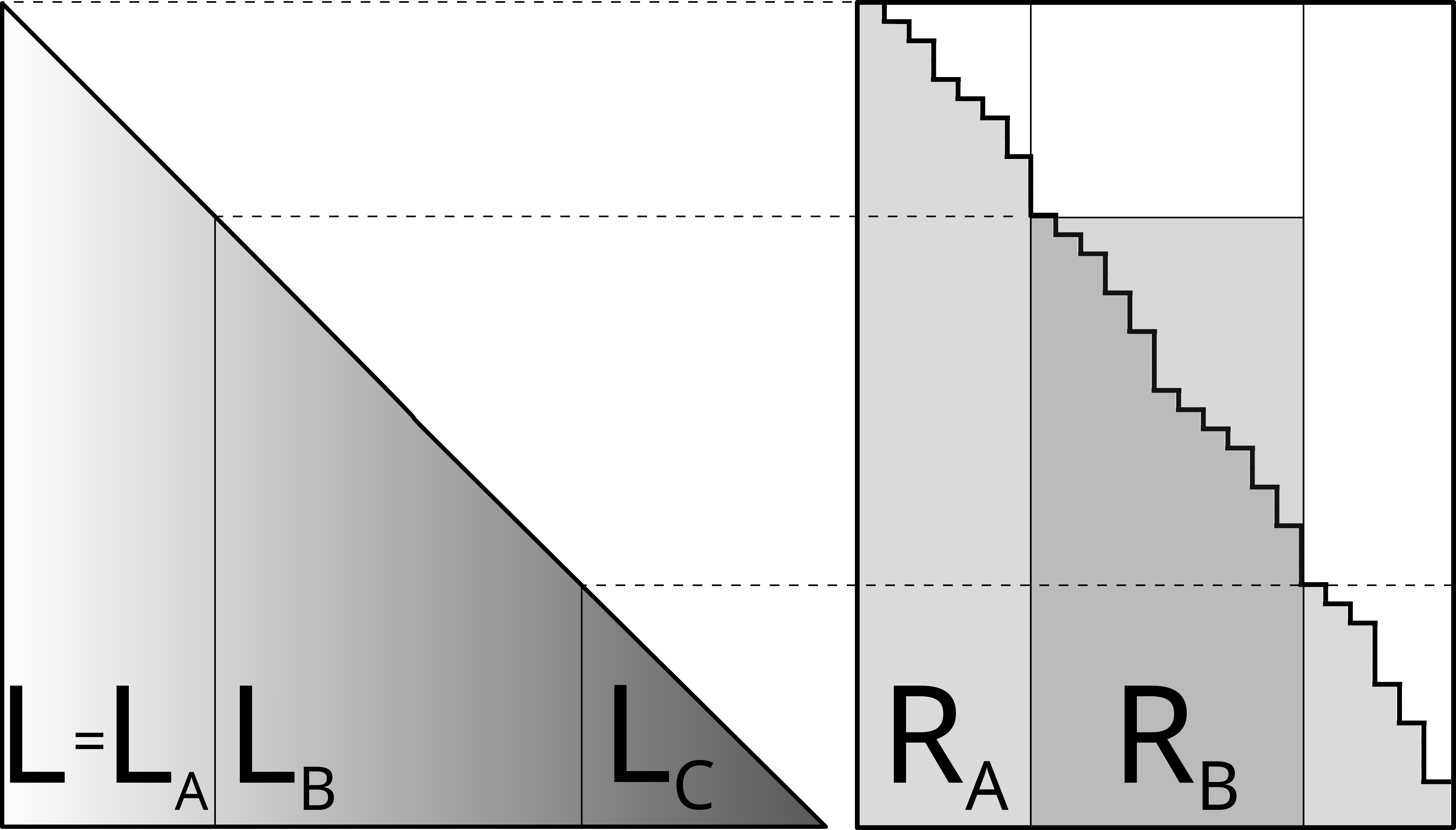}
        \caption{RHS splitting}
        \label{fig: trsm_B}
    \end{subfigure}
    \begin{subfigure}[b]{0.49\columnwidth}
        \centering
        \includegraphics[width=.8\textwidth]{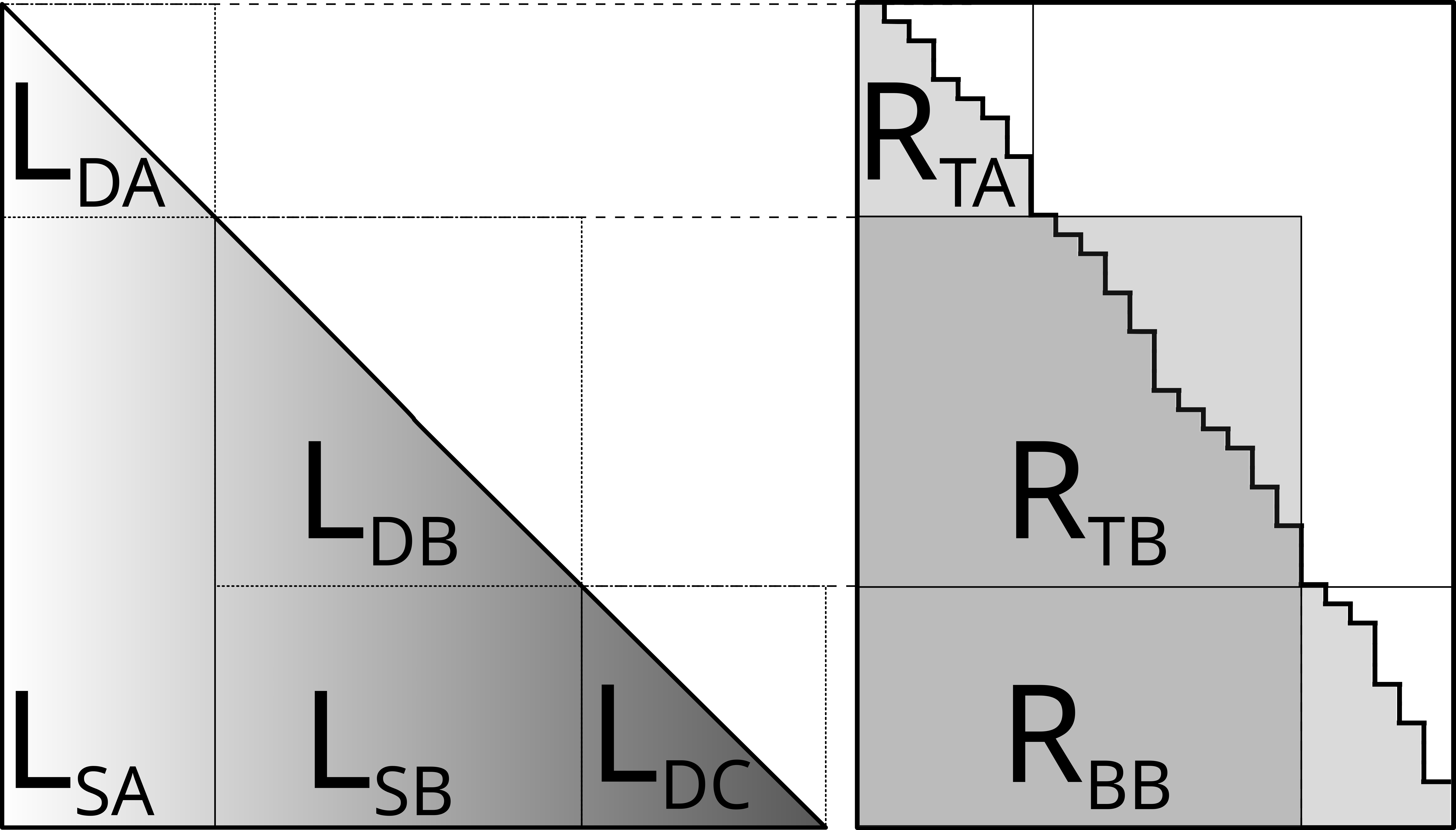}
        \caption{Factor splitting}
        \label{fig: trsm_L}
    \end{subfigure}
    \caption{Illustration of TRSM with sparse matrices $\mL$ and $\tilde{\mB}^\top$.}
    \label{fig: trsm}
\end{figure}

The columns of matrix $\tilde{\mB}_i^\top$ are permuted to create a matrix that resembles an approximately lower triangle (see, e.g., Figure~\ref{fig: trsm}).
The property of this shape is that the \textit{column pivots} (the first non-zeros in each column) go down from left to right,
and the \textit{row trails} (the last non-zeros in each row) go right from top to bottom.
In neighboring rows or columns, equal row trail or column pivot indices are allowed.
We will call this the \textit{stepped} shape. This shape can be easily achieved if the column pivots of $\tilde{\mB}_i^\top$ are approximately uniformly distributed across the rows (which holds, e.g., for permutation provided by Metis).

After the column permutation, the matrices are processed by basic routines provided by high-performance mathematical libraries (BLAS, sparse BLAS), calculating the SC, which we need to permute back in the final phase to match the original ordering of $\tilde{\mB}_i^\top$.
The mentioned mathematical libraries are generally available on most hardware platforms; hence, despite the description's focus on GPU accelerators, the same algorithm can be easily used on the CPU.
The generality is the main advantage compared to the augmented incomplete factorization provided by PARDISO~\cite{pardiso} that requires internal implementation in a given sparse solver.

Our optimization approach is built upon the algorithm described in~\cite{PDSEC} that proved the possibility of efficiently using GPUs to assemble SC and showed the optimal parameters of GPU kernels to get the best performance possible.
The following section briefly describes the original algorithm. Other sections describe its improvement towards the sparsity, i.e., the main contribution of this paper.

\subsection{Original Schur Complement Algorithm}

The original approach, used as a baseline for our improvements, is described in~\cite{PDSEC}. 
SC is computed in the loop across all subdomains in the process's cluster. A one-to-one mapping between GPUs, NUMA domains, processes, and clusters is used to simplify implementation.
This way, processes do not influence each other and do not compete for resources. Although the loop is performed on all processes concurrently, we can focus on optimizing a single process without impacting the overall performance.

After performing the numerical factorization, the algorithm copies factor $\mL_i$ to the GPU, allocates temporal memory (as described below), and starts the computation.
The computation can be performed with two different \textit{factor storage} settings -- \textit{dense} and \textit{sparse}.
It denotes if the TRSM kernel is called with sparse or dense factor, i.e., with TRSM kernel from BLAS or sparse BLAS.
The final SC is computed with the SYRK kernel performing $\mY_i^\top \mY_i$ multiplication.
It has been shown that for SPD systems, it is significantly faster to use this routine instead of the second TRSM and SPMM according to the following scheme:
\begin{equation}
 \tilde{\mF}_i =
 \tilde{\mB}_i \mL_i^{-\top} \mL_i^{-1} \tilde{\mB}_i^\top =
 (\mL_i^{-1} \tilde{\mB}_i^\top)^\top (\mL_i^{-1} \tilde{\mB}_i^\top) =
  \mY_i^\top \mY_i.
\end{equation}

An essential part of the algorithm is memory management, which avoids synchronizing memory allocations and wisely reuses the memory.
It mentally splits the GPU memory into two parts -- \textit{persistent} and \textit{temporary}. All persistent memory is allocated in the initialization phase and deallocated at the end of the program. The persistent memory holds mainly the structures needed outside the SC computation, e.g., the SC matrices $\tilde{\mF}_i$, which are used in each iteration, and persistent workspace buffers needed by the GPU mathematical library's kernels.
The rest of the memory is allocated for use in a special \textit{temporary memory allocator}. It allocates memory needed only for the duration of a specific routine. This includes, e.g., the dense matrix $\mY_i$ that can be discarded when the SC computation finishes. These temporary memory allocations are typically large, and the GPU memory capacity would not suffice if allocated persistently. The temporary memory allocator can reuse memory without calling the GPU library’s memory allocation routines. If there is enough remaining memory in the allocator’s memory pool, memory is assigned and returned immediately. Otherwise, the allocating thread is blocked until enough memory becomes available –- until other threads deallocate sufficient memory.

Fundamental observation from~\cite{PDSEC} is possible significant differences in computational time and memory demands if TRSM and SYRK kernels are called with inappropriate parameters. The optimal parameters depend on the dimension of a solved problem, the size of input matrices, and versions of mathematical libraries.
Despite the reported speed-up, the original algorithm does not benefit from the sparsity of matrix $\tilde{\mB}_i^\top$. In addition, negligible performance differences in dense and sparse TRSM routines for small and medium matrices indicate poor utilization of the sparsity of a factor.
Building on these observations, we describe optimization towards the sparsity for TRSM and SYRK routines in the following sections.

\subsection{Optimization of TRSM}
\label{sec: alg_trsm}

By forward substitution, the TRSM routine solves a triangular system of linear equations with multiple Right-Hand Sides (RHS) stored as a dense matrix. This article discusses TRSM with a lower triangular factor only. We consider TRSM to be an in-place function -- the changes are made directly to the RHS matrix, and after the kernel finishes, it contains the solution.

The inner workings of the classical TRSM kernel can be explained, e.g., by Gaussian elimination. We incorporate the first column into the RHS and then recursively solve the rest of the system, thus progressing column-by-column through the lower triangle. In each column step, the following happens.
First, we incorporate the top diagonal value into the corresponding RHS row with a simple division by that value.
Then, the rest of the column (the subdiagonal values) in the factor is multiplied with the just-updated RHS row, and the resulting matrix is subtracted from RHS.

The observation that is fundamental to our optimization is that during the forward substitution, values in the RHS are propagated downward; therefore, \textbf{zeros above the column pivots are preserved}. In addition, RHS columns do not influence each other in any way.
Obviously, one can skip the evaluation of all unknowns above the column pivots.
In addition, with the stepped shape, it is possible to divide RHS into blocks with similar column pivots. It allows both reducing the system size and preserving the ability to solve multiple RHS columns at once, which can be significantly faster than solving each RHS column individually. 

We developed two techniques to improve the performance of TRSM for the RHS matrix in the stepped shape. The first technique splits RHS into block columns; the second technique splits the triangular factor into blocks, which allows for the removal of zero parts from the factor as well.
Both techniques are described in the following paragraphs.

\paragraph{RHS splitting:}
The first technique is illustrated in Figure~\ref{fig: trsm_B}.
It splits the RHS into several column blocks and solves each separately.
Due to the downward propagation rule, the solution of the column blocks can be optimized. The zeros in the top part of each column block do not affect the bottom; the first few steps inside the TRSM routine are unnecessary. Hence, for each column block, the TRSM routine can be called with only a subfactor of the original triangular factor and an appropriate submatrix of the RHS. Their sizes are set according to the highest column pivot in the column block.

In Figure~\ref{fig: trsm_B}, we can see that the first RHS column block $\mathbf{R}_A$ is solved using the whole factor $\mL_A = \mL$, but the second column block $\mathbf{R}_B$ can be solved using only the subfactor $\mL_B$.

If only dense matrices are used, extracting the submatrix is trivial using pointer arithmetic due to the leading dimension parameter of BLAS routines. We must manually extract the sparse subfactor before each TRSM if we use a sparse factor.

\paragraph{Factor splitting:}
The second technique is shown in Figure~\ref{fig: trsm_L}.
It is based on performing blocking for the TRSM algorithm.
Instead of the single diagonal value, we now have a square matrix on the diagonal, which we incorporate into the corresponding RHS rows (the top RHS block) by a library TRSM call.
The sub-diagonal block in the factor is then incorporated by multiplying it with the top RHS block and subtracting the result from the remaining bottom RHS rows (using the GEMM routine).

Again, because of the zeros in RHS, the inner TRSM and GEMM calls can be optimized. Because the top RHS block contains only a few non-zero columns, we can restrict TRSM and GEMM to these columns only. The TRSM routine does not alter the zeros in any way, and the zero section of the top RHS block will not affect the result of the GEMM routine.

In this variant, the width of the RHS submatrix is dictated by the right-most non-zero in the top RHS block. The RHS submatrix always starts with the first column, and while progressing through the factor's blocks, it gets wider and shorter.

In Figure~\ref{fig: trsm_L}, we can see how the ($\mL_{DB}$, $\mL_{SB}$) part of the factor is incorporated into the RHS. First, the TRSM routine is called with the $\mL_{DB}$ triangle and $\mathbf{R}_{TB}$ part of the RHS. Then we call the GEMM routine $\mathbf{R}_{BB} = \mathbf{R}_{BB} - \mL_{SB} \mathbf{R}_{TB}$.

This factor-splitting technique allows each subfactor to be used in a different format (sparse or dense) as they are non-overlapping. Additionally, GEMM allows for further optimization. Due to the sparsity of the factor, the subdiagonal blocks contain many empty rows. We can, therefore, extract only the non-empty rows and run the GEMM only with them (similarly to  CHOLMOD's supernodal factorization \cite{cholmod}). We call this optimization \textit{pruning}.

\begin{figure}[ht]
    \begin{subfigure}[b]{0.49\columnwidth}
        \centering
        \includegraphics[width=.8\textwidth]{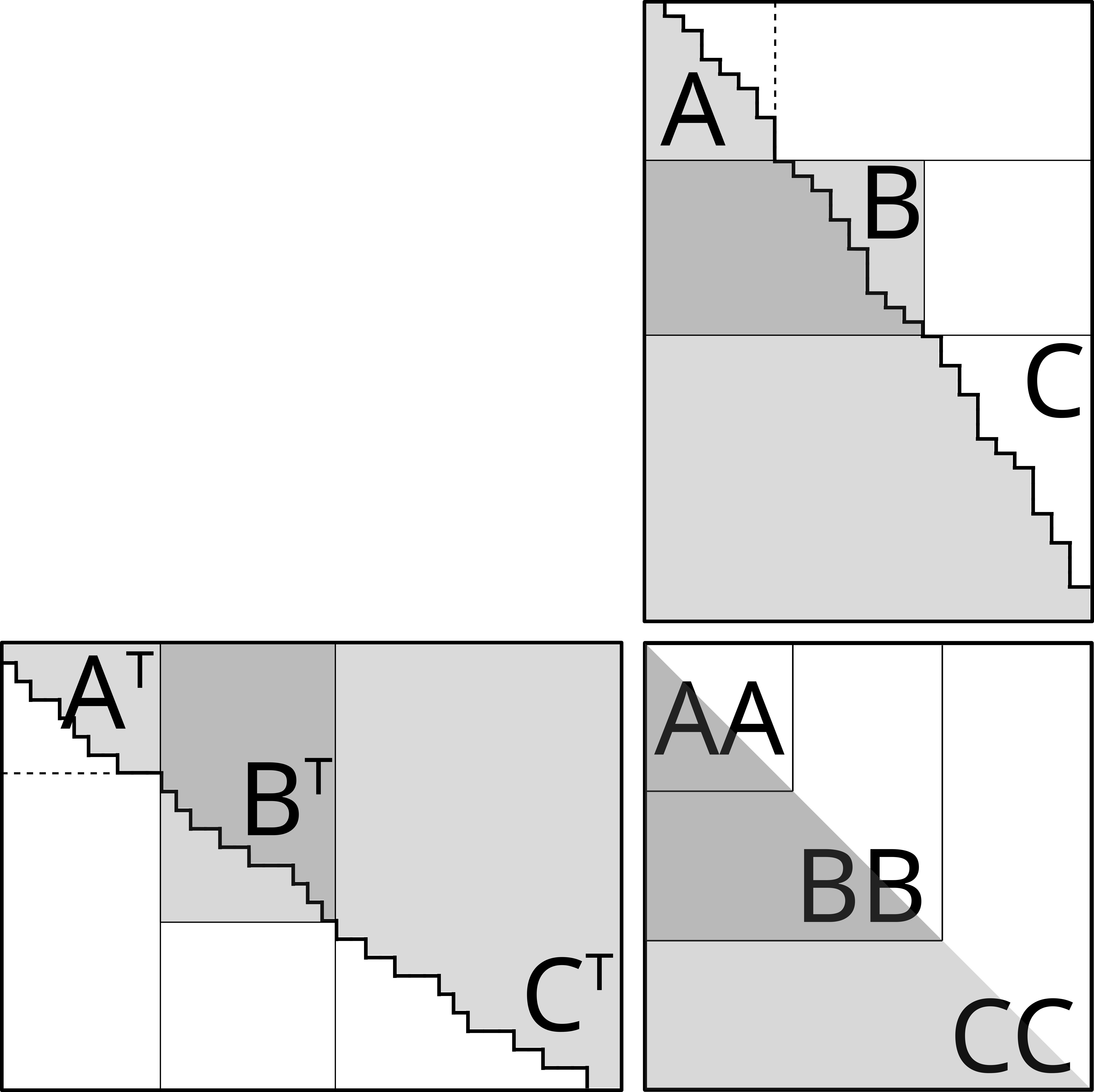}
        \caption{Input matrix splitting}
        \label{fig: syrk_input}
    \end{subfigure}
    \hfill
    \begin{subfigure}[b]{0.49\columnwidth}
        \centering
        \includegraphics[width=.8\textwidth]{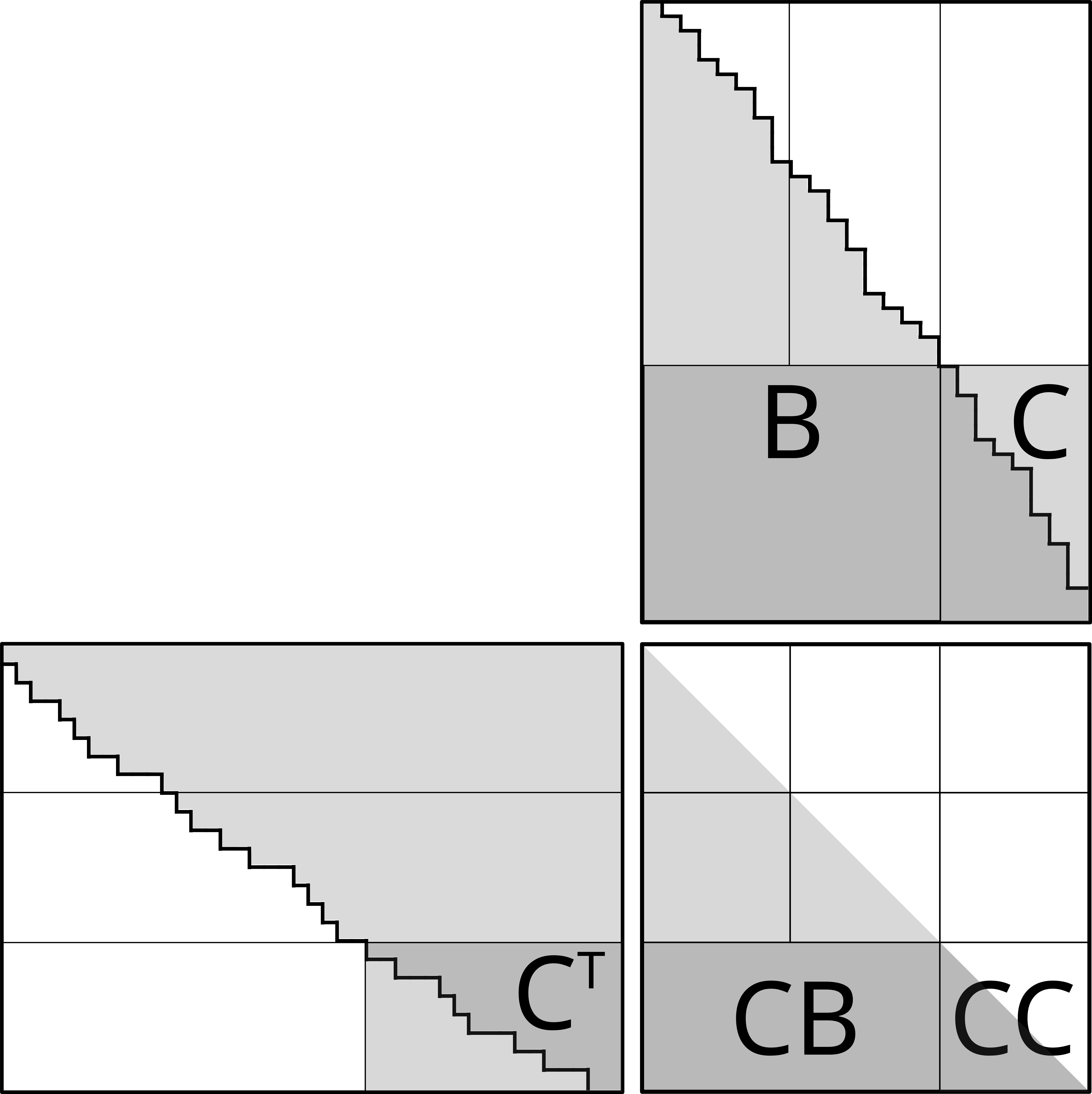}
        \caption{Output matrix splitting}
        \label{fig: syrk_output}
    \end{subfigure}
    \caption{Illustration of SYRK with stepped matrix $\mY$.}
    \label{fig: syrk}
\end{figure}

\subsection{Optimization of SYRK}
\label{sec: alg_syrk}

The SYRK routine is used to compute the multiplication of a matrix with its transposition ($\mF = \mY^\top \mY$ in our case). As the result is always symmetric, only the lower triangle must be computed. 
The input is the solution from the previously described TRSM, which is always a dense matrix in the stepped shape. An example is depicted in Figure~\ref{fig: syrk}.
The optimization again focuses on skipping the computation with zero values by splitting the matrices into blocks.
We consider two possibilities -- split the input matrix into block rows or compute the output matrix by blocks; we either split the input or the output matrix.

\paragraph{Splitting of the input matrix:}
This approach is visualized in Figure~\ref{fig: syrk_input}.
If we imagine the classical dense matrix multiplication algorithm, it contains loops iterating over the $m$, $n$, and $k$ dimensions (output rows, output columns, and the edge of the input, respectively). This approach to SYRK splits the $k$ loop into chunks so that the input matrix $\mY$ is partitioned into block rows. This way, we split the large SYRK operation into several smaller SYRK operations and sum the results together.

Due to the known zero areas in the input matrix $\mY$, the individual inner SYRK operations can be optimized. Each block row contains non-zeros only in the first few columns; the rest will not affect the output. We can, therefore, run the computation only with the left part of the block row, where the non-zeros are located. This way, each inner SYRK operation updates only a square submatrix of the output starting at the top left. The width of each block row is dictated by the right-most non-zero in the block row, i.e., the row trail in its last row.

In Figure~\ref{fig: syrk_input}, we visualize how the second block row of the input matrix is handled. Due to the zero pattern, we only need to consider the shaded submatrix $\mB$, on which we perform the SYRK operation and add the result to the block $\mathbf{BB}$ in the output.

\paragraph{Splitting of the output matrix:}
Output matrix splitting is depicted in Figure~\ref{fig: syrk_output}.
With this approach, we split the computation along the $m$ and $n$ dimensions; but primarily the $m$ dimension, that is, we split the output matrix into block rows and compute each of them separately. In each block row, the diagonal block is computed using an inner SYRK call, whose input is the corresponding block column of the input matrix. The remaining part of the block row located to the left of the diagonal block is then computed using GEMM.

Again, we can save some computation due to the zeros in the input matrix. The $k$ size in inner SYRK and GEMM kernels can be reduced to match the highest column pivot in the input block column above the output diagonal block.

For example, in Figure~\ref{fig: syrk_output}, the diagonal block $\mC \mC$ is computed using the block $\mC$ from the input matrix, and the block $\mC \mB$ is computed by multiplying $\mC^\top$ with $\mB$.

\section{Measurements}
\label{sec:measurement}

Speed-up of the implemented optimizations was measured for a heat transfer problem with varying sizes of subdomains. With decreasing the subdomain size, we increase their count (up to 2000) to have the total number of unknowns approximately constant, mainly to keep the runtime at a reasonable scale\footnote{Note that all GPU operations bring some overhead, and for small subdomains, where the operations themselves are quick, the overheads can dominate the runtime.}. That was around 8.4M unknowns for 2D and 1.1M for 3D problems.
We used a square or cube domain uniformly discretized into triangles or tetrahedra.

The measurements were performed on the GPU partition of the Karolina cluster at IT4Innovations~\cite{karolinadocs}.
Each compute node has two 64-core AMD EPYC 7763 processors with 1 TB of DDR4 memory and 8 NVIDIA A100 GPUs with 40 GB of HBM2 memory.
For most experiments,
we use only a single GPU and a single NUMA domain on the CPU; that is, 16 cores and 128 GB of memory
(it corresponds to 1/8 of available cores on a single node).
This configuration reflects production runs with fully utilized nodes since a single MPI process per NUMA domain with a single GPU usually delivers optimal performance.
Moreover, with this setting, one can scale the application to more MPI processes without influencing single-node performance.
GPU kernels were submitted using 16 CUDA streams, i.e., one stream per OpenMP thread.
The source code and experiment scripts are a part of the publicly available ESPRESO library and can be found in~\cite{espresogithub}.

Tests were performed with Intel MKL 2024.2.0~\cite{intelmkl} on the CPU and the legacy cuSPARSE API in the CUDA toolkit 11.7 on the GPU\footnote{
Despite Nvidia providing a new API, it has been shown that in our context, the performance of cuBLAS is the same, and cuSPARSE is better with legacy API~\cite{PDSEC}.
Hence, we omit the new API to provide graphs that are easier to read.}.
We used two sparse linear solver libraries on the CPU -- CHOLMOD~\cite{cholmod} from SuiteSparse 7.6.0 and PARDISO from Intel MKL.
Both libraries used Metis to reduce fill-in.
PARDISO provides the augmented incomplete factorization to assemble explicit $\tilde{\mF}_i$ via SC on the CPU.
It does not allow the extraction of factors, so it cannot be used for acceleration by the described algorithm.
CHOLMOD provides functions for extracting factors from the solver.

The tests are divided into several categories. We start with the optimal settings of particular routines and choosing the best variant. Then, we compare the speed-up of individual kernels and the speed-up in the context of the whole assembly of SC. 
Section~\ref{sec:overall_performance} then shows the improvement in the context of the FETI solver.

\subsection{Hyperparameters of the kernels}

This section recapitulates parameters found in~\cite{PDSEC} as optimal
and compares new hyperparameters of the optimized routines.

\paragraph{Format of the matrices} 
Since the idea and kernels are the same to be used by the original algorithm, we set parameters already determined in~\cite{PDSEC} equally and provide only recapitulation here without plotting them.
We call the sparse and dense TRSM with a factor in the CSR format and row-major ordering, respectively.
For RHS splitting, we always used the sparse factor. For factor splitting, TRSM and GEMM use the same setting following the recommendation for small matrices -- using factor blocks as sparse in 2D and converting them to dense in 3D. Note that in 3D, pruning played a significant role; without it, sparse GEMM was better.

\begin{figure}[ht]
        \centering
\input{tikz/partition_parameter/trsm_splitfactor.tex}
    \caption{Dependency of SC assembly time on partition parameter for a 3D problem on GPU, factor splitting was used.}
    \label{fig:trsm_partition_parameter}
\end{figure}
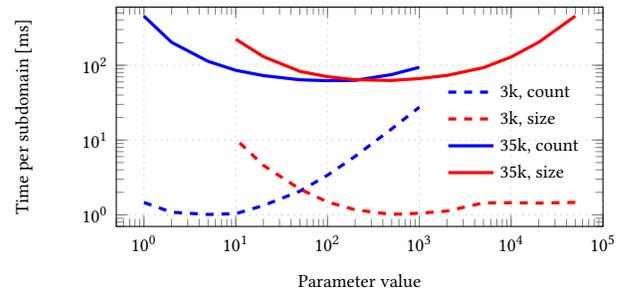

\begin{table}
    \centering
    \small
    \caption{Optimal splitting of the matrices.}
    \label{tab:splitting}
    \begin{tabular}{l|cc|cc}
    \toprule
    algorithm              & CPU 2D   & CPU 3D   & GPU 2D   & GPU 3D   \\
    \midrule
    TRSM, RHS splitting    & S 100    & S 100    & C 1      & S 1000   \\
    TRSM, factor splitting & S 200    & S 200    & S 1000   & S 500    \\
    SYRK, input splitting  & S 200    & C 50     & S 2000   & S 1000   \\
    SYRK, output splitting & S 200    & C 10     & S 200    & S 1000    \\
    \bottomrule
    \end{tabular}
\end{table}

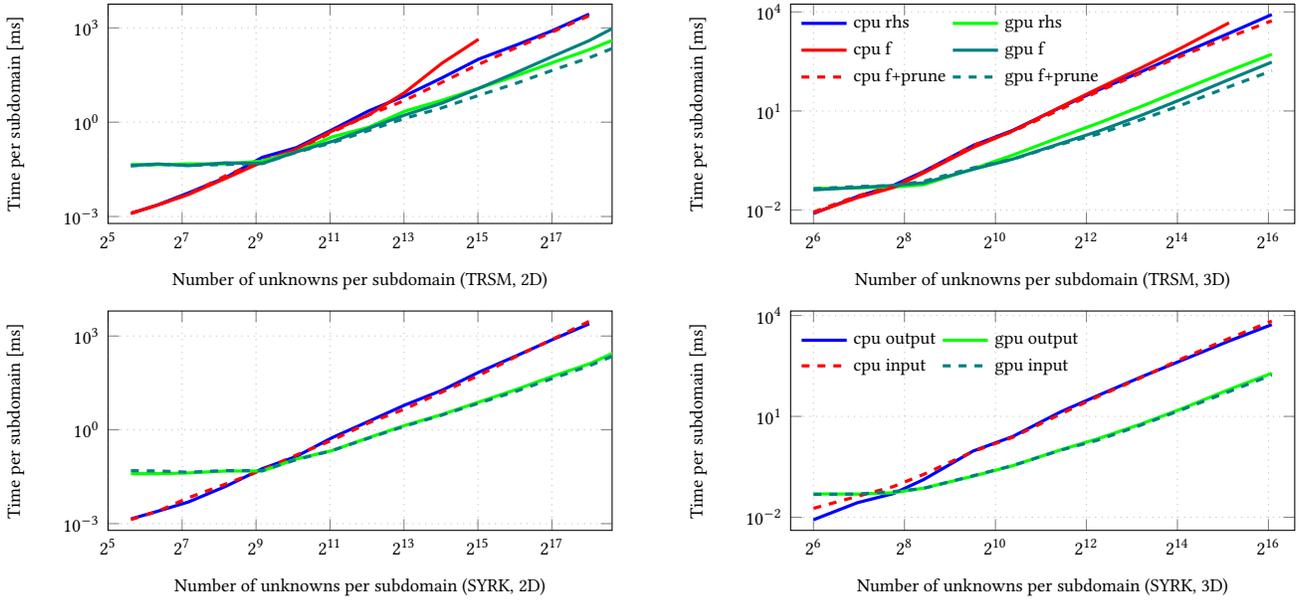
\begin{figure*}[ht]
    \begin{subfigure}[b]{0.49\textwidth}
        \centering
        \input{tikz/strategy_selection/heat_transfer-2D-TRIANGLE3-trsm.tex}
    \end{subfigure}
    \hfill
    \begin{subfigure}[b]{0.49\textwidth}
        \centering
        \input{tikz/strategy_selection/heat_transfer-3D-TETRA4-trsm.tex}
    \end{subfigure}
    
    \begin{subfigure}[b]{0.49\textwidth}
        \centering
        \input{tikz/strategy_selection/heat_transfer-2D-TRIANGLE3-herk.tex}
    \end{subfigure}
    \hfill
    \begin{subfigure}[b]{0.49\textwidth}
        \centering
        \input{tikz/strategy_selection/heat_transfer-3D-TETRA4-herk.tex}
    \end{subfigure}

    \caption{Comparison of TRSM (top) and SYRK (bottom) splitting variants.}
    \label{fig:best_strategy}
\end{figure*}

\paragraph{Splitting the matrices into blocks}
As described in Sections~\ref{sec: alg_trsm} and~\ref{sec: alg_syrk}, we split the input and/or output matrices of the TRSM and SYRK routines into blocks.
Their size (or count) should be set to balance (a) the amount of work saved by omitting zeros and (b) the overhead of calling many small mathematical library kernels over a few large ones.
That is, even though setting the block size to 1 is optimal in terms of the pure number of floating-point operations (FLOPs) performed (because we avoid computation with all the zeros in the stepped-shape matrix), it is counterproductive because of large overheads and because the level 3 BLAS kernels work better with larger matrices, where they are more compute bound.

We always split the matrices into uniformly sized blocks\footnote{One can also split the matrices in a non-uniform way to minimize the theoretical number of FLOPs needed for a given number of blocks. It was tested without observable differences. Hence, here we discuss only this simple setting.}. We tested two partitioning approaches -- (a) fixing the number of blocks and (b) fixing the block size. The difference between them can be observed when varying the subdomain size. By fixing the number of blocks, their size increases with the size of the problem. When fixing the block size, their count increases with the problem size.

In Figure~\ref{fig:trsm_partition_parameter}, we show an example of how the SC assembly time depends on the particular splitting. Solid lines show the dependency for a large subdomain with 35,937 unknowns (35k); dashed lines correspond to a small subdomain with 2,744 unknowns (3k).

For the case shown in the graph, the optimum for fixing the block size (red lines) is around 500 unknowns, independent of the subdomain size.
When fixing the number of blocks (blue lines), the optimal number increases with the subdomain size.
If a subdomain size is divided by the optimal number of chunks, the result is similar to the optimal block size found by the first approach (e.g., $2744 / 5 = 548 \approx 500$). 
Hence, we use the block size parameter and set it to 500.
A similar analysis was performed for RHS splitting, optimized SYRK kernels, and CPU versions; the results are shown in Table~\ref{tab:splitting}, where C stands for block count, and S for block size.

\paragraph{Pruning}
The effect of pruning can be observed in the top two graphs in Figure~\ref{fig:best_strategy}, where factor splitting is tested with (cpu f+prune, gpu f+prune) and without (cpu f, gpu f) the pruning. 
The effect of pruning increases with the size of subdomains (factor).
For large subdomains, pruning always has a positive effect on performance.
In the case of 2D subdomains where sparse routines are always used, the sparsity can be utilized internally in the routine (as the input format defines the sparsity).
However, the positive impact on the performance signalizes poor optimization in CPU and GPU routines towards empty rows.
This can be caused by the natural expectation that used matrices do not contain them.
We also observed a degradation in performance when the factor is split into small blocks.
On the other hand, pruning compensates for the degradation, leading to better performance, especially on the GPU.

\subsection{Splitting variant of optimized kernels}

Figure~\ref{fig:best_strategy} is also used to compare the optimized TRSM and SYRK kernel variants.
We start with the TRSM kernel.
For small subdomains, the performance of both strategies is similar.
For large subdomains, factor splitting with pruning is preferred (due to the positive effect of the pruning, as discussed above). Without pruning, RHS splitting would be better. Hence, we conclude that factor splitting is optimal.

There are only minimal differences in the performance between the variants of the optimized SYRK routine.
In most cases, splitting the input matrix into block rows is preferred. For the CPU with 3D subdomains of size between 1,000 and 16,000 unknowns, output matrix splitting was consistently better.

\subsection{Speedup of TRSM and SYRK routines}

This section measures the speed-up of the individual TRSM and SYRK routines.
The hyperparameters and splitting variants were set according to the previous sections.
The measurement was performed using a single thread. We synchronized with the GPU before and after each TRSM and SYRK.
Hence, the pure routine times were measured (without any possible impact by another code). 

\begin{figure*}[t]
    \begin{subfigure}[b]{0.49\textwidth}
        \centering
        \input{tikz/pure_kernels/heat_transfer-2D-TRIANGLE3-trsm-time.tex}
    \end{subfigure}
    \hfill
    \begin{subfigure}[b]{0.49\textwidth}
        \centering
        \input{tikz/pure_kernels/heat_transfer-3D-TETRA4-trsm-time.tex}
    \end{subfigure}

    \begin{subfigure}[b]{0.49\textwidth}
        \centering
        \input{tikz/pure_kernels/heat_transfer-2D-TRIANGLE3-herk-time.tex}
    \end{subfigure}
    \hfill
    \begin{subfigure}[b]{0.49\textwidth}
        \centering
        \input{tikz/pure_kernels/heat_transfer-3D-TETRA4-herk-time.tex}
    \end{subfigure}

    \begin{subfigure}[b]{0.49\textwidth}
        \centering
        \input{tikz/pure_kernels/heat_transfer-2D-TRIANGLE3-spdp.tex}
    \end{subfigure}
    \hfill
    \begin{subfigure}[b]{0.49\textwidth}
        \centering
        \input{tikz/pure_kernels/heat_transfer-3D-TETRA4-spdp.tex}
    \end{subfigure}
    \caption{Time and speedup of pure TRSM and SYRK kernels.}
    \label{fig:pure_kernels}
\end{figure*}
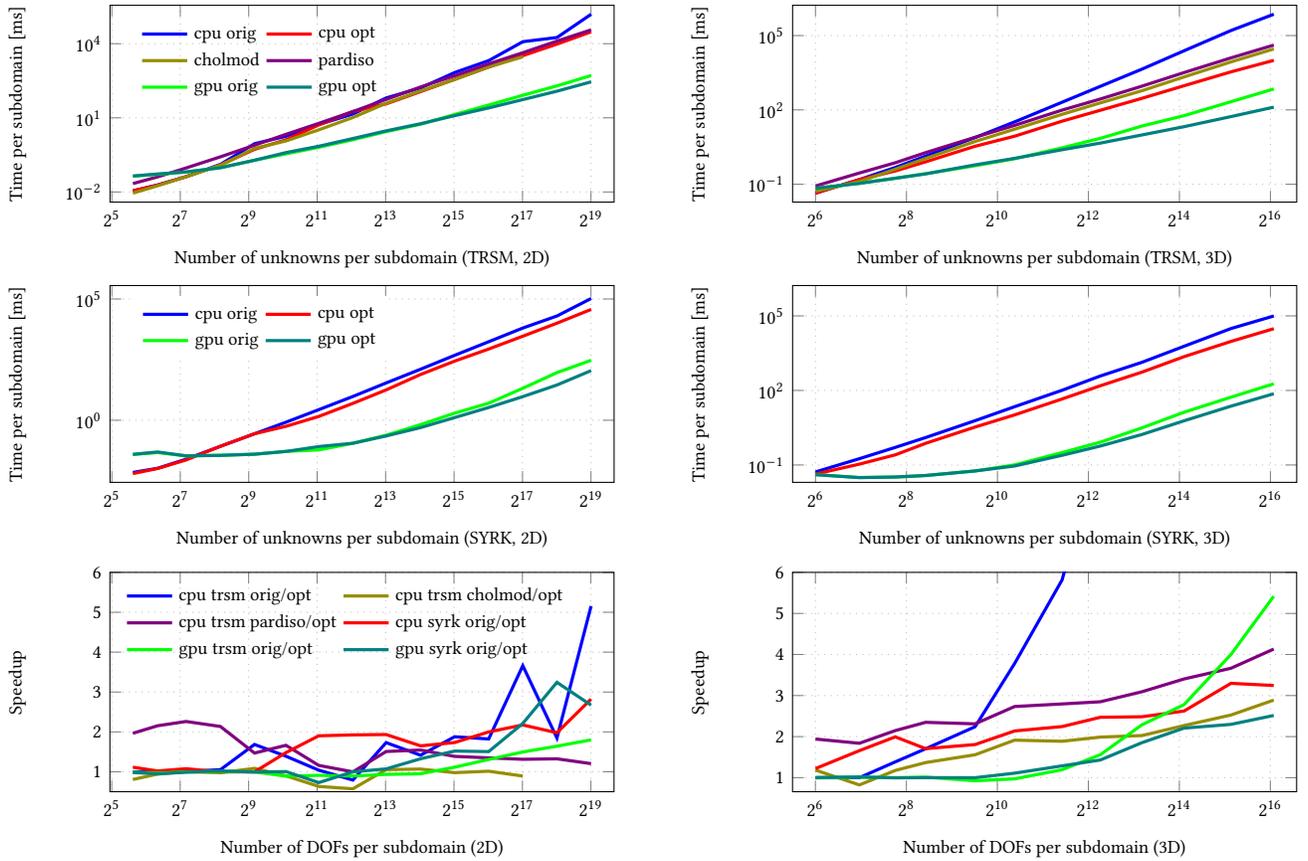

The optimized TRSM and SYRK routines are compared with the original approach (without utilizing the RHS sparsity) in Figure~\ref{fig:pure_kernels}.
The blue lines denote the original approach on the CPU, and the red lines denote our new approach.
On the CPU, we also compare TRSM with the forward substitution provided by PARDISO (violet line) and CHOLMOD (olive line) solvers.
Green and teal lines show the original and optimized approach on the GPU, respectively.
The same colors are used to show the performance of the SYRK routine (middle graphs).
The lower graphs show the speed-up achieved by the optimized version compared to all others.
As can be seen, the highest speed-up was achieved for the largest subdomains.
Such subdomains have factors with large zero subblocks that can be efficiently skipped in computation.
The speed-up of the TRSM routine is better for 3D subdomains.
They have a higher ratio between surface and volume than 2D subdomains; hence, more RHS to solve. In addition, they have a higher number of non-zeros in the factor.
Thus, their processing is generally slower, giving the optimizations a higher impact.
The SYRK routine's speed-up is determined by the overall number of non-zeros that can be omitted in the computation. It is similar for 2D and 3D subdomains, leading to similar speed-up.

In general, the observed speed-up is similar to a theoretical estimation with the assumption of a perfectly triangular RHS (the column pivot in the $i$-th column is located in row $ik/n$); we can obtain the theoretical speedup of the optimized dense TRSM and SYRK routines, which evaluates to 3 for both. We can think of it as a pyramid embedded within a prism with a square base. The ratio of the volume of the prism and the pyramid is 3. The parts of the object outside the pyramid but still within the prism are the computations we avoid by utilizing the non-zero pattern of the RHS matrix. For sparse TRSM, evaluating the theoretical speedup is significantly more complicated, as the number of FLOPs performed in the optimized case heavily depends on the exact non-zero pattern of the factor; hence, we do not provide it here.

An interesting observation is that the optimized version for 3D is faster than the forward substitution provided by PARDISO and CHOLMOD (olive and violet lines). This is caused by solving the full RHS matrix independently to sparsity\footnote{In fact, one can define sparsity of RHS in PARDISO. However, it must be specified during symbolic factorization. Hence, cannot be dynamically changed. CHOLMOD currently supports the sparsity for single RHS only, not for a matrix as in our case.}, i.e., solving a much larger system than the algorithm presented in this paper.

\subsection{Speed-up of the whole Explicit SC Assembly}

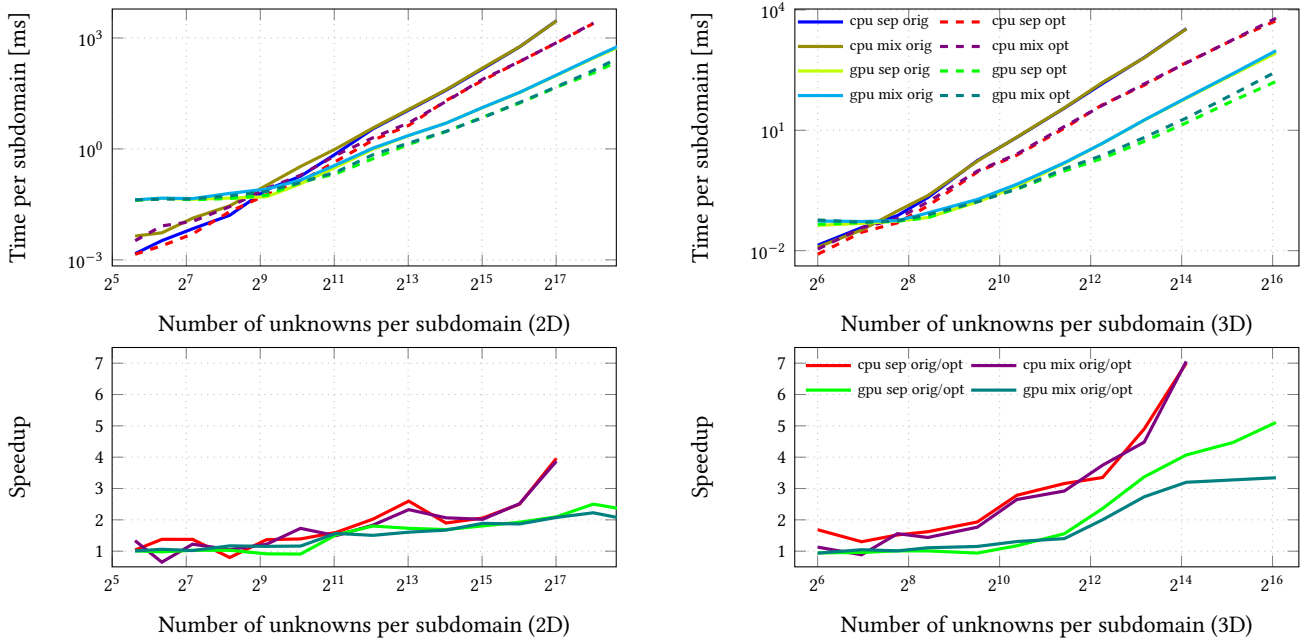
\begin{figure*}[ht]
    \begin{subfigure}[b]{0.49\textwidth}
        \centering
        \input{tikz/compare_to_orig/heat_transfer-2D-TRIANGLE3-time.tex}
    \end{subfigure}
    \hfill
    \begin{subfigure}[b]{0.49\textwidth}
        \centering
        \input{tikz/compare_to_orig/heat_transfer-3D-TETRA4-time.tex}
    \end{subfigure}

        \begin{subfigure}[b]{0.49\textwidth}
        \centering
        \input{tikz/compare_to_orig/heat_transfer-2D-TRIANGLE3-spdp.tex}
    \end{subfigure}
    \hfill
    \begin{subfigure}[b]{0.49\textwidth}
        \centering
        \input{tikz/compare_to_orig/heat_transfer-3D-TETRA4-spdp.tex}
    \end{subfigure}
    \caption{Time and speedup of the assembly of the dual operator.}
    \label{fig:spdpassembly}
\end{figure*}

This section compares the improvements achieved in the explicit assembly of the SC.
As described in Section~\ref{sec:feti}, the computation of SC is performed in the loop with numerical factorization.
We measured the SC assembly in two configurations. In the first configuration, 
we first perform numerical factorization to compute the factors $\mL_i$ for all subdomains. Then, we measure the operations used to assemble the SC using the already precomputed factors.
It differs from the measurements in the previous subsection by parallel processing of multiple subdomains concurrently, combining the TRSM and SYRK kernels, and it also includes the required permutation of both the columns of $\tilde{\mB}_i$ and the final SC $\tilde{\mF}_i$.
The second configuration measures the whole SC assembly, including the numerical factorization and extraction of factors from the sparse solver. Due to the asynchronicity of GPU operations, we achieve CPU-GPU computation overlap after the first batch of subdomains is factorized.

The performance of explicit evaluation of SC in both configurations is shown in Figure~\ref{fig:spdpassembly}.
The upper graphs show the performance of the first configuration (sep, the separated factorization) and the second configuration (mix, both factorization and SC mixed together). The lower graphs show the achieved speedup. 
On the CPU, the speed-up for both configurations is similar since they perform the same operations, and their order is irrelevant. The speed-up of the approach utilizing the sparsity is up to 4 and 7 for 2D and 3D subdomains, respectively.

On the GPU, for large subdomains, the speed-up of the first configuration is higher than the speed-up of the second configuration.
This is partly caused by the delayed start of GPU computations, which can start only after the factorization of the first subdomains finishes.
For 3D problems, without factorization (the first configuration), we achieved speedup in the GPU code of up to 5.1. 
If we include the factorization (second configuration), the speedup is up to 3.3. For 2D problems, we achieved a speedup above 2.

\section{Overall Performance of FETI Solver}
\label{sec:overall_performance}

\begin{table}
    \centering
    \small
    \caption{Tested approaches for the dual operator.}
    \label{tab:dual_approaches}
    \begin{tabular}{ll}
    \toprule
    approach           & description           \\
    \midrule
    impl\_mkl          & the MKL PARDISO solver on CPU \\
    impl\_cholmod      & the CHOLMOD solver on CPU\\
    \midrule
    expl\_mkl          & aug. incomplete fact. from MKL PARDISO on CPU \\
    expl\_cholmod      & TRSM with the CHOLMOD solver on CPU \\
    expl\_cuda         & CUDA with factors from CHOLMOD \\
    expl\_cpu\_opt     & optimized TRSM and SYRK on CPU \\
    expl\_gpu\_opt     & optimized TRSM and SYRK on GPU \\
    \midrule
    expl\_hybrid       & assembly expl\_mkl, application GPU \\
    \bottomrule
    \end{tabular}
\end{table}

This section describes the impact of the improvements in the context of the FETI solver. That is a comparison with other implicit and explicit dual-operator approaches. The tested approaches are summarized in Table~\ref{tab:dual_approaches}. 
Best-known approaches for explicitly evaluating SC on the CPU are \textit{expl\_mkl} and \textit{expl\_hybrid} using incomplete augmented factorization from PARDISO.
The previously best-known approach using the GPU is \textit{expl\_cuda}.
It is the implementation of the algorithm in~\cite{PDSEC}.
Its optimized version for sparse matrices (described by this paper) is \textit{expl\_gpu\_opt}.

\begin{figure*}[ht]
    \begin{subfigure}[b]{0.47\textwidth}
        \centering
        \input{tikz/update-2D}
    \end{subfigure}
    \hfill
    \begin{subfigure}[b]{0.47\textwidth}
        \centering
        \input{tikz/update-3D}
    \end{subfigure}
    \caption{Comparison of preprocessing time between multiple dual operator approaches.}
    \label{fig:compare_assembly}
\end{figure*}
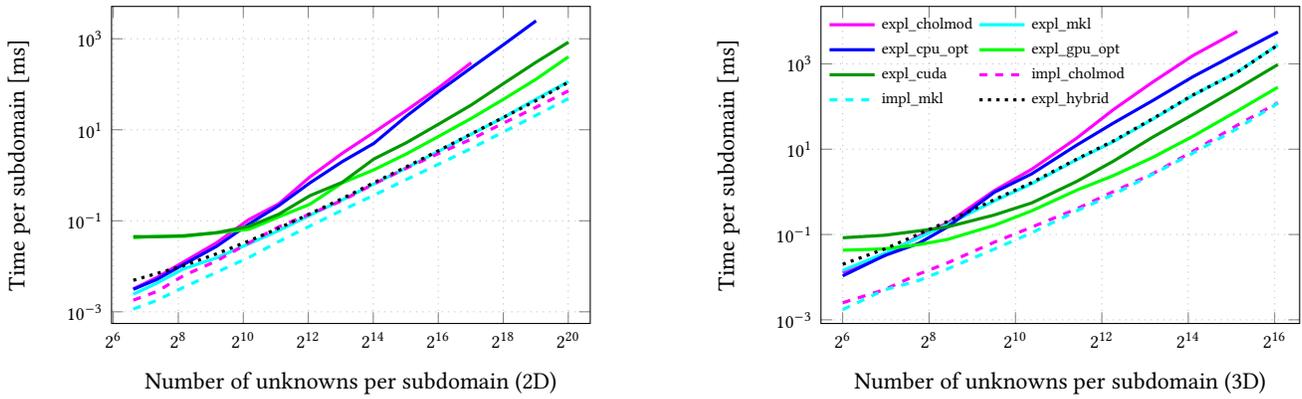

Figure~\ref{fig:compare_assembly} compares different FETI preprocessing on 2D and 3D heat transfer examples. 
The fastest variants are implicit approaches that contain only numerical factorization. Their disadvantage is the slower dual operator (SC) application in the iterative solver. They are included in the figure to show the slow-down of the explicit approaches discussed in this paper.
The fastest factorization is provided by the PARDISO from the MKL library. It is significantly faster for 2D subdomains. For the large 3D subdomains, the performance of the Cholmod and PARDISO is similar.
Since only Cholmod allows extraction of factors, \textit{impl\_cholmod} is the baseline for CUDA-based explicit approaches.
If PARDISO allowed the extraction of factors, it would be possible to speed up all CUDA-based approaches where PARDISO is faster than Cholmod.

Despite our improvements, augmented incomplete factorization from PARDISO is still the fastest way to assemble SC for 2D subdomains explicitly. It is caused by the high sparsity of the factors and a relatively small number of RHS. Hence, the augmented matrix $[ K, B^\top; B, O ]$ is of similar size to the original matrix $K$, making the incomplete factorization very fast. For some subdomain sizes, \textit{expl\_mkl} is faster than \textit{impl\_cholmod}, i.e., explicit assembly by PARDISO is faster than numerical factorization by Cholmod. That indicates the poor performance of Cholmod for 2D subdomains.

Omitting very small 3D subdomains with high CUDA kernel launch overhead, \textit{expl\_gpu\_opt} is the fastest explicit approach.
Our new approach is up to 9.8x faster than \textit{expl\_mkl}, and explicit preprocessing on large subdomains is only 2.3x slower than implicit preprocessing.
Optimizations toward sparsity efficiently decrease the computational time.
In addition, the relative slow-down decreases with the subdomain size.
That is highly important since the speed-up of the application of SC in iterative solver increases with increasing subdomain size,
making this acceleration approach very effective and beneficial early.

\begin{figure*}[ht]
    \begin{subfigure}[b]{0.47\textwidth}
        \centering
        \input{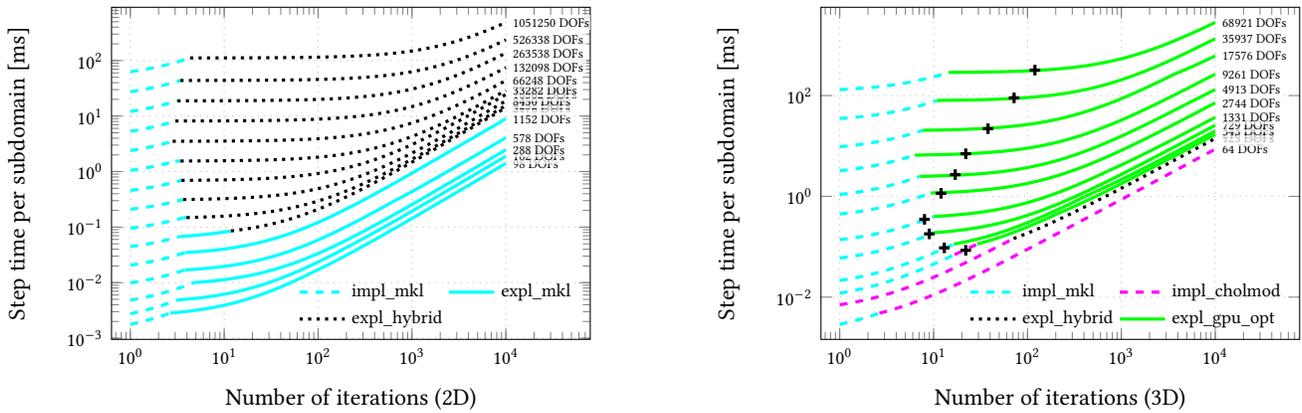}
    \end{subfigure}
    \hfill
    \begin{subfigure}[b]{0.47\textwidth}
        \centering
        \input{tikz/amort-3D}
    \end{subfigure}
    \caption{Overall time spent in the single run of the FETI solver w.r.t. dual operator.}
    \label{fig:compare_amort}
\end{figure*}

The impact on the amortization points is depicted in Figure~\ref{fig:compare_amort}. It shows the overall time spent in the FETI dual operator for several subdomain sizes and a given number of iterations (determined by the type of solver problem and conditional number).
The line style denotes the best approach for a given setting. The amortization point lies where the dashed (implicit CPU) line transitions into the solid (explicit GPU) line (or dotted lines in the case of 2D problems).
The amortization points reported in~\cite{PDSEC} are for 3D subdomains depicted as black plus signs. As we can observe, the newly implemented optimizations significantly decrease the amortization points for large 3D subdomains.
Now, their amortization points are at about 10 iterations for subdomain sizes ranging from 1,000 to 70,000 unknowns.
As problems that can be solved in such a small number of iterations are rare, the acceleration benefits a wide spectrum of problems.

\section{Conclusion}

The paper presents several approaches for using sparsity in the explicit computation of Schur complement matrices in a FETI solver.
With appropriate settings, the effectivity of the presented optimizations increases with the subdomain sizes.
It is in conjunction with the increasing speed up of an iterative solver where the explicitly assembled SC is used.

Using the Karolina GPU node with two 64-core AMD Zen3 CPUs and 8 Nvidia A100 GPUs, we achieved a speed-up of up to 5.1 in the GPU-only code and a speed-up of up to 3.3 in the context of the whole FETI solver preprocessing for large 3D subdomains compared to previously the best GPU approach. 

With the optimized algorithm, amortization points are about 10 for 3D subdomains, with the size ranging from 1,000 unknowns to 70,000 unknowns.
With amortization points equal for such a wide range of subdomain sizes, one can set the subdomain sizes to be optimal from the FETI scalability point of view, i.e., to set the optimal ratio between subdomain sizes and their overall number as was shown in~\cite{espreso-pasc}.
It allows us to compute the solution as fast as possible, utilizing GPU nodes efficiently and scaling to the required number of nodes.

Despite the approach being tested for the FETI solver, it can be successfully used in other methods where SC of the form of $\mB\mK^{-1}\mB^\top$ are computed, and matrix $\mB^\top$ can be reordered into approximately lower triangular form. The approach can also benefit from the sparsity of matrix $\mK$.

Since the algorithm requires the basic mathematical routines available on most hardware platforms, it can be easily adopted by other tools or implemented to support other GPU accelerators from AMD and Intel.

\clearpage

\bibliographystyle{unsrt}
\bibliography{references}

\clearpage

\end{document}

%% file: tikz/partition_parameter/trsm_splitfactor.tex
\begin{tikzpicture}
  \begin{loglogaxis}[
    width=.95\columnwidth,
    height=4.5cm,
    log basis x=10,
    log basis y=10,
    xlabel={Parameter value},
    ylabel={Time per subdomain [ms]},
    xmajorgrids=true,
    ymajorgrids=true,
    every axis plot/.append style={line width=1.2pt},
    grid style=dotted,
    xmin=0.5,
    xmax=100000,
    ymin=0.7,
    ymax=600,
    label style={font=\footnotesize},
    tick label style={font=\footnotesize},
    legend style={font=\footnotesize, draw=none, fill=none, at={(0.82, .7)}, anchor=north},
    legend cell align={left}
  ]
    \addplot[dashed, color={blue}, mark=none] coordinates {
        ( 1, 1.463808594 )
        ( 2, 1.085082031 )
        ( 5, 1.012361328 )
        ( 10, 1.036791016 )
        ( 20, 1.331589844 )
        ( 50, 2.07359375 )
        ( 100, 3.391873047 )
        ( 200, 6.059601563 )
        ( 500, 14.11987305 )
        ( 1000, 27.55772461 )
    };
    \addplot[dashed, color={red}, mark=none] coordinates {
        ( 50000, 1.467216797 )
        ( 20000, 1.437396484 )
        ( 10000, 1.451853516 )
        ( 5000, 1.445638672 )
        ( 2000, 1.124185547 )
        ( 1000, 1.050673828 )
        ( 500, 1.014994141 )
        ( 200, 1.167269531 )
        ( 100, 1.487205078 )
        ( 50, 2.209876953 )
        ( 20, 4.594392578 )
        ( 10, 10.30415039 )
    };

    \addplot[solid, color={blue}, mark=none] coordinates {
        ( 1, 454.9992188 )
        ( 2, 202.8679375 )
        ( 5, 112.847875 )
        ( 10, 85.63596875 )
        ( 20, 72.7559375 )
        ( 50, 63.95021875 )
        ( 100, 62.29115625 )
        ( 200, 62.85396875 )
        ( 500, 75.13875 )
        ( 1000, 94.018375 )
    };
    \addplot[solid, color={red}, mark=none] coordinates {
        ( 50000, 455.9356563 )
        ( 20000, 203.4772188 )
        ( 10000, 129.105375 )
        ( 5000, 92.93234375 )
        ( 2000, 73.2176875 )
        ( 1000, 66.4011875 )
        ( 500, 62.665375 )
        ( 200, 64.02975 )
        ( 100, 70.34765625 )
        ( 50, 82.93546875 )
        ( 20, 130.6307813 )
        ( 10, 223.1800625 )
    };
    \addlegendentry{3k, count}
    \addlegendentry{3k, size}
    \addlegendentry{35k, count}
    \addlegendentry{35k, size}
  \end{loglogaxis}
\end{tikzpicture}

%% file: tikz/strategy_selection/heat_transfer-2D-TRIANGLE3-trsm.tex
\begin{tikzpicture}
  \begin{loglogaxis}[
    width=.95\columnwidth,
    height=4.5cm,
    log basis x=2,
    log basis y=10,
    xlabel={Number of unknowns per subdomain (TRSM, 2D)},
    ylabel={Time per subdomain [ms]},
    xmajorgrids=true,
    ymajorgrids=true,
    every axis plot/.append style={line width=1.2pt},
    grid style=dotted,
    xmin=3.189e+01,
    xmax=4.043e+05,
    ymin=5.992e-04,
    ymax=5.677e+03,
    label style={font=\footnotesize},
    tick label style={font=\footnotesize},
    legend style={font=\footnotesize, draw=none, fill=none, at={(0.01,0.9)}, anchor=north},
    legend columns=2, legend cell align={left}
  ]
    \addplot[solid, color=blue, mark=none] coordinates {
      ( 4.900e+01 , 1.261e-03 )
      ( 8.100e+01 , 2.338e-03 )
      ( 1.440e+02 , 5.630e-03 )
      ( 2.890e+02 , 1.727e-02 )
      ( 5.760e+02 , 7.645e-02 )
      ( 1.089e+03 , 1.547e-01 )
      ( 2.116e+03 , 5.670e-01 )
      ( 4.225e+03 , 2.283e+00 )
      ( 8.281e+03 , 6.836e+00 )
      ( 1.664e+04 , 2.550e+01 )
      ( 3.312e+04 , 1.008e+02 )
      ( 6.605e+04 , 2.777e+02 )
      ( 1.318e+05 , 8.100e+02 )
      ( 2.632e+05 , 2.735e+03 )
      ( 5.256e+05 , nan )
    };
    \addplot[solid, color=red, mark=none] coordinates {
      ( 4.900e+01 , 1.244e-03 )
      ( 8.100e+01 , 2.304e-03 )
      ( 1.440e+02 , 4.992e-03 )
      ( 2.890e+02 , 1.624e-02 )
      ( 5.760e+02 , 5.438e-02 )
      ( 1.089e+03 , 1.392e-01 )
      ( 2.116e+03 , 5.170e-01 )
      ( 4.225e+03 , 1.698e+00 )
      ( 8.281e+03 , 8.710e+00 )
      ( 1.664e+04 , 7.459e+01 )
      ( 3.312e+04 , 4.329e+02 )
      ( 6.605e+04 , nan )
      ( 1.318e+05 , nan )
      ( 2.632e+05 , nan )
      ( 5.256e+05 , nan )
    };
    \addplot[dashed, color=red, mark=none] coordinates {
      ( 4.900e+01 , 1.274e-03 )
      ( 8.100e+01 , 2.307e-03 )
      ( 1.440e+02 , 5.088e-03 )
      ( 2.890e+02 , 1.971e-02 )
      ( 5.760e+02 , 6.117e-02 )
      ( 1.089e+03 , 1.295e-01 )
      ( 2.116e+03 , 4.729e-01 )
      ( 4.225e+03 , 1.700e+00 )
      ( 8.281e+03 , 4.890e+00 )
      ( 1.664e+04 , 1.846e+01 )
      ( 3.312e+04 , 6.899e+01 )
      ( 6.605e+04 , 2.293e+02 )
      ( 1.318e+05 , 7.367e+02 )
      ( 2.632e+05 , 2.432e+03 )
      ( 5.256e+05 , nan )
    };
    \addplot[solid, color=green, mark=none] coordinates {
      ( 4.900e+01 , 4.488e-02 )
      ( 8.100e+01 , 4.447e-02 )
      ( 1.440e+02 , 4.710e-02 )
      ( 2.890e+02 , 4.678e-02 )
      ( 5.760e+02 , 5.793e-02 )
      ( 1.089e+03 , 1.130e-01 )
      ( 2.116e+03 , 3.377e-01 )
      ( 4.225e+03 , 6.871e-01 )
      ( 8.281e+03 , 2.223e+00 )
      ( 1.664e+04 , 4.923e+00 )
      ( 3.312e+04 , 1.188e+01 )
      ( 6.605e+04 , 2.932e+01 )
      ( 1.318e+05 , 7.776e+01 )
      ( 2.632e+05 , 2.023e+02 )
      ( 5.256e+05 , 5.971e+02 )
    };
    \addplot[solid, color=teal, mark=none] coordinates {
      ( 4.900e+01 , 4.150e-02 )
      ( 8.100e+01 , 4.691e-02 )
      ( 1.440e+02 , 4.135e-02 )
      ( 2.890e+02 , 5.089e-02 )
      ( 5.760e+02 , 4.834e-02 )
      ( 1.089e+03 , 1.136e-01 )
      ( 2.116e+03 , 2.419e-01 )
      ( 4.225e+03 , 6.350e-01 )
      ( 8.281e+03 , 1.681e+00 )
      ( 1.664e+04 , 4.015e+00 )
      ( 3.312e+04 , 1.186e+01 )
      ( 6.605e+04 , 3.695e+01 )
      ( 1.318e+05 , 1.219e+02 )
      ( 2.632e+05 , 3.919e+02 )
      ( 5.256e+05 , 1.607e+03 )
    };
    \addplot[dashed, color=teal, mark=none] coordinates {
      ( 4.900e+01 , 4.080e-02 )
      ( 8.100e+01 , 4.583e-02 )
      ( 1.440e+02 , 4.219e-02 )
      ( 2.890e+02 , 4.491e-02 )
      ( 5.760e+02 , 4.907e-02 )
      ( 1.089e+03 , 1.106e-01 )
      ( 2.116e+03 , 2.193e-01 )
      ( 4.225e+03 , 5.440e-01 )
      ( 8.281e+03 , 1.317e+00 )
      ( 1.664e+04 , 2.851e+00 )
      ( 3.312e+04 , 7.075e+00 )
      ( 6.605e+04 , 1.691e+01 )
      ( 1.318e+05 , 4.446e+01 )
      ( 2.632e+05 , 1.115e+02 )
      ( 5.256e+05 , 3.317e+02 )
    };
  \end{loglogaxis}
\end{tikzpicture}

%% file: tikz/strategy_selection/heat_transfer-3D-TETRA4-trsm.tex
\begin{tikzpicture}
  \begin{loglogaxis}[
    width=.95\columnwidth,
    height=4.5cm,
    log basis x=2,
    log basis y=10,
    xlabel={Number of unknowns per subdomain (TRSM, 3D)},
    ylabel={Time per subdomain [ms]},
    xmajorgrids=true,
    ymajorgrids=true,
    every axis plot/.append style={line width=1.2pt},
    grid style=dotted,
    xmin=4.514e+01,
    xmax=9.771e+04,
    ymin=3.926e-03,
    ymax=1.719e+04,
    label style={font=\footnotesize},
    tick label style={font=\footnotesize},
    legend style={font=\footnotesize, draw=none, fill=none, at={(0.32,1)}, anchor=north},
    legend columns=2, legend cell align={left}
  ]
    \addplot[solid, color=blue, mark=none] coordinates {
      ( 6.400e+01 , 7.868e-03 )
      ( 1.250e+02 , 2.581e-02 )
      ( 2.160e+02 , 5.583e-02 )
      ( 3.430e+02 , 1.477e-01 )
      ( 7.290e+02 , 9.033e-01 )
      ( 1.331e+03 , 2.595e+00 )
      ( 2.744e+03 , 1.269e+01 )
      ( 4.913e+03 , 4.326e+01 )
      ( 9.261e+03 , 1.489e+02 )
      ( 1.758e+04 , 5.579e+02 )
      ( 3.594e+04 , 2.250e+03 )
      ( 6.892e+04 , 8.318e+03 )
    };
    \addplot[solid, color=green, mark=none] coordinates {
      ( 6.400e+01 , 4.514e-02 )
      ( 1.250e+02 , 4.718e-02 )
      ( 2.160e+02 , 4.976e-02 )
      ( 3.430e+02 , 5.886e-02 )
      ( 7.290e+02 , 1.772e-01 )
      ( 1.331e+03 , 4.602e-01 )
      ( 2.744e+03 , 1.619e+00 )
      ( 4.913e+03 , 4.300e+00 )
      ( 9.261e+03 , 1.329e+01 )
      ( 1.758e+04 , 4.378e+01 )
      ( 3.594e+04 , 1.639e+02 )
      ( 6.892e+04 , 5.214e+02 )
    };
    \addplot[solid, color=red, mark=none] coordinates {
      ( 6.400e+01 , 8.129e-03 )
      ( 1.250e+02 , 2.367e-02 )
      ( 2.160e+02 , 4.806e-02 )
      ( 3.430e+02 , 1.334e-01 )
      ( 7.290e+02 , 7.876e-01 )
      ( 1.331e+03 , 2.480e+00 )
      ( 2.744e+03 , 1.311e+01 )
      ( 4.913e+03 , 4.818e+01 )
      ( 9.261e+03 , 1.953e+02 )
      ( 1.758e+04 , 8.377e+02 )
      ( 3.594e+04 , 4.666e+03 )
      ( 6.892e+04 , nan )
    };
    \addplot[solid, color=teal, mark=none] coordinates {
      ( 6.400e+01 , 4.085e-02 )
      ( 1.250e+02 , 4.806e-02 )
      ( 2.160e+02 , 5.571e-02 )
      ( 3.430e+02 , 6.677e-02 )
      ( 7.290e+02 , 1.714e-01 )
      ( 1.331e+03 , 3.489e-01 )
      ( 2.744e+03 , 1.030e+00 )
      ( 4.913e+03 , 2.446e+00 )
      ( 9.261e+03 , 7.126e+00 )
      ( 1.758e+04 , 2.297e+01 )
      ( 3.594e+04 , 8.932e+01 )
      ( 6.892e+04 , 3.012e+02 )
    };
    \addplot[dashed, color=red, mark=none] coordinates {
      ( 6.400e+01 , 8.714e-03 )
      ( 1.250e+02 , 2.690e-02 )
      ( 2.160e+02 , 5.073e-02 )
      ( 3.430e+02 , 1.422e-01 )
      ( 7.290e+02 , 9.143e-01 )
      ( 1.331e+03 , 2.390e+00 )
      ( 2.744e+03 , 1.113e+01 )
      ( 4.913e+03 , 3.942e+01 )
      ( 9.261e+03 , 1.368e+02 )
      ( 1.758e+04 , 4.730e+02 )
      ( 3.594e+04 , 1.729e+03 )
      ( 6.892e+04 , 5.365e+03 )
    };
    \addplot[dashed, color=teal, mark=none] coordinates {
      ( 6.400e+01 , 4.408e-02 )
      ( 1.250e+02 , 5.167e-02 )
      ( 2.160e+02 , 5.527e-02 )
      ( 3.430e+02 , 7.448e-02 )
      ( 7.290e+02 , 1.897e-01 )
      ( 1.331e+03 , 3.476e-01 )
      ( 2.744e+03 , 9.731e-01 )
      ( 4.913e+03 , 2.025e+00 )
      ( 9.261e+03 , 5.307e+00 )
      ( 1.758e+04 , 1.540e+01 )
      ( 3.594e+04 , 5.411e+01 )
      ( 6.892e+04 , 1.644e+02 )
    };
    
    \addlegendentry{cpu rhs}
    \addlegendentry{gpu rhs}
    \addlegendentry{cpu f}
    \addlegendentry{gpu f}
    \addlegendentry{cpu f+prune}
    \addlegendentry{gpu f+prune}
  \end{loglogaxis}
\end{tikzpicture}

%% file: tikz/strategy_selection/heat_transfer-2D-TRIANGLE3-herk.tex
\begin{tikzpicture}
  \begin{loglogaxis}[
    width=.95\columnwidth,
    height=4.5cm,
    log basis x=2,
    log basis y=10,
    xlabel={Number of unknowns per subdomain (SYRK, 2D)},
    ylabel={Time per subdomain [ms]},
    xmajorgrids=true,
    ymajorgrids=true,
    every axis plot/.append style={line width=1.2pt},
    grid style=dotted,
    xmin=3.189e+01,
    xmax=4.043e+05,
    ymin=6.097e-04,
    ymax=6.309e+03,
    label style={font=\footnotesize},
    tick label style={font=\footnotesize},
    legend style={font=\footnotesize, draw=none, at={(0.5,-0.1)}, anchor=north},
    legend cell align={left}
  ]
    \addplot[solid, color=blue, mark=none] coordinates {
      ( 4.900e+01 , 1.414e-03 )
      ( 8.100e+01 , 2.495e-03 )
      ( 1.440e+02 , 4.924e-03 )
      ( 2.890e+02 , 1.551e-02 )
      ( 5.760e+02 , 5.746e-02 )
      ( 1.089e+03 , 1.409e-01 )
      ( 2.116e+03 , 5.552e-01 )
      ( 4.225e+03 , 1.869e+00 )
      ( 8.281e+03 , 6.158e+00 )
      ( 1.664e+04 , 1.831e+01 )
      ( 3.312e+04 , 6.910e+01 )
      ( 6.605e+04 , 2.257e+02 )
      ( 1.318e+05 , 7.629e+02 )
      ( 2.632e+05 , 2.459e+03 )
      ( 5.256e+05 , nan )
    };
    \addplot[dashed, color=red, mark=none] coordinates {
      ( 4.900e+01 , 1.351e-03 )
      ( 8.100e+01 , 2.478e-03 )
      ( 1.440e+02 , 6.738e-03 )
      ( 2.890e+02 , 1.841e-02 )
      ( 5.760e+02 , 5.204e-02 )
      ( 1.089e+03 , 1.561e-01 )
      ( 2.116e+03 , 4.637e-01 )
      ( 4.225e+03 , 1.668e+00 )
      ( 8.281e+03 , 4.607e+00 )
      ( 1.664e+04 , 1.605e+01 )
      ( 3.312e+04 , 5.423e+01 )
      ( 6.605e+04 , 2.151e+02 )
      ( 1.318e+05 , 7.830e+02 )
      ( 2.632e+05 , 2.952e+03 )
      ( 5.256e+05 , nan )
    };

    \addplot[solid, color=green, mark=none] coordinates {
      ( 4.900e+01 , 3.945e-02 )
      ( 8.100e+01 , 3.976e-02 )
      ( 1.440e+02 , 4.176e-02 )
      ( 2.890e+02 , 4.822e-02 )
      ( 5.760e+02 , 4.829e-02 )
      ( 1.089e+03 , 1.170e-01 )
      ( 2.116e+03 , 2.156e-01 )
      ( 4.225e+03 , 5.620e-01 )
      ( 8.281e+03 , 1.372e+00 )
      ( 1.664e+04 , 3.077e+00 )
      ( 3.312e+04 , 7.681e+00 )
      ( 6.605e+04 , 1.907e+01 )
      ( 1.318e+05 , 5.104e+01 )
      ( 2.632e+05 , 1.283e+02 )
      ( 5.256e+05 , 4.317e+02 )
    };
    \addplot[dashed, color=teal, mark=none] coordinates {
      ( 4.900e+01 , 4.935e-02 )
      ( 8.100e+01 , 4.849e-02 )
      ( 1.440e+02 , 4.408e-02 )
      ( 2.890e+02 , 4.988e-02 )
      ( 5.760e+02 , 4.954e-02 )
      ( 1.089e+03 , 1.110e-01 )
      ( 2.116e+03 , 2.161e-01 )
      ( 4.225e+03 , 5.425e-01 )
      ( 8.281e+03 , 1.308e+00 )
      ( 1.664e+04 , 2.955e+00 )
      ( 3.312e+04 , 7.092e+00 )
      ( 6.605e+04 , 1.680e+01 )
      ( 1.318e+05 , 4.357e+01 )
      ( 2.632e+05 , 1.110e+02 )
      ( 5.256e+05 , 3.439e+02 )
    };
  \end{loglogaxis}
\end{tikzpicture}

%% file: tikz/strategy_selection/heat_transfer-3D-TETRA4-herk.tex
\begin{tikzpicture}
  \begin{loglogaxis}[
    width=.95\columnwidth,
    height=4.5cm,
    log basis x=2,
    log basis y=10,
    xlabel={Number of unknowns per subdomain (SYRK, 3D)},
    ylabel={Time per subdomain [ms]},
    xmajorgrids=true,
    ymajorgrids=true,
    every axis plot/.append style={line width=1.2pt},
    grid style=dotted,
    xmin=4.514e+01,
    xmax=9.771e+04,
    ymin=4.076e-03,
    ymax=1.364e+04,
    label style={font=\footnotesize},
    tick label style={font=\footnotesize},
    legend style={font=\footnotesize, draw=none, fill=none, at={(0.3,0.95)}, anchor=north},
    legend columns=2,legend cell align={left}
  ]
    \addplot[solid, color=blue, mark=none] coordinates {
      ( 6.400e+01 , 8.426e-03 )
      ( 1.250e+02 , 2.717e-02 )
      ( 2.160e+02 , 5.051e-02 )
      ( 3.430e+02 , 1.324e-01 )
      ( 7.290e+02 , 9.336e-01 )
      ( 1.331e+03 , 2.515e+00 )
      ( 2.744e+03 , 1.347e+01 )
      ( 4.913e+03 , 4.243e+01 )
      ( 9.261e+03 , 1.442e+02 )
      ( 1.758e+04 , 4.681e+02 )
      ( 3.594e+04 , 1.745e+03 )
      ( 6.892e+04 , 5.378e+03 )
    };
    \addplot[solid, color=green, mark=none] coordinates {
      ( 6.400e+01 , 4.963e-02 )
      ( 1.250e+02 , 4.871e-02 )
      ( 2.160e+02 , 5.363e-02 )
      ( 3.430e+02 , 7.329e-02 )
      ( 7.290e+02 , 1.669e-01 )
      ( 1.331e+03 , 3.413e-01 )
      ( 2.744e+03 , 9.990e-01 )
      ( 4.913e+03 , 2.194e+00 )
      ( 9.261e+03 , 5.780e+00 )
      ( 1.758e+04 , 1.706e+01 )
      ( 3.594e+04 , 6.296e+01 )
      ( 6.892e+04 , 1.920e+02 )
    };
    
    \addplot[dashed, color=red, mark=none] coordinates {
      ( 6.400e+01 , 1.838e-02 )
      ( 1.250e+02 , 4.207e-02 )
      ( 2.160e+02 , 8.176e-02 )
      ( 3.430e+02 , 1.892e-01 )
      ( 7.290e+02 , 9.010e-01 )
      ( 1.331e+03 , 2.385e+00 )
      ( 2.744e+03 , 1.162e+01 )
      ( 4.913e+03 , 3.957e+01 )
      ( 9.261e+03 , 1.378e+02 )
      ( 1.758e+04 , 5.236e+02 )
      ( 3.594e+04 , 2.082e+03 )
      ( 6.892e+04 , 6.837e+03 )
    };

    \addplot[dashed, color=teal, mark=none] coordinates {
      ( 6.400e+01 , 4.774e-02 )
      ( 1.250e+02 , 4.908e-02 )
      ( 2.160e+02 , 5.789e-02 )
      ( 3.430e+02 , 7.238e-02 )
      ( 7.290e+02 , 1.717e-01 )
      ( 1.331e+03 , 3.441e-01 )
      ( 2.744e+03 , 9.798e-01 )
      ( 4.913e+03 , 1.993e+00 )
      ( 9.261e+03 , 5.376e+00 )
      ( 1.758e+04 , 1.593e+01 )
      ( 3.594e+04 , 5.456e+01 )
      ( 6.892e+04 , 1.693e+02 )
    };
    \addlegendentry{cpu output}
    \addlegendentry{gpu output}
    \addlegendentry{cpu input}
    \addlegendentry{gpu input}
  \end{loglogaxis}
\end{tikzpicture}

%% file: tikz/pure_kernels/heat_transfer-2D-TRIANGLE3-trsm-time.tex
\begin{tikzpicture}
  \begin{loglogaxis}[
    width=.95\columnwidth,
    height=4.2cm,
    log basis x=2,
    log basis y=10,
    xlabel={Number of unknowns per subdomain (TRSM, 2D)},
    ylabel={Time per subdomain [ms]},
    xmajorgrids=true,
    ymajorgrids=true,
    every axis plot/.append style={line width=1.2pt},
    grid style=dotted,
    xmin=3.081e+01,
    xmax=8.360e+05,
    ymin=3.891e-03,
    ymax=3.590e+05,
    label style={font=\footnotesize},
    tick label style={font=\footnotesize},
    legend style={font=\footnotesize, draw=none, fill=none, at={(0.3,0.95)}, anchor=north},
    legend columns=2, legend cell align={left}
  ]
    \addplot[solid, color=blue, mark=none] coordinates {
      ( 4.900e+01 , 1.106e-02 )
      ( 8.100e+01 , 1.928e-02 )
      ( 1.440e+02 , 4.166e-02 )
      ( 2.890e+02 , 1.307e-01 )
      ( 5.760e+02 , 8.974e-01 )
      ( 1.089e+03 , 1.800e+00 )
      ( 2.116e+03 , 5.385e+00 )
      ( 4.225e+03 , 1.429e+01 )
      ( 8.281e+03 , 6.378e+01 )
      ( 1.664e+04 , 1.628e+02 )
      ( 3.312e+04 , 6.809e+02 )
      ( 6.605e+04 , 2.071e+03 )
      ( 1.318e+05 , 1.207e+04 )
      ( 2.632e+05 , 1.783e+04 )
      ( 5.256e+05 , 1.560e+05 )
    };
    \addplot[solid, color=red, mark=none] coordinates {
      ( 4.900e+01 , 1.110e-02 )
      ( 8.100e+01 , 1.897e-02 )
      ( 1.440e+02 , 4.139e-02 )
      ( 2.890e+02 , 1.242e-01 )
      ( 5.760e+02 , 5.327e-01 )
      ( 1.089e+03 , 1.295e+00 )
      ( 2.116e+03 , 5.180e+00 )
      ( 4.225e+03 , 1.803e+01 )
      ( 8.281e+03 , 3.684e+01 )
      ( 1.664e+04 , 1.150e+02 )
      ( 3.312e+04 , 3.626e+02 )
      ( 6.605e+04 , 1.136e+03 )
      ( 1.318e+05 , 3.296e+03 )
      ( 2.632e+05 , 9.651e+03 )
      ( 5.256e+05 , 3.025e+04 )
    };
    \addplot[solid, color=olive, mark=none] coordinates {
      ( 4.900e+01 , 8.955e-03 )
      ( 8.100e+01 , 1.786e-02 )
      ( 1.440e+02 , 4.139e-02 )
      ( 2.890e+02 , 1.212e-01 )
      ( 5.760e+02 , 5.783e-01 )
      ( 1.089e+03 , 1.163e+00 )
      ( 2.116e+03 , 3.276e+00 )
      ( 4.225e+03 , 1.034e+01 )
      ( 8.281e+03 , 3.881e+01 )
      ( 1.664e+04 , 1.228e+02 )
      ( 3.312e+04 , 3.547e+02 )
      ( 6.605e+04 , 1.152e+03 )
      ( 1.318e+05 , 2.943e+03 )
      ( 2.632e+05 , nan )
      ( 5.256e+05 , nan )
    };
    \addplot[solid, color=violet, mark=none] coordinates {
      ( 4.900e+01 , 2.182e-02 )
      ( 8.100e+01 , 4.086e-02 )
      ( 1.440e+02 , 9.359e-02 )
      ( 2.890e+02 , 2.653e-01 )
      ( 5.760e+02 , 7.826e-01 )
      ( 1.089e+03 , 2.155e+00 )
      ( 2.116e+03 , 6.006e+00 )
      ( 4.225e+03 , 1.799e+01 )
      ( 8.281e+03 , 5.554e+01 )
      ( 1.664e+04 , 1.778e+02 )
      ( 3.312e+04 , 5.024e+02 )
      ( 6.605e+04 , 1.532e+03 )
      ( 1.318e+05 , 4.329e+03 )
      ( 2.632e+05 , 1.279e+04 )
      ( 5.256e+05 , 3.642e+04 )
    };

    \addplot[solid, color=green, mark=none] coordinates {
      ( 4.900e+01 , 4.364e-02 )
      ( 8.100e+01 , 5.169e-02 )
      ( 1.440e+02 , 6.441e-02 )
      ( 2.890e+02 , 9.647e-02 )
      ( 5.760e+02 , 1.922e-01 )
      ( 1.089e+03 , 3.501e-01 )
      ( 2.116e+03 , 6.456e-01 )
      ( 4.225e+03 , 1.305e+00 )
      ( 8.281e+03 , 2.778e+00 )
      ( 1.664e+04 , 5.503e+00 )
      ( 3.312e+04 , 1.364e+01 )
      ( 6.605e+04 , 3.303e+01 )
      ( 1.318e+05 , 8.194e+01 )
      ( 2.632e+05 , 1.970e+02 )
      ( 5.256e+05 , 5.160e+02 )
    };
    \addplot[solid, color=teal, mark=none] coordinates {
      ( 4.900e+01 , 4.411e-02 )
      ( 8.100e+01 , 5.355e-02 )
      ( 1.440e+02 , 6.459e-02 )
      ( 2.890e+02 , 9.566e-02 )
      ( 5.760e+02 , 1.925e-01 )
      ( 1.089e+03 , 3.936e-01 )
      ( 2.116e+03 , 7.093e-01 )
      ( 4.225e+03 , 1.461e+00 )
      ( 8.281e+03 , 2.975e+00 )
      ( 1.664e+04 , 5.790e+00 )
      ( 3.312e+04 , 1.224e+01 )
      ( 6.605e+04 , 2.523e+01 )
      ( 1.318e+05 , 5.477e+01 )
      ( 2.632e+05 , 1.200e+02 )
      ( 5.256e+05 , 2.866e+02 )
    };
    \addlegendentry{cpu orig}
    \addlegendentry{cpu opt}
    \addlegendentry{cholmod}
    \addlegendentry{pardiso}
    \addlegendentry{gpu orig}
    \addlegendentry{gpu opt}
  \end{loglogaxis}
\end{tikzpicture}

%% file: tikz/pure_kernels/heat_transfer-3D-TETRA4-trsm-time.tex
\begin{tikzpicture}
  \begin{loglogaxis}[
    width=.95\columnwidth,
    height=4.2cm,
    log basis x=2,
    log basis y=10,
    xlabel={Number of unknowns per subdomain (TRSM, 3D)},
    ylabel={Time per subdomain [ms]},
    xmajorgrids=true,
    ymajorgrids=true,
    every axis plot/.append style={line width=1.2pt},
    grid style=dotted,
    xmin=4.514e+01,
    xmax=9.771e+04,
    ymin=1.899e-02,
    ymax=1.667e+06,
    label style={font=\footnotesize},
    tick label style={font=\footnotesize},
    legend style={font=\footnotesize, draw=none, fill=none, at={(0.3,0.95)}, anchor=north},
    legend columns=2, legend cell align={left}
  ]
    \addplot[solid, color=blue, mark=none] coordinates {
      ( 6.400e+01 , 4.360e-02 )
      ( 1.250e+02 , 1.545e-01 )
      ( 2.160e+02 , 4.753e-01 )
      ( 3.430e+02 , 1.371e+00 )
      ( 7.290e+02 , 7.648e+00 )
      ( 1.331e+03 , 3.287e+01 )
      ( 2.744e+03 , 2.047e+02 )
      ( 4.913e+03 , 8.955e+02 )
      ( 9.261e+03 , 4.427e+03 )
      ( 1.758e+04 , 2.437e+04 )
      ( 3.594e+04 , 1.600e+05 )
      ( 6.892e+04 , 7.259e+05 )
    };
    \addplot[solid, color=red, mark=none] coordinates {
      ( 6.400e+01 , 4.370e-02 )
      ( 1.250e+02 , 1.545e-01 )
      ( 2.160e+02 , 3.432e-01 )
      ( 3.430e+02 , 8.041e-01 )
      ( 7.290e+02 , 3.412e+00 )
      ( 1.331e+03 , 8.693e+00 )
      ( 2.744e+03 , 3.518e+01 )
      ( 4.913e+03 , 9.678e+01 )
      ( 9.261e+03 , 2.941e+02 )
      ( 1.758e+04 , 9.445e+02 )
      ( 3.594e+04 , 3.432e+03 )
      ( 6.892e+04 , 9.972e+03 )
    };
    \addplot[solid, color=olive, mark=none] coordinates {
      ( 6.400e+01 , 5.181e-02 )
      ( 1.250e+02 , 1.273e-01 )
      ( 2.160e+02 , 4.037e-01 )
      ( 3.430e+02 , 1.100e+00 )
      ( 7.290e+02 , 5.317e+00 )
      ( 1.331e+03 , 1.664e+01 )
      ( 2.744e+03 , 6.640e+01 )
      ( 4.913e+03 , 1.924e+02 )
      ( 9.261e+03 , 5.960e+02 )
      ( 1.758e+04 , 2.143e+03 )
      ( 3.594e+04 , 8.669e+03 )
      ( 6.892e+04 , 2.881e+04 )
    };
    \addplot[solid, color=violet, mark=none] coordinates {
      ( 6.400e+01 , 8.481e-02 )
      ( 1.250e+02 , 2.840e-01 )
      ( 2.160e+02 , 7.368e-01 )
      ( 3.430e+02 , 1.888e+00 )
      ( 7.290e+02 , 7.885e+00 )
      ( 1.331e+03 , 2.375e+01 )
      ( 2.744e+03 , 9.836e+01 )
      ( 4.913e+03 , 2.757e+02 )
      ( 9.261e+03 , 9.097e+02 )
      ( 1.758e+04 , 3.215e+03 )
      ( 3.594e+04 , 1.258e+04 )
      ( 6.892e+04 , 4.121e+04 )
    };

    \addplot[solid, color=green, mark=none] coordinates {
      ( 6.400e+01 , 6.838e-02 )
      ( 1.250e+02 , 1.097e-01 )
      ( 2.160e+02 , 1.739e-01 )
      ( 3.430e+02 , 2.653e-01 )
      ( 7.290e+02 , 5.601e-01 )
      ( 1.331e+03 , 1.075e+00 )
      ( 2.744e+03 , 2.949e+00 )
      ( 4.913e+03 , 7.140e+00 )
      ( 9.261e+03 , 2.234e+01 )
      ( 1.758e+04 , 5.843e+01 )
      ( 3.594e+04 , 2.146e+02 )
      ( 6.892e+04 , 6.984e+02 )
    };
    \addplot[solid, color=teal, mark=none] coordinates {
      ( 6.400e+01 , 6.862e-02 )
      ( 1.250e+02 , 1.080e-01 )
      ( 2.160e+02 , 1.743e-01 )
      ( 3.430e+02 , 2.620e-01 )
      ( 7.290e+02 , 6.061e-01 )
      ( 1.331e+03 , 1.107e+00 )
      ( 2.744e+03 , 2.470e+00 )
      ( 4.913e+03 , 4.577e+00 )
      ( 9.261e+03 , 9.762e+00 )
      ( 1.758e+04 , 2.102e+01 )
      ( 3.594e+04 , 5.365e+01 )
      ( 6.892e+04 , 1.289e+02 )
    };
  \end{loglogaxis}
\end{tikzpicture}

%% file: tikz/pure_kernels/heat_transfer-2D-TRIANGLE3-herk-time.tex
\begin{tikzpicture}
  \begin{loglogaxis}[
    width=.95\columnwidth,
    height=4.2cm,
    log basis x=2,
    log basis y=10,
    xlabel={Number of unknowns per subdomain (SYRK, 2D)},
    ylabel={Time per subdomain [ms]},
    xmajorgrids=true,
    ymajorgrids=true,
    every axis plot/.append style={line width=1.2pt},
    grid style=dotted,
    xmin=3.081e+01,
    xmax=8.360e+05,
    ymin=2.748e-03,
    ymax=3.590e+05,
    label style={font=\footnotesize},
    tick label style={font=\footnotesize},
    legend style={font=\footnotesize, draw=none, at={(0.3,0.95)}, anchor=north},
    legend columns=2, legend cell align={left}
  ]
    \addplot[solid, color=blue, mark=none] coordinates {
      ( 4.900e+01 , 7.040e-03 )
      ( 8.100e+01 , 1.046e-02 )
      ( 1.440e+02 , 2.481e-02 )
      ( 2.890e+02 , 8.447e-02 )
      ( 5.760e+02 , 2.773e-01 )
      ( 1.089e+03 , 8.319e-01 )
      ( 2.116e+03 , 2.738e+00 )
      ( 4.225e+03 , 9.581e+00 )
      ( 8.281e+03 , 3.458e+01 )
      ( 1.664e+04 , 1.275e+02 )
      ( 3.312e+04 , 4.760e+02 )
      ( 6.605e+04 , 1.734e+03 )
      ( 1.318e+05 , 6.366e+03 )
      ( 2.632e+05 , 1.996e+04 )
      ( 5.256e+05 , 1.050e+05 )
    };
    \addplot[solid, color=red, mark=none] coordinates {
      ( 4.900e+01 , 6.311e-03 )
      ( 8.100e+01 , 1.026e-02 )
      ( 1.440e+02 , 2.304e-02 )
      ( 2.890e+02 , 8.480e-02 )
      ( 5.760e+02 , 2.780e-01 )
      ( 1.089e+03 , 5.623e-01 )
      ( 2.116e+03 , 1.440e+00 )
      ( 4.225e+03 , 4.978e+00 )
      ( 8.281e+03 , 1.788e+01 )
      ( 1.664e+04 , 7.730e+01 )
      ( 3.312e+04 , 2.746e+02 )
      ( 6.605e+04 , 8.675e+02 )
      ( 1.318e+05 , 2.928e+03 )
      ( 2.632e+05 , 1.009e+04 )
      ( 5.256e+05 , 3.720e+04 )
    };

    \addplot[solid, color=green, mark=none] coordinates {
      ( 4.900e+01 , 3.865e-02 )
      ( 8.100e+01 , 4.658e-02 )
      ( 1.440e+02 , 3.364e-02 )
      ( 2.890e+02 , 3.612e-02 )
      ( 5.760e+02 , 3.957e-02 )
      ( 1.089e+03 , 5.194e-02 )
      ( 2.116e+03 , 5.990e-02 )
      ( 4.225e+03 , 1.125e-01 )
      ( 8.281e+03 , 2.414e-01 )
      ( 1.664e+04 , 6.606e-01 )
      ( 3.312e+04 , 1.958e+00 )
      ( 6.605e+04 , 5.060e+00 )
      ( 1.318e+05 , 2.088e+01 )
      ( 2.632e+05 , 9.232e+01 )
      ( 5.256e+05 , 2.979e+02 )
    };
    \addplot[solid, color=teal, mark=none] coordinates {
      ( 4.900e+01 , 3.934e-02 )
      ( 8.100e+01 , 4.913e-02 )
      ( 1.440e+02 , 3.401e-02 )
      ( 2.890e+02 , 3.542e-02 )
      ( 5.760e+02 , 3.989e-02 )
      ( 1.089e+03 , 5.168e-02 )
      ( 2.116e+03 , 8.225e-02 )
      ( 4.225e+03 , 1.126e-01 )
      ( 8.281e+03 , 2.252e-01 )
      ( 1.664e+04 , 4.981e-01 )
      ( 3.312e+04 , 1.289e+00 )
      ( 6.605e+04 , 3.361e+00 )
      ( 1.318e+05 , 9.441e+00 )
      ( 2.632e+05 , 2.846e+01 )
      ( 5.256e+05 , 1.116e+02 )
    };
    \addlegendentry{cpu orig}
    \addlegendentry{cpu opt}
    \addlegendentry{gpu orig}
    \addlegendentry{gpu opt}
  \end{loglogaxis}
\end{tikzpicture}

%% file: tikz/pure_kernels/heat_transfer-3D-TETRA4-herk-time.tex
\begin{tikzpicture}
  \begin{loglogaxis}[
    width=.95\columnwidth,
    height=4.2cm,
    log basis x=2,
    log basis y=10,
    xlabel={Number of unknowns per subdomain (SYRK, 3D)},
    ylabel={Time per subdomain [ms]},
    xmajorgrids=true,
    ymajorgrids=true,
    every axis plot/.append style={line width=1.2pt},
    grid style=dotted,
    xmin=4.514e+01,
    xmax=9.771e+04,
    ymin=1.999e-02,
    ymax=1.667e+06,
    label style={font=\footnotesize},
    tick label style={font=\footnotesize},
    legend style={font=\footnotesize, draw=none, fill=none, at={(0.3,0.95)}, anchor=north},
    legend columns=2, legend cell align={left}
  ]
    \addplot[solid, color=blue, mark=none] coordinates {
      ( 6.400e+01 , 5.250e-02 )
      ( 1.250e+02 , 1.794e-01 )
      ( 2.160e+02 , 5.105e-01 )
      ( 3.430e+02 , 1.272e+00 )
      ( 7.290e+02 , 6.082e+00 )
      ( 1.331e+03 , 2.221e+01 )
      ( 2.744e+03 , 1.020e+02 )
      ( 4.913e+03 , 3.828e+02 )
      ( 9.261e+03 , 1.359e+03 )
      ( 1.758e+04 , 6.002e+03 )
      ( 3.594e+04 , 3.104e+04 )
      ( 6.892e+04 , 9.922e+04 )
    };
    \addplot[solid, color=red, mark=none] coordinates {
      ( 6.400e+01 , 4.299e-02 )
      ( 1.250e+02 , 1.082e-01 )
      ( 2.160e+02 , 2.560e-01 )
      ( 3.430e+02 , 7.451e-01 )
      ( 7.290e+02 , 3.365e+00 )
      ( 1.331e+03 , 1.039e+01 )
      ( 2.744e+03 , 4.545e+01 )
      ( 4.913e+03 , 1.549e+02 )
      ( 9.261e+03 , 5.473e+02 )
      ( 1.758e+04 , 2.289e+03 )
      ( 3.594e+04 , 9.412e+03 )
      ( 6.892e+04 , 3.058e+04 )
    };

    \addplot[solid, color=green, mark=none] coordinates {
      ( 6.400e+01 , 4.085e-02 )
      ( 1.250e+02 , 3.137e-02 )
      ( 2.160e+02 , 3.262e-02 )
      ( 3.430e+02 , 3.769e-02 )
      ( 7.290e+02 , 5.743e-02 )
      ( 1.331e+03 , 1.009e-01 )
      ( 2.744e+03 , 3.162e-01 )
      ( 4.913e+03 , 8.284e-01 )
      ( 9.261e+03 , 3.133e+00 )
      ( 1.758e+04 , 1.325e+01 )
      ( 3.594e+04 , 5.355e+01 )
      ( 6.892e+04 , 1.846e+02 )
    };
    \addplot[solid, color=teal, mark=none] coordinates {
      ( 6.400e+01 , 4.072e-02 )
      ( 1.250e+02 , 3.088e-02 )
      ( 2.160e+02 , 3.275e-02 )
      ( 3.430e+02 , 3.767e-02 )
      ( 7.290e+02 , 5.746e-02 )
      ( 1.331e+03 , 9.094e-02 )
      ( 2.744e+03 , 2.455e-01 )
      ( 4.913e+03 , 5.791e-01 )
      ( 9.261e+03 , 1.690e+00 )
      ( 1.758e+04 , 6.004e+00 )
      ( 3.594e+04 , 2.331e+01 )
      ( 6.892e+04 , 7.341e+01 )
    };
  \end{loglogaxis}
\end{tikzpicture}

%% file: tikz/pure_kernels/heat_transfer-2D-TRIANGLE3-spdp.tex
\begin{tikzpicture}
  \begin{semilogxaxis}[
    width=.95\columnwidth,
    height=4.5cm,
    log basis x=2,
    xlabel={Number of DOFs per subdomain (2D)},
    ylabel={Speedup},
    xmajorgrids=true,
    ymajorgrids=true,
    ytick={1, 2, 3, 4, 5, 6},
    every axis plot/.append style={line width=1.2pt},
    grid style=dotted,
    xmin=3.081e+01,
    xmax=8.360e+05,
    ymin=5.000e-01,
    ymax=6.000e+00,
    label style={font=\footnotesize},
    tick label style={font=\footnotesize},
    legend style={font=\footnotesize, draw=none, fill=none, at={(0.47,0.98)}, anchor=north},
    legend columns=2, legend cell align={left}
  ]
    \addplot[solid, color=blue, mark=none] coordinates {
      ( 4.900e+01 , 9.962e-01 )
      ( 8.100e+01 , 1.016e+00 )
      ( 1.440e+02 , 1.007e+00 )
      ( 2.890e+02 , 1.052e+00 )
      ( 5.760e+02 , 1.685e+00 )
      ( 1.089e+03 , 1.390e+00 )
      ( 2.116e+03 , 1.040e+00 )
      ( 4.225e+03 , 7.925e-01 )
      ( 8.281e+03 , 1.731e+00 )
      ( 1.664e+04 , 1.415e+00 )
      ( 3.312e+04 , 1.878e+00 )
      ( 6.605e+04 , 1.823e+00 )
      ( 1.318e+05 , 3.661e+00 )
      ( 2.632e+05 , 1.847e+00 )
      ( 5.256e+05 , 5.155e+00 )
    };
    \addplot[solid, color=olive, mark=none] coordinates {
      ( 4.900e+01 , 8.066e-01 )
      ( 8.100e+01 , 9.414e-01 )
      ( 1.440e+02 , 1.000e+00 )
      ( 2.890e+02 , 9.759e-01 )
      ( 5.760e+02 , 1.086e+00 )
      ( 1.089e+03 , 8.984e-01 )
      ( 2.116e+03 , 6.325e-01 )
      ( 4.225e+03 , 5.734e-01 )
      ( 8.281e+03 , 1.054e+00 )
      ( 1.664e+04 , 1.068e+00 )
      ( 3.312e+04 , 9.782e-01 )
      ( 6.605e+04 , 1.014e+00 )
      ( 1.318e+05 , 8.929e-01 )
      ( 2.632e+05 , nan )
      ( 5.256e+05 , nan )
    };
    \addplot[solid, color=violet, mark=none] coordinates {
      ( 4.900e+01 , 1.965e+00 )
      ( 8.100e+01 , 2.154e+00 )
      ( 1.440e+02 , 2.261e+00 )
      ( 2.890e+02 , 2.136e+00 )
      ( 5.760e+02 , 1.469e+00 )
      ( 1.089e+03 , 1.664e+00 )
      ( 2.116e+03 , 1.160e+00 )
      ( 4.225e+03 , 9.977e-01 )
      ( 8.281e+03 , 1.508e+00 )
      ( 1.664e+04 , 1.545e+00 )
      ( 3.312e+04 , 1.386e+00 )
      ( 6.605e+04 , 1.349e+00 )
      ( 1.318e+05 , 1.313e+00 )
      ( 2.632e+05 , 1.325e+00 )
      ( 5.256e+05 , 1.204e+00 )
    };

    \addplot[solid, color=red, mark=none] coordinates {
      ( 4.900e+01 , 1.115e+00 )
      ( 8.100e+01 , 1.020e+00 )
      ( 1.440e+02 , 1.077e+00 )
      ( 2.890e+02 , 9.962e-01 )
      ( 5.760e+02 , 9.976e-01 )
      ( 1.089e+03 , 1.480e+00 )
      ( 2.116e+03 , 1.902e+00 )
      ( 4.225e+03 , 1.925e+00 )
      ( 8.281e+03 , 1.934e+00 )
      ( 1.664e+04 , 1.650e+00 )
      ( 3.312e+04 , 1.733e+00 )
      ( 6.605e+04 , 1.999e+00 )
      ( 1.318e+05 , 2.174e+00 )
      ( 2.632e+05 , 1.978e+00 )
      ( 5.256e+05 , 2.823e+00 )
    };

    \addplot[solid, color=green, mark=none] coordinates {
      ( 4.900e+01 , 9.892e-01 )
      ( 8.100e+01 , 9.653e-01 )
      ( 1.440e+02 , 9.972e-01 )
      ( 2.890e+02 , 1.008e+00 )
      ( 5.760e+02 , 9.984e-01 )
      ( 1.089e+03 , 8.893e-01 )
      ( 2.116e+03 , 9.103e-01 )
      ( 4.225e+03 , 8.933e-01 )
      ( 8.281e+03 , 9.339e-01 )
      ( 1.664e+04 , 9.504e-01 )
      ( 3.312e+04 , 1.114e+00 )
      ( 6.605e+04 , 1.309e+00 )
      ( 1.318e+05 , 1.496e+00 )
      ( 2.632e+05 , 1.642e+00 )
      ( 5.256e+05 , 1.800e+00 )
    };
    \addplot[solid, color=teal, mark=none] coordinates {
      ( 4.900e+01 , 9.825e-01 )
      ( 8.100e+01 , 9.482e-01 )
      ( 1.440e+02 , 9.893e-01 )
      ( 2.890e+02 , 1.020e+00 )
      ( 5.760e+02 , 9.920e-01 )
      ( 1.089e+03 , 1.005e+00 )
      ( 2.116e+03 , 7.284e-01 )
      ( 4.225e+03 , 9.984e-01 )
      ( 8.281e+03 , 1.072e+00 )
      ( 1.664e+04 , 1.326e+00 )
      ( 3.312e+04 , 1.519e+00 )
      ( 6.605e+04 , 1.506e+00 )
      ( 1.318e+05 , 2.212e+00 )
      ( 2.632e+05 , 3.243e+00 )
      ( 5.256e+05 , 2.670e+00 )
    };
    \addlegendentry{cpu trsm orig/opt}
    \addlegendentry{cpu trsm cholmod/opt}
    \addlegendentry{cpu trsm pardiso/opt}
    \addlegendentry{cpu syrk orig/opt}
    \addlegendentry{gpu trsm orig/opt}
    \addlegendentry{gpu syrk orig/opt}
  \end{semilogxaxis}
\end{tikzpicture}

%% file: tikz/pure_kernels/heat_transfer-3D-TETRA4-spdp.tex
\begin{tikzpicture}
  \begin{semilogxaxis}[
    width=.95\columnwidth,
    height=4.5cm,
    log basis x=2,
    log basis y=10,
    xlabel={Number of DOFs per subdomain (3D)},
    ylabel={Speedup},
    xmajorgrids=true,
    ymajorgrids=true,
    ytick={1, 2, 3, 4, 5, 6},
    every axis plot/.append style={line width=1.2pt},
    grid style=dotted,
    xmin=4.514e+01,
    xmax=9.771e+04,
    ymin=6.583e-01,
    ymax=6.000e+00,
    label style={font=\footnotesize},
    tick label style={font=\footnotesize},
    legend style={font=\footnotesize, draw=none, at={(0.5,-0.1)}, anchor=north},
    legend cell align={left}
  ]
    \addplot[solid, color=blue, mark=none] coordinates {
      ( 6.400e+01 , 9.978e-01 )
      ( 1.250e+02 , 1.000e+00 )
      ( 2.160e+02 , 1.385e+00 )
      ( 3.430e+02 , 1.704e+00 )
      ( 7.290e+02 , 2.241e+00 )
      ( 1.331e+03 , 3.781e+00 )
      ( 2.744e+03 , 5.818e+00 )
      ( 4.913e+03 , 9.252e+00 )
      ( 9.261e+03 , 1.505e+01 )
      ( 1.758e+04 , 2.580e+01 )
      ( 3.594e+04 , 4.661e+01 )
      ( 6.892e+04 , 7.279e+01 )
    };
    \addplot[solid, color=olive, mark=none] coordinates {
      ( 6.400e+01 , 1.186e+00 )
      ( 1.250e+02 , 8.236e-01 )
      ( 2.160e+02 , 1.176e+00 )
      ( 3.430e+02 , 1.368e+00 )
      ( 7.290e+02 , 1.558e+00 )
      ( 1.331e+03 , 1.914e+00 )
      ( 2.744e+03 , 1.887e+00 )
      ( 4.913e+03 , 1.988e+00 )
      ( 9.261e+03 , 2.026e+00 )
      ( 1.758e+04 , 2.269e+00 )
      ( 3.594e+04 , 2.526e+00 )
      ( 6.892e+04 , 2.889e+00 )
    };
    \addplot[solid, color=violet, mark=none] coordinates {
      ( 6.400e+01 , 1.941e+00 )
      ( 1.250e+02 , 1.838e+00 )
      ( 2.160e+02 , 2.147e+00 )
      ( 3.430e+02 , 2.348e+00 )
      ( 7.290e+02 , 2.311e+00 )
      ( 1.331e+03 , 2.733e+00 )
      ( 2.744e+03 , 2.796e+00 )
      ( 4.913e+03 , 2.849e+00 )
      ( 9.261e+03 , 3.093e+00 )
      ( 1.758e+04 , 3.404e+00 )
      ( 3.594e+04 , 3.665e+00 )
      ( 6.892e+04 , 4.132e+00 )
    };

    \addplot[solid, color=red, mark=none] coordinates {
      ( 6.400e+01 , 1.221e+00 )
      ( 1.250e+02 , 1.658e+00 )
      ( 2.160e+02 , 1.994e+00 )
      ( 3.430e+02 , 1.707e+00 )
      ( 7.290e+02 , 1.807e+00 )
      ( 1.331e+03 , 2.138e+00 )
      ( 2.744e+03 , 2.245e+00 )
      ( 4.913e+03 , 2.471e+00 )
      ( 9.261e+03 , 2.484e+00 )
      ( 1.758e+04 , 2.623e+00 )
      ( 3.594e+04 , 3.298e+00 )
      ( 6.892e+04 , 3.245e+00 )
    };

    \addplot[solid, color=green, mark=none] coordinates {
      ( 6.400e+01 , 9.966e-01 )
      ( 1.250e+02 , 1.016e+00 )
      ( 2.160e+02 , 9.979e-01 )
      ( 3.430e+02 , 1.013e+00 )
      ( 7.290e+02 , 9.241e-01 )
      ( 1.331e+03 , 9.712e-01 )
      ( 2.744e+03 , 1.194e+00 )
      ( 4.913e+03 , 1.560e+00 )
      ( 9.261e+03 , 2.288e+00 )
      ( 1.758e+04 , 2.779e+00 )
      ( 3.594e+04 , 4.001e+00 )
      ( 6.892e+04 , 5.418e+00 )
    };

    \addplot[solid, color=teal, mark=none] coordinates {
      ( 6.400e+01 , 1.003e+00 )
      ( 1.250e+02 , 1.016e+00 )
      ( 2.160e+02 , 9.961e-01 )
      ( 3.430e+02 , 1.001e+00 )
      ( 7.290e+02 , 9.994e-01 )
      ( 1.331e+03 , 1.110e+00 )
      ( 2.744e+03 , 1.288e+00 )
      ( 4.913e+03 , 1.431e+00 )
      ( 9.261e+03 , 1.853e+00 )
      ( 1.758e+04 , 2.207e+00 )
      ( 3.594e+04 , 2.298e+00 )
      ( 6.892e+04 , 2.515e+00 )
    };
  \end{semilogxaxis}
\end{tikzpicture}

%% file: tikz/compare_to_orig/heat_transfer-2D-TRIANGLE3-time.tex
\begin{tikzpicture}
  \begin{loglogaxis}[
    width=.95\columnwidth,
    height=5cm,
    log basis x=2,
    log basis y=10,
    xlabel={Number of unknowns per subdomain (2D)},
    ylabel={Time per subdomain [ms]},
    xmajorgrids=true,
    ymajorgrids=true,
    every axis plot/.append style={line width=1.2pt},
    grid style=dotted,
    xmin=3.189e+01,
    xmax=4.043e+05,
    ymin=6.843e-04,
    ymax=6.069e+03,
    tick label style={font=\footnotesize},
    legend style={font=\scriptsize, draw=none, fill=none, at={(0.28,1)}, anchor=north},
    legend columns=2, legend cell align={left}
  ]
    \addplot[solid, color=blue, mark=none] coordinates {
      ( 4.900e+01 , 1.469e-03 )
      ( 8.100e+01 , 3.317e-03 )
      ( 1.440e+02 , 6.933e-03 )
      ( 2.890e+02 , 1.610e-02 )
      ( 5.760e+02 , 7.910e-02 )
      ( 1.089e+03 , 1.830e-01 )
      ( 2.116e+03 , 7.424e-01 )
      ( 4.225e+03 , 3.482e+00 )
      ( 8.281e+03 , 1.127e+01 )
      ( 1.664e+04 , 3.858e+01 )
      ( 3.312e+04 , 1.471e+02 )
      ( 6.605e+04 , 5.755e+02 )
      ( 1.318e+05 , 2.933e+03 )
      ( 2.632e+05 , nan )
      ( 5.256e+05 , nan )
    };
    \addplot[dashed, color=red, mark=none] coordinates {
      ( 4.900e+01 , 1.416e-03 )
      ( 8.100e+01 , 2.407e-03 )
      ( 1.440e+02 , 5.035e-03 )
      ( 2.890e+02 , 2.018e-02 )
      ( 5.760e+02 , 5.785e-02 )
      ( 1.089e+03 , 1.316e-01 )
      ( 2.116e+03 , 4.645e-01 )
      ( 4.225e+03 , 1.723e+00 )
      ( 8.281e+03 , 4.338e+00 )
      ( 1.664e+04 , 2.032e+01 )
      ( 3.312e+04 , 7.146e+01 )
      ( 6.605e+04 , 2.301e+02 )
      ( 1.318e+05 , 7.402e+02 )
      ( 2.632e+05 , 2.473e+03 )
      ( 5.256e+05 , nan )
    };

    \addplot[solid, color=olive, mark=none] coordinates {
      ( 4.900e+01 , 4.395e-03 )
      ( 8.100e+01 , 5.391e-03 )
      ( 1.440e+02 , 1.339e-02 )
      ( 2.890e+02 , 2.892e-02 )
      ( 5.760e+02 , 1.041e-01 )
      ( 1.089e+03 , 3.357e-01 )
      ( 2.116e+03 , 1.014e+00 )
      ( 4.225e+03 , 3.679e+00 )
      ( 8.281e+03 , 1.179e+01 )
      ( 1.664e+04 , 3.990e+01 )
      ( 3.312e+04 , 1.558e+02 )
      ( 6.605e+04 , 5.803e+02 )
      ( 1.318e+05 , 2.893e+03 )
      ( 2.632e+05 , nan )
      ( 5.256e+05 , nan )
    };
    \addplot[dashed, color=violet, mark=none] coordinates {
      ( 4.900e+01 , 3.292e-03 )
      ( 8.100e+01 , 8.293e-03 )
      ( 1.440e+02 , 1.094e-02 )
      ( 2.890e+02 , 2.752e-02 )
      ( 5.760e+02 , 8.531e-02 )
      ( 1.089e+03 , 1.942e-01 )
      ( 2.116e+03 , 6.790e-01 )
      ( 4.225e+03 , 2.010e+00 )
      ( 8.281e+03 , 5.066e+00 )
      ( 1.664e+04 , 1.932e+01 )
      ( 3.312e+04 , 7.727e+01 )
      ( 6.605e+04 , 2.308e+02 )
      ( 1.318e+05 , 7.471e+02 )
      ( 2.632e+05 , 2.544e+03 )
      ( 5.256e+05 , nan )
    };

    \addplot[solid, color=lime, mark=none] coordinates {
      ( 4.900e+01 , 4.096e-02 )
      ( 8.100e+01 , 4.606e-02 )
      ( 1.440e+02 , 4.335e-02 )
      ( 2.890e+02 , 4.696e-02 )
      ( 5.760e+02 , 5.087e-02 )
      ( 1.089e+03 , 1.140e-01 )
      ( 2.116e+03 , 3.240e-01 )
      ( 4.225e+03 , 9.811e-01 )
      ( 8.281e+03 , 2.293e+00 )
      ( 1.664e+04 , 4.997e+00 )
      ( 3.312e+04 , 1.314e+01 )
      ( 6.605e+04 , 3.340e+01 )
      ( 1.318e+05 , 9.718e+01 )
      ( 2.632e+05 , 2.837e+02 )
      ( 5.256e+05 , 7.786e+02 )
    };
    \addplot[dashed, color=green, mark=none] coordinates {
      ( 4.900e+01 , 4.057e-02 )
      ( 8.100e+01 , 4.700e-02 )
      ( 1.440e+02 , 4.221e-02 )
      ( 2.890e+02 , 4.565e-02 )
      ( 5.760e+02 , 5.561e-02 )
      ( 1.089e+03 , 1.255e-01 )
      ( 2.116e+03 , 2.133e-01 )
      ( 4.225e+03 , 5.441e-01 )
      ( 8.281e+03 , 1.325e+00 )
      ( 1.664e+04 , 2.959e+00 )
      ( 3.312e+04 , 7.271e+00 )
      ( 6.605e+04 , 1.736e+01 )
      ( 1.318e+05 , 4.636e+01 )
      ( 2.632e+05 , 1.134e+02 )
      ( 5.256e+05 , 3.383e+02 )
    };

    \addplot[solid, color=cyan, mark=none] coordinates {
      ( 4.900e+01 , 4.204e-02 )
      ( 8.100e+01 , 4.655e-02 )
      ( 1.440e+02 , 4.540e-02 )
      ( 2.890e+02 , 6.277e-02 )
      ( 5.760e+02 , 8.205e-02 )
      ( 1.089e+03 , 1.438e-01 )
      ( 2.116e+03 , 3.705e-01 )
      ( 4.225e+03 , 1.049e+00 )
      ( 8.281e+03 , 2.320e+00 )
      ( 1.664e+04 , 4.977e+00 )
      ( 3.312e+04 , 1.327e+01 )
      ( 6.605e+04 , 3.387e+01 )
      ( 1.318e+05 , 9.954e+01 )
      ( 2.632e+05 , 2.951e+02 )
      ( 5.256e+05 , 8.429e+02 )
    };
    \addplot[dashed, color=teal, mark=none] coordinates {
      ( 4.900e+01 , 4.169e-02 )
      ( 8.100e+01 , 4.388e-02 )
      ( 1.440e+02 , 4.453e-02 )
      ( 2.890e+02 , 5.365e-02 )
      ( 5.760e+02 , 7.112e-02 )
      ( 1.089e+03 , 1.236e-01 )
      ( 2.116e+03 , 2.365e-01 )
      ( 4.225e+03 , 6.970e-01 )
      ( 8.281e+03 , 1.444e+00 )
      ( 1.664e+04 , 2.976e+00 )
      ( 3.312e+04 , 7.027e+00 )
      ( 6.605e+04 , 1.812e+01 )
      ( 1.318e+05 , 4.792e+01 )
      ( 2.632e+05 , 1.326e+02 )
      ( 5.256e+05 , 4.225e+02 )
    };
  \end{loglogaxis}
\end{tikzpicture}

%% file: tikz/compare_to_orig/heat_transfer-3D-TETRA4-time.tex
\begin{tikzpicture}
  \begin{loglogaxis}[
    width=.95\columnwidth,
    height=5cm,
    log basis x=2,
    log basis y=10,
    xlabel={Number of unknowns per subdomain (3D)},
    ylabel={Time per subdomain [ms]},
    xmajorgrids=true,
    ymajorgrids=true,
    every axis plot/.append style={line width=1.2pt},
    grid style=dotted,
    xmin=4.514e+01,
    xmax=9.771e+04,
    ymin=4.158e-03,
    ymax=1.052e+04,
    tick label style={font=\footnotesize},
    legend style={font=\scriptsize, draw=none, fill=none, at={(0.28,1.02)}, anchor=north},
    legend columns=2, legend cell align={left}
  ]
    \addplot[solid, color=blue, mark=none] coordinates {
      ( 6.400e+01 , 1.369e-02 )
      ( 1.250e+02 , 3.753e-02 )
      ( 2.160e+02 , 7.514e-02 )
      ( 3.430e+02 , 2.114e-01 )
      ( 7.290e+02 , 1.767e+00 )
      ( 1.331e+03 , 6.673e+00 )
      ( 2.744e+03 , 3.571e+01 )
      ( 4.913e+03 , 1.465e+02 )
      ( 9.261e+03 , 6.527e+02 )
      ( 1.758e+04 , 3.415e+03 )
      ( 3.594e+04 , nan )
      ( 6.892e+04 , nan )
    };
    \addplot[dashed, color=red, mark=none] coordinates {
      ( 6.400e+01 , 8.126e-03 )
      ( 1.250e+02 , 2.883e-02 )
      ( 2.160e+02 , 4.948e-02 )
      ( 3.430e+02 , 1.305e-01 )
      ( 7.290e+02 , 9.145e-01 )
      ( 1.331e+03 , 2.397e+00 )
      ( 2.744e+03 , 1.130e+01 )
      ( 4.913e+03 , 4.372e+01 )
      ( 9.261e+03 , 1.332e+02 )
      ( 1.758e+04 , 4.876e+02 )
      ( 3.594e+04 , 1.798e+03 )
      ( 6.892e+04 , 5.384e+03 )
    };

    \addplot[solid, color=olive, mark=none] coordinates {
      ( 6.400e+01 , 1.229e-02 )
      ( 1.250e+02 , 3.280e-02 )
      ( 2.160e+02 , 9.787e-02 )
      ( 3.430e+02 , 2.365e-01 )
      ( 7.290e+02 , 1.708e+00 )
      ( 1.331e+03 , 6.779e+00 )
      ( 2.744e+03 , 3.635e+01 )
      ( 4.913e+03 , 1.574e+02 )
      ( 9.261e+03 , 6.421e+02 )
      ( 1.758e+04 , 3.356e+03 )
      ( 3.594e+04 , nan )
      ( 6.892e+04 , nan )
    };
    \addplot[dashed, color=violet, mark=none] coordinates {
      ( 6.400e+01 , 1.088e-02 )
      ( 1.250e+02 , 3.711e-02 )
      ( 2.160e+02 , 6.279e-02 )
      ( 3.430e+02 , 1.645e-01 )
      ( 7.290e+02 , 9.673e-01 )
      ( 1.331e+03 , 2.559e+00 )
      ( 2.744e+03 , 1.245e+01 )
      ( 4.913e+03 , 4.206e+01 )
      ( 9.261e+03 , 1.434e+02 )
      ( 1.758e+04 , 4.756e+02 )
      ( 3.594e+04 , 1.834e+03 )
      ( 6.892e+04 , 6.197e+03 )
    };

    \addplot[solid, color=lime, mark=none] coordinates {
      ( 6.400e+01 , 4.273e-02 )
      ( 1.250e+02 , 4.789e-02 )
      ( 2.160e+02 , 5.451e-02 )
      ( 3.430e+02 , 6.703e-02 )
      ( 7.290e+02 , 1.629e-01 )
      ( 1.331e+03 , 3.994e-01 )
      ( 2.744e+03 , 1.506e+00 )
      ( 4.913e+03 , 4.739e+00 )
      ( 9.261e+03 , 1.785e+01 )
      ( 1.758e+04 , 6.256e+01 )
      ( 3.594e+04 , 2.458e+02 )
      ( 6.892e+04 , 8.394e+02 )
    };
    \addplot[dashed, color=green, mark=none] coordinates {
      ( 6.400e+01 , 4.524e-02 )
      ( 1.250e+02 , 4.990e-02 )
      ( 2.160e+02 , 5.397e-02 )
      ( 3.430e+02 , 6.642e-02 )
      ( 7.290e+02 , 1.733e-01 )
      ( 1.331e+03 , 3.404e-01 )
      ( 2.744e+03 , 9.671e-01 )
      ( 4.913e+03 , 2.008e+00 )
      ( 9.261e+03 , 5.298e+00 )
      ( 1.758e+04 , 1.537e+01 )
      ( 3.594e+04 , 5.497e+01 )
      ( 6.892e+04 , 1.643e+02 )
    };

    \addplot[solid, color=cyan, mark=none] coordinates {
      ( 6.400e+01 , 5.457e-02 )
      ( 1.250e+02 , 5.337e-02 )
      ( 2.160e+02 , 5.595e-02 )
      ( 3.430e+02 , 8.714e-02 )
      ( 7.290e+02 , 1.891e-01 )
      ( 1.331e+03 , 4.529e-01 )
      ( 2.744e+03 , 1.552e+00 )
      ( 4.913e+03 , 4.774e+00 )
      ( 9.261e+03 , 1.805e+01 )
      ( 1.758e+04 , 6.455e+01 )
      ( 3.594e+04 , 2.613e+02 )
      ( 6.892e+04 , 9.617e+02 )
    };
    \addplot[dashed, color=teal, mark=none] coordinates {
      ( 6.400e+01 , 5.826e-02 )
      ( 1.250e+02 , 5.098e-02 )
      ( 2.160e+02 , 5.538e-02 )
      ( 3.430e+02 , 7.859e-02 )
      ( 7.290e+02 , 1.643e-01 )
      ( 1.331e+03 , 3.463e-01 )
      ( 2.744e+03 , 1.108e+00 )
      ( 4.913e+03 , 2.385e+00 )
      ( 9.261e+03 , 6.598e+00 )
      ( 1.758e+04 , 2.018e+01 )
      ( 3.594e+04 , 7.979e+01 )
      ( 6.892e+04 , 2.877e+02 )
    };
    \addlegendentry{cpu sep orig}
    \addlegendentry{cpu sep opt}
    \addlegendentry{cpu mix orig}
    \addlegendentry{cpu mix opt}
    \addlegendentry{gpu sep orig}
    \addlegendentry{gpu sep opt}
    \addlegendentry{gpu mix orig}
    \addlegendentry{gpu mix opt}
  \end{loglogaxis}
\end{tikzpicture}

%% file: tikz/compare_to_orig/heat_transfer-2D-TRIANGLE3-spdp.tex
\begin{tikzpicture}
  \begin{semilogxaxis}[
    width=.95\columnwidth,
    height=4.5cm,
    log basis x=2,
    xlabel={Number of unknowns per subdomain (2D)},
    ylabel={Speedup},
    xmajorgrids=true,
    ymajorgrids=true,
    ytick={1, 2, 3, 4, 5, 6, 7},
    every axis plot/.append style={line width=1.2pt},
    grid style=dotted,
    xmin=3.189e+01,
    xmax=4.043e+05,
    ymin=5.000e-01,
    ymax=7.500e+00,
    tick label style={font=\footnotesize},
    legend style={font=\scriptsize, draw=none, at={(0.5,-0.1)}, anchor=north},
    legend cell align={left}
  ]
    \addplot[solid, color=red, mark=none] coordinates {
      ( 4.900e+01 , 1.037e+00 )
      ( 8.100e+01 , 1.378e+00 )
      ( 1.440e+02 , 1.377e+00 )
      ( 2.890e+02 , 7.978e-01 )
      ( 5.760e+02 , 1.367e+00 )
      ( 1.089e+03 , 1.391e+00 )
      ( 2.116e+03 , 1.598e+00 )
      ( 4.225e+03 , 2.021e+00 )
      ( 8.281e+03 , 2.598e+00 )
      ( 1.664e+04 , 1.899e+00 )
      ( 3.312e+04 , 2.058e+00 )
      ( 6.605e+04 , 2.501e+00 )
      ( 1.318e+05 , 3.962e+00 )
      ( 2.632e+05 , nan )
      ( 5.256e+05 , nan )
    };
    \addplot[solid, color=violet, mark=none] coordinates {
      ( 4.900e+01 , 1.335e+00 )
      ( 8.100e+01 , 6.500e-01 )
      ( 1.440e+02 , 1.224e+00 )
      ( 2.890e+02 , 1.051e+00 )
      ( 5.760e+02 , 1.221e+00 )
      ( 1.089e+03 , 1.729e+00 )
      ( 2.116e+03 , 1.494e+00 )
      ( 4.225e+03 , 1.831e+00 )
      ( 8.281e+03 , 2.326e+00 )
      ( 1.664e+04 , 2.065e+00 )
      ( 3.312e+04 , 2.017e+00 )
      ( 6.605e+04 , 2.514e+00 )
      ( 1.318e+05 , 3.872e+00 )
      ( 2.632e+05 , nan )
      ( 5.256e+05 , nan )
    };
    \addplot[solid, color=green, mark=none] coordinates {
      ( 4.900e+01 , 1.009e+00 )
      ( 8.100e+01 , 9.801e-01 )
      ( 1.440e+02 , 1.027e+00 )
      ( 2.890e+02 , 1.029e+00 )
      ( 5.760e+02 , 9.148e-01 )
      ( 1.089e+03 , 9.082e-01 )
      ( 2.116e+03 , 1.519e+00 )
      ( 4.225e+03 , 1.803e+00 )
      ( 8.281e+03 , 1.730e+00 )
      ( 1.664e+04 , 1.689e+00 )
      ( 3.312e+04 , 1.807e+00 )
      ( 6.605e+04 , 1.924e+00 )
      ( 1.318e+05 , 2.096e+00 )
      ( 2.632e+05 , 2.502e+00 )
      ( 5.256e+05 , 2.301e+00 )
    };
    \addplot[solid, color=teal, mark=none] coordinates {
      ( 4.900e+01 , 1.009e+00 )
      ( 8.100e+01 , 1.061e+00 )
      ( 1.440e+02 , 1.020e+00 )
      ( 2.890e+02 , 1.170e+00 )
      ( 5.760e+02 , 1.154e+00 )
      ( 1.089e+03 , 1.163e+00 )
      ( 2.116e+03 , 1.567e+00 )
      ( 4.225e+03 , 1.505e+00 )
      ( 8.281e+03 , 1.607e+00 )
      ( 1.664e+04 , 1.672e+00 )
      ( 3.312e+04 , 1.888e+00 )
      ( 6.605e+04 , 1.869e+00 )
      ( 1.318e+05 , 2.077e+00 )
      ( 2.632e+05 , 2.225e+00 )
      ( 5.256e+05 , 1.995e+00 )
    };
  \end{semilogxaxis}
\end{tikzpicture}

%% file: tikz/compare_to_orig/heat_transfer-3D-TETRA4-spdp.tex
\begin{tikzpicture}
  \begin{semilogxaxis}[
    width=.95\columnwidth,
    height=4.5cm,
    log basis x=2,
    xlabel={Number of unknowns per subdomain (3D)},
    ylabel={Speedup},
    xmajorgrids=true,
    ymajorgrids=true,
    ytick={1, 2, 3, 4, 5, 6, 7},
    every axis plot/.append style={line width=1.2pt},
    grid style=dotted,
    xmin=4.514e+01,
    xmax=9.771e+04,
    ymin=5.000e-01,
    ymax=7.500e+00,
    tick label style={font=\footnotesize},
    legend style={font=\scriptsize, draw=none, fill=none, at={(0.35,1)}, anchor=north},
    legend columns=2, legend cell align={left}
  ]
    \addplot[solid, color=red, mark=none] coordinates {
      ( 6.400e+01 , 1.685e+00 )
      ( 1.250e+02 , 1.302e+00 )
      ( 2.160e+02 , 1.519e+00 )
      ( 3.430e+02 , 1.621e+00 )
      ( 7.290e+02 , 1.932e+00 )
      ( 1.331e+03 , 2.784e+00 )
      ( 2.744e+03 , 3.159e+00 )
      ( 4.913e+03 , 3.351e+00 )
      ( 9.261e+03 , 4.900e+00 )
      ( 1.758e+04 , 7.003e+00 )
      ( 3.594e+04 , nan )
      ( 6.892e+04 , nan )
    };
    \addplot[solid, color=violet, mark=none] coordinates {
      ( 6.400e+01 , 1.130e+00 )
      ( 1.250e+02 , 8.840e-01 )
      ( 2.160e+02 , 1.559e+00 )
      ( 3.430e+02 , 1.437e+00 )
      ( 7.290e+02 , 1.766e+00 )
      ( 1.331e+03 , 2.649e+00 )
      ( 2.744e+03 , 2.920e+00 )
      ( 4.913e+03 , 3.743e+00 )
      ( 9.261e+03 , 4.479e+00 )
      ( 1.758e+04 , 7.056e+00 )
      ( 3.594e+04 , nan )
      ( 6.892e+04 , nan )
    };
    \addplot[solid, color=green, mark=none] coordinates {
      ( 6.400e+01 , 9.445e-01 )
      ( 1.250e+02 , 9.599e-01 )
      ( 2.160e+02 , 1.010e+00 )
      ( 3.430e+02 , 1.009e+00 )
      ( 7.290e+02 , 9.398e-01 )
      ( 1.331e+03 , 1.173e+00 )
      ( 2.744e+03 , 1.557e+00 )
      ( 4.913e+03 , 2.360e+00 )
      ( 9.261e+03 , 3.370e+00 )
      ( 1.758e+04 , 4.070e+00 )
      ( 3.594e+04 , 4.471e+00 )
      ( 6.892e+04 , 5.108e+00 )
    };
    \addplot[solid, color=teal, mark=none] coordinates {
      ( 6.400e+01 , 9.366e-01 )
      ( 1.250e+02 , 1.047e+00 )
      ( 2.160e+02 , 1.010e+00 )
      ( 3.430e+02 , 1.109e+00 )
      ( 7.290e+02 , 1.151e+00 )
      ( 1.331e+03 , 1.308e+00 )
      ( 2.744e+03 , 1.401e+00 )
      ( 4.913e+03 , 2.002e+00 )
      ( 9.261e+03 , 2.735e+00 )
      ( 1.758e+04 , 3.199e+00 )
      ( 3.594e+04 , 3.275e+00 )
      ( 6.892e+04 , 3.343e+00 )
    };
    \addlegendentry{cpu sep orig/opt}
    \addlegendentry{cpu mix orig/opt}
    \addlegendentry{gpu sep orig/opt}
    \addlegendentry{gpu mix orig/opt}
  \end{semilogxaxis}
\end{tikzpicture}

%% file: tikz/update-2D.tex
\begin{tikzpicture}
  \begin{loglogaxis}[
    width=.95\columnwidth,
    height=5.8cm,
    log basis x=2,
    log basis y=10,
    xlabel={Number of unknowns per subdomain (2D)},
    ylabel={Time per subdomain [ms]},
    xmajorgrids=true,
    ymajorgrids=true,
    every axis plot/.append style={line width=1.2pt},
    grid style=dotted,
    xmin=6.162e+01,
    xmax=1.672e+06,
    ymin=5.536e-04,
    ymax=5.140e+03,
    tick label style={font=\footnotesize},
    legend style={font=\scriptsize, draw=none, at={(0.5,-0.1)}, anchor=north},
    legend cell align={left}
  ]
    \addplot[solid, color={rgb,255:red,255;green,0;blue,255}, mark=none] coordinates {
      ( 9.800e+01 , 3.185e-03 )
      ( 1.620e+02 , 5.681e-03 )
      ( 2.880e+02 , 1.253e-02 )
      ( 5.780e+02 , 3.287e-02 )
      ( 1.152e+03 , 1.063e-01 )
      ( 2.178e+03 , 2.355e-01 )
      ( 4.232e+03 , 9.196e-01 )
      ( 8.450e+03 , 3.046e+00 )
      ( 1.656e+04 , 8.754e+00 )
      ( 3.328e+04 , 2.685e+01 )
      ( 6.625e+04 , 8.477e+01 )
      ( 1.321e+05 , 2.979e+02 )
      ( 2.635e+05 , nan )
      ( 5.263e+05 , nan )
      ( 1.051e+06 , nan )
    };
    \addplot[solid, color={rgb,255:red,0;green,255;blue,255}, mark=none] coordinates {
      ( 9.800e+01 , 2.436e-03 )
      ( 1.620e+02 , 4.259e-03 )
      ( 2.880e+02 , 8.851e-03 )
      ( 5.780e+02 , 1.530e-02 )
      ( 1.152e+03 , 3.126e-02 )
      ( 2.178e+03 , 6.091e-02 )
      ( 4.232e+03 , 1.320e-01 )
      ( 8.450e+03 , 2.868e-01 )
      ( 1.656e+04 , 6.436e-01 )
      ( 3.328e+04 , 1.455e+00 )
      ( 6.625e+04 , 3.398e+00 )
      ( 1.321e+05 , 7.967e+00 )
      ( 2.635e+05 , 1.859e+01 )
      ( 5.263e+05 , 4.731e+01 )
      ( 1.051e+06 , 1.146e+02 )
    };
    \addplot[solid, color=blue, mark=none] coordinates {
      ( 9.800e+01 , 3.113e-03 )
      ( 1.620e+02 , 5.158e-03 )
      ( 2.880e+02 , 1.105e-02 )
      ( 5.780e+02 , 2.738e-02 )
      ( 1.152e+03 , 8.479e-02 )
      ( 2.178e+03 , 2.155e-01 )
      ( 4.232e+03 , 6.794e-01 )
      ( 8.450e+03 , 1.998e+00 )
      ( 1.656e+04 , 5.022e+00 )
      ( 3.328e+04 , 2.006e+01 )
      ( 6.625e+04 , 7.019e+01 )
      ( 1.321e+05 , 2.298e+02 )
      ( 2.635e+05 , 7.446e+02 )
      ( 5.263e+05 , 2.479e+03 )
      ( 1.051e+06 , nan )
    };
    \addplot[solid, color=green, mark=none] coordinates {
      ( 9.800e+01 , 4.239e-02 )
      ( 1.620e+02 , 4.563e-02 )
      ( 2.880e+02 , 4.533e-02 )
      ( 5.780e+02 , 5.456e-02 )
      ( 1.152e+03 , 6.488e-02 )
      ( 2.178e+03 , 1.216e-01 )
      ( 4.232e+03 , 2.308e-01 )
      ( 8.450e+03 , 6.817e-01 )
      ( 1.656e+04 , 1.318e+00 )
      ( 3.328e+04 , 2.961e+00 )
      ( 6.625e+04 , 7.204e+00 )
      ( 1.321e+05 , 1.775e+01 )
      ( 2.635e+05 , 4.695e+01 )
      ( 5.263e+05 , 1.292e+02 )
      ( 1.051e+06 , 4.017e+02 )
    };
    \addplot[solid, color={green!60!black}, mark=none] coordinates {
      ( 9.800e+01 , 4.497e-02 )
      ( 1.620e+02 , 4.463e-02 )
      ( 2.880e+02 , 4.696e-02 )
      ( 5.780e+02 , 5.469e-02 )
      ( 1.152e+03 , 7.531e-02 )
      ( 2.178e+03 , 1.392e-01 )
      ( 4.232e+03 , 3.560e-01 )
      ( 8.450e+03 , 7.011e-01 )
      ( 1.656e+04 , 2.277e+00 )
      ( 3.328e+04 , 5.228e+00 )
      ( 6.625e+04 , 1.339e+01 )
      ( 1.321e+05 , 3.501e+01 )
      ( 2.635e+05 , 1.041e+02 )
      ( 5.263e+05 , 3.055e+02 )
      ( 1.051e+06 , 8.415e+02 )
    };
    \addplot[dashed, color={rgb,255:red,255;green,0;blue,255}, mark=none] coordinates {
      ( 9.800e+01 , 1.801e-03 )
      ( 1.620e+02 , 2.794e-03 )
      ( 2.880e+02 , 6.283e-03 )
      ( 5.780e+02 , 1.310e-02 )
      ( 1.152e+03 , 3.262e-02 )
      ( 2.178e+03 , 7.361e-02 )
      ( 4.232e+03 , 1.407e-01 )
      ( 8.450e+03 , 2.786e-01 )
      ( 1.656e+04 , 6.442e-01 )
      ( 3.328e+04 , 1.439e+00 )
      ( 6.625e+04 , 3.033e+00 )
      ( 1.321e+05 , 6.213e+00 )
      ( 2.635e+05 , 1.409e+01 )
      ( 5.263e+05 , 3.123e+01 )
      ( 1.051e+06 , 7.166e+01 )
    };
    \addplot[dashed, color={rgb,255:red,0;green,255;blue,255}, mark=none] coordinates {
      ( 9.800e+01 , 1.148e-03 )
      ( 1.620e+02 , 1.782e-03 )
      ( 2.880e+02 , 3.483e-03 )
      ( 5.780e+02 , 7.371e-03 )
      ( 1.152e+03 , 1.562e-02 )
      ( 2.178e+03 , 3.495e-02 )
      ( 4.232e+03 , 7.673e-02 )
      ( 8.450e+03 , 1.711e-01 )
      ( 1.656e+04 , 3.638e-01 )
      ( 3.328e+04 , 8.101e-01 )
      ( 6.625e+04 , 1.765e+00 )
      ( 1.321e+05 , 3.890e+00 )
      ( 2.635e+05 , 8.988e+00 )
      ( 5.263e+05 , 2.080e+01 )
      ( 1.051e+06 , 4.835e+01 )
    };
    \addplot[dotted, color=black, mark=none] coordinates {
      ( 9.800e+01 , 4.926e-03 )
      ( 1.620e+02 , 6.955e-03 )
      ( 2.880e+02 , 1.036e-02 )
      ( 5.780e+02 , 1.879e-02 )
      ( 1.152e+03 , 3.593e-02 )
      ( 2.178e+03 , 6.877e-02 )
      ( 4.232e+03 , 1.427e-01 )
      ( 8.450e+03 , 3.077e-01 )
      ( 1.656e+04 , 6.865e-01 )
      ( 3.328e+04 , 1.542e+00 )
      ( 6.625e+04 , 3.508e+00 )
      ( 1.321e+05 , 8.139e+00 )
      ( 2.635e+05 , 1.881e+01 )
      ( 5.263e+05 , 4.375e+01 )
      ( 1.051e+06 , 1.114e+02 )
    };
  \end{loglogaxis}
\end{tikzpicture}

%% file: tikz/update-3D.tex
\begin{tikzpicture}
  \begin{loglogaxis}[
    width=.95\columnwidth,
    height=5.8cm,
    log basis x=2,
    log basis y=10,
    xlabel={Number of unknowns per subdomain (3D)},
    ylabel={Time per subdomain [ms]},
    xmajorgrids=true,
    ymajorgrids=true,
    every axis plot/.append style={line width=1.2pt},
    grid style=dotted,
    xmin=4.514e+01,
    xmax=9.771e+04,
    ymin=8.199e-04,
    ymax=2.225e+04,
    tick label style={font=\footnotesize},
    legend style={font=\scriptsize, draw=none, fill=none, at={(0.33,1)}, anchor=north},
    legend columns=2, legend cell align={left}
  ]
    \addplot[solid, color={rgb,255:red,255;green,0;blue,255}, mark=none] coordinates {
      ( 6.400e+01 , 1.294e-02 )
      ( 1.250e+02 , 3.166e-02 )
      ( 2.160e+02 , 9.111e-02 )
      ( 3.430e+02 , 1.858e-01 )
      ( 7.290e+02 , 1.070e+00 )
      ( 1.331e+03 , 3.399e+00 )
      ( 2.744e+03 , 1.834e+01 )
      ( 4.913e+03 , 8.431e+01 )
      ( 9.261e+03 , 3.753e+02 )
      ( 1.758e+04 , 1.555e+03 )
      ( 3.594e+04 , 5.781e+03 )
      ( 6.892e+04 , nan )
    };
    \addplot[solid, color={rgb,255:red,0;green,255;blue,255}, mark=none] coordinates {
      ( 6.400e+01 , 1.441e-02 )
      ( 1.250e+02 , 3.567e-02 )
      ( 2.160e+02 , 8.406e-02 )
      ( 3.430e+02 , 1.713e-01 )
      ( 7.290e+02 , 6.084e-01 )
      ( 1.331e+03 , 1.552e+00 )
      ( 2.744e+03 , 5.718e+00 )
      ( 4.913e+03 , 1.493e+01 )
      ( 9.261e+03 , 5.031e+01 )
      ( 1.758e+04 , 1.814e+02 )
      ( 3.594e+04 , 6.443e+02 )
      ( 6.892e+04 , 2.783e+03 )
    };
    \addplot[solid, color=blue, mark=none] coordinates {
      ( 6.400e+01 , 1.071e-02 )
      ( 1.250e+02 , 3.202e-02 )
      ( 2.160e+02 , 6.170e-02 )
      ( 3.430e+02 , 1.489e-01 )
      ( 7.290e+02 , 9.791e-01 )
      ( 1.331e+03 , 2.651e+00 )
      ( 2.744e+03 , 1.257e+01 )
      ( 4.913e+03 , 4.079e+01 )
      ( 9.261e+03 , 1.382e+02 )
      ( 1.758e+04 , 4.970e+02 )
      ( 3.594e+04 , 1.778e+03 )
      ( 6.892e+04 , 5.600e+03 )
    };
    \addplot[solid, color=green, mark=none] coordinates {
      ( 6.400e+01 , 4.277e-02 )
      ( 1.250e+02 , 4.602e-02 )
      ( 2.160e+02 , 5.774e-02 )
      ( 3.430e+02 , 7.636e-02 )
      ( 7.290e+02 , 1.640e-01 )
      ( 1.331e+03 , 3.597e-01 )
      ( 2.744e+03 , 1.101e+00 )
      ( 4.913e+03 , 2.396e+00 )
      ( 9.261e+03 , 6.415e+00 )
      ( 1.758e+04 , 1.992e+01 )
      ( 3.594e+04 , 7.859e+01 )
      ( 6.892e+04 , 2.837e+02 )
    };
    \addplot[solid, color={green!60!black}, mark=none] coordinates {
      ( 6.400e+01 , 8.378e-02 )
      ( 1.250e+02 , 9.591e-02 )
      ( 2.160e+02 , 1.211e-01 )
      ( 3.430e+02 , 1.501e-01 )
      ( 7.290e+02 , 2.870e-01 )
      ( 1.331e+03 , 5.510e-01 )
      ( 2.744e+03 , 1.767e+00 )
      ( 4.913e+03 , 5.215e+00 )
      ( 9.261e+03 , 1.920e+01 )
      ( 1.758e+04 , 6.489e+01 )
      ( 3.594e+04 , 2.622e+02 )
      ( 6.892e+04 , 9.636e+02 )
    };
    \addplot[dashed, color={rgb,255:red,255;green,0;blue,255}, mark=none] coordinates {
      ( 6.400e+01 , 2.508e-03 )
      ( 1.250e+02 , 5.100e-03 )
      ( 2.160e+02 , 1.176e-02 )
      ( 3.430e+02 , 2.152e-02 )
      ( 7.290e+02 , 6.607e-02 )
      ( 1.331e+03 , 1.492e-01 )
      ( 2.744e+03 , 4.024e-01 )
      ( 4.913e+03 , 1.011e+00 )
      ( 9.261e+03 , 2.664e+00 )
      ( 1.758e+04 , 8.893e+00 )
      ( 3.594e+04 , 3.657e+01 )
      ( 6.892e+04 , 1.228e+02 )
    };
    \addplot[dashed, color={rgb,255:red,0;green,255;blue,255}, mark=none] coordinates {
      ( 6.400e+01 , 1.737e-03 )
      ( 1.250e+02 , 5.018e-03 )
      ( 2.160e+02 , 8.286e-03 )
      ( 3.430e+02 , 1.554e-02 )
      ( 7.290e+02 , 4.550e-02 )
      ( 1.331e+03 , 1.094e-01 )
      ( 2.744e+03 , 3.611e-01 )
      ( 4.913e+03 , 8.620e-01 )
      ( 9.261e+03 , 2.582e+00 )
      ( 1.758e+04 , 8.020e+00 )
      ( 3.594e+04 , 3.044e+01 )
      ( 6.892e+04 , 1.197e+02 )
    };
    \addplot[dotted, color=black, mark=none] coordinates {
      ( 6.400e+01 , 2.007e-02 )
      ( 1.250e+02 , 4.548e-02 )
      ( 2.160e+02 , 1.017e-01 )
      ( 3.430e+02 , 2.014e-01 )
      ( 7.290e+02 , 6.567e-01 )
      ( 1.331e+03 , 1.656e+00 )
      ( 2.744e+03 , 5.910e+00 )
      ( 4.913e+03 , 1.560e+01 )
      ( 9.261e+03 , 5.202e+01 )
      ( 1.758e+04 , 1.859e+02 )
      ( 3.594e+04 , 6.395e+02 )
      ( 6.892e+04 , 2.771e+03 )
    };
    \addlegendentry{expl\_cholmod}
    \addlegendentry{expl\_mkl}
    \addlegendentry{expl\_cpu\_opt}
    \addlegendentry{expl\_gpu\_opt}
    \addlegendentry{expl\_cuda}
    \addlegendentry{impl\_cholmod}
    \addlegendentry{impl\_mkl}
    \addlegendentry{expl\_hybrid}
  \end{loglogaxis}
\end{tikzpicture}

%% file: tikz/amort-3D.tex
\begin{tikzpicture}
  \begin{loglogaxis}[
    width=.95\columnwidth, 
    height=6cm,
    log basis x=10,
    log basis y=10,
    xlabel={Number of iterations (3D)},
    ylabel={Step time per subdomain [ms]},
    xmajorgrids=true,
    ymajorgrids=true,
    every axis plot/.append style={line width=1.2pt},
    grid style=dotted,
    xmin=6.310e-01,
    xmax=7.924e+04,
    ymin=1.452e-03,
    ymax=5.643e+03,
    tick label style={font=\footnotesize},
    legend style={font=\footnotesize, draw=none, fill=none, at={(0.64,0.2)}, anchor=north},
    legend columns=2, legend cell align={left}
  ]
    \addplot[dashed, color={rgb,255:red,0;green,255;blue,255}, mark=none] coordinates {
      ( 1.000e+00 , 2.894e-03 )
      ( 1.093e+00 , 3.001e-03 )
      ( 1.194e+00 , 3.118e-03 )
      ( 1.304e+00 , 3.245e-03 )
      ( 1.425e+00 , 3.385e-03 )
      ( 1.557e+00 , 3.537e-03 )
      ( 1.701e+00 , 3.704e-03 )
      ( 1.858e+00 , 3.886e-03 )
      ( 2.030e+00 , 4.085e-03 )
      ( 2.218e+00 , 4.302e-03 )
      ( 2.424e+00 , 4.540e-03 )
      ( 2.648e+00 , 4.799e-03 )
    };
    \addplot[dashed, color={rgb,255:red,255;green,0;blue,255}, mark=none] coordinates {
      ( 2.648e+00 , 4.799e-03 )
      ( 2.911e+00 , 5.027e-03 )
      ( 3.200e+00 , 5.277e-03 )
      ( 3.518e+00 , 5.552e-03 )
      ( 3.867e+00 , 5.854e-03 )
      ( 4.251e+00 , 6.186e-03 )
      ( 4.673e+00 , 6.551e-03 )
      ( 5.137e+00 , 6.953e-03 )
      ( 5.648e+00 , 7.394e-03 )
      ( 6.208e+00 , 7.879e-03 )
      ( 6.825e+00 , 8.412e-03 )
      ( 7.503e+00 , 8.999e-03 )
      ( 8.248e+00 , 9.643e-03 )
      ( 9.067e+00 , 1.035e-02 )
      ( 9.967e+00 , 1.113e-02 )
      ( 1.096e+01 , 1.199e-02 )
      ( 1.204e+01 , 1.293e-02 )
      ( 1.324e+01 , 1.396e-02 )
      ( 1.456e+01 , 1.510e-02 )
      ( 1.600e+01 , 1.635e-02 )
      ( 1.759e+01 , 1.772e-02 )
      ( 1.934e+01 , 1.924e-02 )
      ( 2.126e+01 , 2.090e-02 )
      ( 2.337e+01 , 2.272e-02 )
      ( 2.569e+01 , 2.473e-02 )
      ( 2.824e+01 , 2.694e-02 )
      ( 3.104e+01 , 2.936e-02 )
      ( 3.412e+01 , 3.203e-02 )
      ( 3.751e+01 , 3.496e-02 )
      ( 4.124e+01 , 3.818e-02 )
      ( 4.533e+01 , 4.173e-02 )
      ( 4.983e+01 , 4.562e-02 )
      ( 5.478e+01 , 4.990e-02 )
      ( 6.022e+01 , 5.461e-02 )
      ( 6.620e+01 , 5.978e-02 )
      ( 7.278e+01 , 6.547e-02 )
      ( 8.000e+01 , 7.172e-02 )
      ( 8.795e+01 , 7.859e-02 )
      ( 9.668e+01 , 8.615e-02 )
      ( 1.063e+02 , 9.445e-02 )
      ( 1.168e+02 , 1.036e-01 )
      ( 1.284e+02 , 1.136e-01 )
      ( 1.412e+02 , 1.246e-01 )
      ( 1.552e+02 , 1.368e-01 )
      ( 1.706e+02 , 1.501e-01 )
      ( 1.876e+02 , 1.648e-01 )
      ( 2.062e+02 , 1.809e-01 )
      ( 2.267e+02 , 1.986e-01 )
      ( 2.492e+02 , 2.181e-01 )
      ( 2.739e+02 , 2.395e-01 )
      ( 3.011e+02 , 2.630e-01 )
      ( 3.310e+02 , 2.889e-01 )
      ( 3.639e+02 , 3.173e-01 )
      ( 4.000e+02 , 3.486e-01 )
      ( 4.397e+02 , 3.829e-01 )
      ( 4.834e+02 , 4.207e-01 )
      ( 5.314e+02 , 4.622e-01 )
      ( 5.842e+02 , 5.079e-01 )
      ( 6.422e+02 , 5.580e-01 )
      ( 7.059e+02 , 6.132e-01 )
      ( 7.760e+02 , 6.739e-01 )
      ( 8.531e+02 , 7.405e-01 )
      ( 9.378e+02 , 8.138e-01 )
      ( 1.031e+03 , 8.944e-01 )
      ( 1.133e+03 , 9.829e-01 )
      ( 1.246e+03 , 1.080e+00 )
      ( 1.370e+03 , 1.187e+00 )
      ( 1.506e+03 , 1.305e+00 )
      ( 1.655e+03 , 1.434e+00 )
      ( 1.819e+03 , 1.576e+00 )
      ( 2.000e+03 , 1.733e+00 )
      ( 2.199e+03 , 1.905e+00 )
      ( 2.417e+03 , 2.093e+00 )
      ( 2.657e+03 , 2.301e+00 )
      ( 2.921e+03 , 2.529e+00 )
      ( 3.211e+03 , 2.780e+00 )
      ( 3.530e+03 , 3.056e+00 )
      ( 3.880e+03 , 3.359e+00 )
      ( 4.265e+03 , 3.693e+00 )
      ( 4.689e+03 , 4.059e+00 )
      ( 5.155e+03 , 4.462e+00 )
      ( 5.666e+03 , 4.905e+00 )
      ( 6.229e+03 , 5.391e+00 )
      ( 6.848e+03 , 5.926e+00 )
      ( 7.528e+03 , 6.515e+00 )
      ( 8.275e+03 , 7.161e+00 )
      ( 9.097e+03 , 7.872e+00 )
      ( 1.000e+04 , 8.654e+00 )
    };
    \addplot[dashed, color={rgb,255:red,255;green,0;blue,255}, mark=none, forget plot] coordinates {
      ( 1.000e+00 , 7.072e-03 )
      ( 1.099e+00 , 7.268e-03 )
      ( 1.209e+00 , 7.484e-03 )
      ( 1.329e+00 , 7.721e-03 )
      ( 1.461e+00 , 7.981e-03 )
      ( 1.606e+00 , 8.268e-03 )
      ( 1.766e+00 , 8.583e-03 )
      ( 1.941e+00 , 8.929e-03 )
      ( 2.134e+00 , 9.309e-03 )
      ( 2.346e+00 , 9.728e-03 )
      ( 2.579e+00 , 1.019e-02 )
      ( 2.835e+00 , 1.069e-02 )
      ( 3.117e+00 , 1.125e-02 )
      ( 3.427e+00 , 1.186e-02 )
      ( 3.768e+00 , 1.253e-02 )
      ( 4.142e+00 , 1.327e-02 )
      ( 4.554e+00 , 1.408e-02 )
      ( 5.006e+00 , 1.498e-02 )
      ( 5.504e+00 , 1.596e-02 )
      ( 6.051e+00 , 1.704e-02 )
      ( 6.652e+00 , 1.822e-02 )
      ( 7.313e+00 , 1.953e-02 )
      ( 8.040e+00 , 2.096e-02 )
      ( 8.839e+00 , 2.254e-02 )
      ( 9.717e+00 , 2.427e-02 )
      ( 1.068e+01 , 2.617e-02 )
      ( 1.174e+01 , 2.827e-02 )
      ( 1.291e+01 , 3.057e-02 )
      ( 1.419e+01 , 3.310e-02 )
      ( 1.561e+01 , 3.588e-02 )
      ( 1.716e+01 , 3.894e-02 )
      ( 1.886e+01 , 4.231e-02 )
      ( 2.074e+01 , 4.600e-02 )
      ( 2.280e+01 , 5.007e-02 )
      ( 2.506e+01 , 5.454e-02 )
      ( 2.755e+01 , 5.945e-02 )
      ( 3.029e+01 , 6.485e-02 )
      ( 3.330e+01 , 7.079e-02 )
      ( 3.661e+01 , 7.732e-02 )
      ( 4.025e+01 , 8.450e-02 )
      ( 4.425e+01 , 9.239e-02 )
      ( 4.864e+01 , 1.011e-01 )
      ( 5.348e+01 , 1.106e-01 )
      ( 5.879e+01 , 1.211e-01 )
      ( 6.464e+01 , 1.326e-01 )
      ( 7.106e+01 , 1.453e-01 )
    };
    \addplot[dotted, color=black, mark=none] coordinates {
      ( 7.106e+01 , 1.453e-01 )
      ( 7.815e+01 , 1.552e-01 )
      ( 8.595e+01 , 1.662e-01 )
      ( 9.453e+01 , 1.782e-01 )
      ( 1.040e+02 , 1.915e-01 )
      ( 1.143e+02 , 2.061e-01 )
      ( 1.258e+02 , 2.221e-01 )
      ( 1.383e+02 , 2.397e-01 )
      ( 1.521e+02 , 2.591e-01 )
      ( 1.673e+02 , 2.804e-01 )
      ( 1.840e+02 , 3.039e-01 )
      ( 2.023e+02 , 3.297e-01 )
      ( 2.225e+02 , 3.580e-01 )
      ( 2.447e+02 , 3.892e-01 )
      ( 2.692e+02 , 4.235e-01 )
      ( 2.960e+02 , 4.613e-01 )
      ( 3.256e+02 , 5.028e-01 )
      ( 3.581e+02 , 5.484e-01 )
      ( 3.938e+02 , 5.986e-01 )
      ( 4.331e+02 , 6.538e-01 )
      ( 4.763e+02 , 7.145e-01 )
      ( 5.239e+02 , 7.813e-01 )
      ( 5.762e+02 , 8.547e-01 )
      ( 6.337e+02 , 9.355e-01 )
      ( 6.969e+02 , 1.024e+00 )
      ( 7.665e+02 , 1.122e+00 )
      ( 8.430e+02 , 1.229e+00 )
      ( 9.271e+02 , 1.348e+00 )
      ( 1.020e+03 , 1.478e+00 )
      ( 1.121e+03 , 1.620e+00 )
      ( 1.233e+03 , 1.778e+00 )
      ( 1.356e+03 , 1.950e+00 )
      ( 1.492e+03 , 2.141e+00 )
      ( 1.641e+03 , 2.350e+00 )
      ( 1.804e+03 , 2.580e+00 )
      ( 1.984e+03 , 2.833e+00 )
      ( 2.183e+03 , 3.111e+00 )
      ( 2.400e+03 , 3.417e+00 )
      ( 2.640e+03 , 3.753e+00 )
      ( 2.903e+03 , 4.123e+00 )
      ( 3.193e+03 , 4.530e+00 )
      ( 3.512e+03 , 4.978e+00 )
      ( 3.862e+03 , 5.470e+00 )
      ( 4.248e+03 , 6.011e+00 )
      ( 4.672e+03 , 6.607e+00 )
      ( 5.138e+03 , 7.262e+00 )
      ( 5.651e+03 , 7.982e+00 )
      ( 6.215e+03 , 8.774e+00 )
      ( 6.835e+03 , 9.645e+00 )
      ( 7.517e+03 , 1.060e+01 )
      ( 8.267e+03 , 1.166e+01 )
      ( 9.093e+03 , 1.282e+01 )
      ( 1.000e+04 , 1.409e+01 )
    };
    \addplot[dashed, color={rgb,255:red,0;green,255;blue,255}, mark=none, forget plot] coordinates {
      ( 1.000e+00 , 1.197e-02 )
      ( 1.098e+00 , 1.233e-02 )
      ( 1.206e+00 , 1.273e-02 )
      ( 1.325e+00 , 1.316e-02 )
      ( 1.456e+00 , 1.364e-02 )
      ( 1.599e+00 , 1.417e-02 )
      ( 1.756e+00 , 1.475e-02 )
      ( 1.929e+00 , 1.539e-02 )
      ( 2.119e+00 , 1.608e-02 )
      ( 2.327e+00 , 1.685e-02 )
      ( 2.556e+00 , 1.769e-02 )
      ( 2.808e+00 , 1.862e-02 )
      ( 3.084e+00 , 1.964e-02 )
      ( 3.387e+00 , 2.075e-02 )
      ( 3.721e+00 , 2.198e-02 )
      ( 4.087e+00 , 2.333e-02 )
      ( 4.489e+00 , 2.481e-02 )
      ( 4.931e+00 , 2.643e-02 )
      ( 5.416e+00 , 2.822e-02 )
      ( 5.949e+00 , 3.018e-02 )
      ( 6.534e+00 , 3.233e-02 )
      ( 7.177e+00 , 3.470e-02 )
      ( 7.883e+00 , 3.730e-02 )
      ( 8.659e+00 , 4.015e-02 )
      ( 9.511e+00 , 4.329e-02 )
      ( 1.045e+01 , 4.673e-02 )
      ( 1.147e+01 , 5.052e-02 )
      ( 1.260e+01 , 5.467e-02 )
      ( 1.384e+01 , 5.924e-02 )
      ( 1.521e+01 , 6.425e-02 )
      ( 1.670e+01 , 6.976e-02 )
    };
    \addplot[dashed, color={rgb,255:red,255;green,0;blue,255}, mark=none, forget plot] coordinates {
      ( 1.670e+01 , 6.976e-02 )
      ( 1.833e+01 , 7.542e-02 )
      ( 2.013e+01 , 8.164e-02 )
      ( 2.209e+01 , 8.847e-02 )
      ( 2.425e+01 , 9.597e-02 )
      ( 2.662e+01 , 1.042e-01 )
      ( 2.922e+01 , 1.132e-01 )
    };
    \addplot[solid, color=green, mark=none] coordinates {
      ( 2.922e+01 , 1.132e-01 )
      ( 3.206e+01 , 1.186e-01 )
      ( 3.517e+01 , 1.245e-01 )
      ( 3.859e+01 , 1.310e-01 )
      ( 4.233e+01 , 1.381e-01 )
      ( 4.644e+01 , 1.459e-01 )
      ( 5.095e+01 , 1.545e-01 )
      ( 5.589e+01 , 1.639e-01 )
      ( 6.132e+01 , 1.742e-01 )
      ( 6.727e+01 , 1.855e-01 )
      ( 7.380e+01 , 1.979e-01 )
      ( 8.096e+01 , 2.115e-01 )
      ( 8.882e+01 , 2.264e-01 )
      ( 9.744e+01 , 2.428e-01 )
      ( 1.069e+02 , 2.607e-01 )
      ( 1.173e+02 , 2.805e-01 )
      ( 1.287e+02 , 3.021e-01 )
      ( 1.412e+02 , 3.258e-01 )
      ( 1.549e+02 , 3.518e-01 )
      ( 1.699e+02 , 3.804e-01 )
      ( 1.864e+02 , 4.117e-01 )
      ( 2.045e+02 , 4.460e-01 )
    };
    \addplot[solid, color=green, mark=none, forget plot] coordinates {
      ( 2.045e+02 , 4.460e-01 )
      ( 2.248e+02 , 4.803e-01 )
      ( 2.472e+02 , 5.180e-01 )
      ( 2.718e+02 , 5.594e-01 )
      ( 2.988e+02 , 6.049e-01 )
      ( 3.286e+02 , 6.550e-01 )
      ( 3.613e+02 , 7.101e-01 )
      ( 3.972e+02 , 7.706e-01 )
      ( 4.368e+02 , 8.372e-01 )
      ( 4.802e+02 , 9.104e-01 )
      ( 5.280e+02 , 9.909e-01 )
      ( 5.806e+02 , 1.079e+00 )
      ( 6.384e+02 , 1.177e+00 )
      ( 7.019e+02 , 1.284e+00 )
      ( 7.718e+02 , 1.401e+00 )
      ( 8.486e+02 , 1.531e+00 )
      ( 9.330e+02 , 1.673e+00 )
      ( 1.026e+03 , 1.829e+00 )
      ( 1.128e+03 , 2.001e+00 )
      ( 1.240e+03 , 2.190e+00 )
      ( 1.364e+03 , 2.398e+00 )
      ( 1.499e+03 , 2.626e+00 )
      ( 1.649e+03 , 2.878e+00 )
      ( 1.813e+03 , 3.154e+00 )
      ( 1.993e+03 , 3.458e+00 )
      ( 2.191e+03 , 3.792e+00 )
      ( 2.410e+03 , 4.159e+00 )
      ( 2.649e+03 , 4.563e+00 )
      ( 2.913e+03 , 5.007e+00 )
      ( 3.203e+03 , 5.495e+00 )
      ( 3.522e+03 , 6.032e+00 )
      ( 3.872e+03 , 6.622e+00 )
      ( 4.258e+03 , 7.271e+00 )
      ( 4.681e+03 , 7.984e+00 )
      ( 5.147e+03 , 8.769e+00 )
      ( 5.659e+03 , 9.631e+00 )
      ( 6.223e+03 , 1.058e+01 )
      ( 6.842e+03 , 1.162e+01 )
      ( 7.523e+03 , 1.277e+01 )
      ( 8.272e+03 , 1.403e+01 )
      ( 9.095e+03 , 1.542e+01 )
      ( 1.000e+04 , 1.694e+01 )
    };
    \addplot[dashed, color={rgb,255:red,0;green,255;blue,255}, mark=none, forget plot] coordinates {
      ( 1.000e+00 , 2.148e-02 )
      ( 1.098e+00 , 2.206e-02 )
      ( 1.205e+00 , 2.269e-02 )
      ( 1.323e+00 , 2.339e-02 )
      ( 1.452e+00 , 2.416e-02 )
      ( 1.595e+00 , 2.501e-02 )
      ( 1.750e+00 , 2.593e-02 )
      ( 1.922e+00 , 2.695e-02 )
      ( 2.110e+00 , 2.806e-02 )
      ( 2.316e+00 , 2.929e-02 )
      ( 2.543e+00 , 3.063e-02 )
      ( 2.791e+00 , 3.211e-02 )
      ( 3.064e+00 , 3.373e-02 )
      ( 3.364e+00 , 3.551e-02 )
      ( 3.693e+00 , 3.746e-02 )
      ( 4.054e+00 , 3.961e-02 )
      ( 4.451e+00 , 4.196e-02 )
      ( 4.886e+00 , 4.455e-02 )
      ( 5.364e+00 , 4.738e-02 )
      ( 5.888e+00 , 5.050e-02 )
      ( 6.464e+00 , 5.392e-02 )
      ( 7.097e+00 , 5.767e-02 )
      ( 7.791e+00 , 6.179e-02 )
      ( 8.553e+00 , 6.632e-02 )
      ( 9.389e+00 , 7.128e-02 )
      ( 1.031e+01 , 7.673e-02 )
      ( 1.132e+01 , 8.272e-02 )
      ( 1.242e+01 , 8.929e-02 )
      ( 1.364e+01 , 9.650e-02 )
      ( 1.497e+01 , 1.044e-01 )
      ( 1.644e+01 , 1.131e-01 )
    };
    \addplot[solid, color=green, mark=none, forget plot] coordinates {
      ( 1.644e+01 , 1.131e-01 )
      ( 1.806e+01 , 1.167e-01 )
      ( 1.985e+01 , 1.207e-01 )
      ( 2.181e+01 , 1.251e-01 )
      ( 2.396e+01 , 1.299e-01 )
      ( 2.633e+01 , 1.352e-01 )
      ( 2.893e+01 , 1.411e-01 )
      ( 3.179e+01 , 1.475e-01 )
      ( 3.494e+01 , 1.545e-01 )
      ( 3.839e+01 , 1.622e-01 )
      ( 4.219e+01 , 1.707e-01 )
      ( 4.636e+01 , 1.800e-01 )
      ( 5.094e+01 , 1.903e-01 )
      ( 5.597e+01 , 2.015e-01 )
      ( 6.150e+01 , 2.139e-01 )
      ( 6.758e+01 , 2.275e-01 )
      ( 7.426e+01 , 2.424e-01 )
      ( 8.160e+01 , 2.589e-01 )
      ( 8.967e+01 , 2.769e-01 )
      ( 9.853e+01 , 2.967e-01 )
      ( 1.083e+02 , 3.185e-01 )
      ( 1.190e+02 , 3.424e-01 )
      ( 1.307e+02 , 3.687e-01 )
      ( 1.437e+02 , 3.976e-01 )
      ( 1.579e+02 , 4.294e-01 )
      ( 1.735e+02 , 4.643e-01 )
      ( 1.906e+02 , 5.026e-01 )
      ( 2.095e+02 , 5.448e-01 )
      ( 2.302e+02 , 5.911e-01 )
      ( 2.529e+02 , 6.420e-01 )
      ( 2.779e+02 , 6.979e-01 )
      ( 3.054e+02 , 7.593e-01 )
      ( 3.356e+02 , 8.268e-01 )
      ( 3.687e+02 , 9.010e-01 )
      ( 4.052e+02 , 9.825e-01 )
    };
    \addplot[solid, color=green, mark=none] coordinates {
      ( 4.052e+02 , 9.825e-01 )
      ( 4.445e+02 , 1.063e+00 )
      ( 4.877e+02 , 1.152e+00 )
      ( 5.350e+02 , 1.249e+00 )
    };
    \addplot[solid, color=green, mark=none, forget plot] coordinates {
      ( 5.350e+02 , 1.249e+00 )
      ( 5.880e+02 , 1.353e+00 )
      ( 6.463e+02 , 1.467e+00 )
      ( 7.103e+02 , 1.593e+00 )
      ( 7.807e+02 , 1.730e+00 )
      ( 8.580e+02 , 1.882e+00 )
      ( 9.430e+02 , 2.048e+00 )
      ( 1.036e+03 , 2.231e+00 )
      ( 1.139e+03 , 2.432e+00 )
      ( 1.252e+03 , 2.653e+00 )
      ( 1.376e+03 , 2.896e+00 )
      ( 1.512e+03 , 3.163e+00 )
      ( 1.662e+03 , 3.457e+00 )
      ( 1.827e+03 , 3.779e+00 )
      ( 2.008e+03 , 4.133e+00 )
      ( 2.206e+03 , 4.523e+00 )
      ( 2.425e+03 , 4.951e+00 )
      ( 2.665e+03 , 5.421e+00 )
      ( 2.929e+03 , 5.938e+00 )
      ( 3.219e+03 , 6.507e+00 )
      ( 3.538e+03 , 7.131e+00 )
      ( 3.889e+03 , 7.818e+00 )
      ( 4.274e+03 , 8.572e+00 )
      ( 4.697e+03 , 9.401e+00 )
      ( 5.163e+03 , 1.031e+01 )
      ( 5.674e+03 , 1.131e+01 )
      ( 6.236e+03 , 1.241e+01 )
      ( 6.854e+03 , 1.362e+01 )
      ( 7.533e+03 , 1.495e+01 )
      ( 8.279e+03 , 1.642e+01 )
      ( 9.099e+03 , 1.802e+01 )
      ( 1.000e+04 , 1.979e+01 )
    };
    \addplot[dashed, color={rgb,255:red,0;green,255;blue,255}, mark=none, forget plot] coordinates {
      ( 1.000e+00 , 5.962e-02 )
      ( 1.098e+00 , 6.100e-02 )
      ( 1.205e+00 , 6.251e-02 )
      ( 1.322e+00 , 6.417e-02 )
      ( 1.451e+00 , 6.599e-02 )
      ( 1.593e+00 , 6.799e-02 )
      ( 1.748e+00 , 7.019e-02 )
      ( 1.919e+00 , 7.259e-02 )
      ( 2.106e+00 , 7.524e-02 )
      ( 2.312e+00 , 7.814e-02 )
      ( 2.538e+00 , 8.133e-02 )
      ( 2.785e+00 , 8.482e-02 )
      ( 3.057e+00 , 8.866e-02 )
      ( 3.355e+00 , 9.287e-02 )
      ( 3.683e+00 , 9.749e-02 )
      ( 4.042e+00 , 1.026e-01 )
      ( 4.437e+00 , 1.081e-01 )
      ( 4.870e+00 , 1.143e-01 )
      ( 5.345e+00 , 1.210e-01 )
      ( 5.866e+00 , 1.283e-01 )
      ( 6.439e+00 , 1.364e-01 )
      ( 7.067e+00 , 1.453e-01 )
      ( 7.757e+00 , 1.550e-01 )
      ( 8.514e+00 , 1.657e-01 )
      ( 9.345e+00 , 1.774e-01 )
      ( 1.026e+01 , 1.903e-01 )
    };
    \addplot[solid, color=green, mark=none, forget plot] coordinates {
      ( 1.026e+01 , 1.903e-01 )
      ( 1.127e+01 , 1.929e-01 )
      ( 1.239e+01 , 1.958e-01 )
      ( 1.361e+01 , 1.989e-01 )
      ( 1.496e+01 , 2.024e-01 )
      ( 1.643e+01 , 2.062e-01 )
      ( 1.806e+01 , 2.103e-01 )
      ( 1.984e+01 , 2.149e-01 )
      ( 2.181e+01 , 2.200e-01 )
      ( 2.396e+01 , 2.255e-01 )
      ( 2.633e+01 , 2.316e-01 )
      ( 2.893e+01 , 2.383e-01 )
      ( 3.180e+01 , 2.456e-01 )
      ( 3.494e+01 , 2.537e-01 )
      ( 3.839e+01 , 2.625e-01 )
      ( 4.219e+01 , 2.723e-01 )
      ( 4.636e+01 , 2.830e-01 )
      ( 5.094e+01 , 2.948e-01 )
      ( 5.598e+01 , 3.077e-01 )
      ( 6.151e+01 , 3.219e-01 )
      ( 6.760e+01 , 3.375e-01 )
      ( 7.428e+01 , 3.547e-01 )
      ( 8.162e+01 , 3.735e-01 )
      ( 8.969e+01 , 3.942e-01 )
      ( 9.856e+01 , 4.170e-01 )
      ( 1.083e+02 , 4.420e-01 )
      ( 1.190e+02 , 4.695e-01 )
      ( 1.308e+02 , 4.997e-01 )
      ( 1.437e+02 , 5.329e-01 )
      ( 1.579e+02 , 5.694e-01 )
      ( 1.735e+02 , 6.094e-01 )
      ( 1.907e+02 , 6.535e-01 )
      ( 2.095e+02 , 7.019e-01 )
      ( 2.303e+02 , 7.551e-01 )
      ( 2.530e+02 , 8.135e-01 )
      ( 2.780e+02 , 8.777e-01 )
      ( 3.055e+02 , 9.483e-01 )
      ( 3.357e+02 , 1.026e+00 )
      ( 3.689e+02 , 1.111e+00 )
      ( 4.054e+02 , 1.205e+00 )
      ( 4.455e+02 , 1.308e+00 )
      ( 4.895e+02 , 1.421e+00 )
      ( 5.379e+02 , 1.545e+00 )
      ( 5.911e+02 , 1.681e+00 )
      ( 6.495e+02 , 1.831e+00 )
      ( 7.137e+02 , 1.996e+00 )
      ( 7.843e+02 , 2.177e+00 )
      ( 8.619e+02 , 2.376e+00 )
      ( 9.471e+02 , 2.595e+00 )
      ( 1.041e+03 , 2.836e+00 )
      ( 1.144e+03 , 3.100e+00 )
      ( 1.257e+03 , 3.390e+00 )
      ( 1.381e+03 , 3.709e+00 )
      ( 1.517e+03 , 4.059e+00 )
      ( 1.667e+03 , 4.444e+00 )
      ( 1.832e+03 , 4.868e+00 )
      ( 2.013e+03 , 5.333e+00 )
      ( 2.212e+03 , 5.844e+00 )
      ( 2.431e+03 , 6.405e+00 )
      ( 2.672e+03 , 7.022e+00 )
      ( 2.936e+03 , 7.700e+00 )
      ( 3.226e+03 , 8.445e+00 )
      ( 3.545e+03 , 9.264e+00 )
      ( 3.895e+03 , 1.016e+01 )
      ( 4.281e+03 , 1.115e+01 )
      ( 4.704e+03 , 1.224e+01 )
      ( 5.169e+03 , 1.343e+01 )
      ( 5.680e+03 , 1.474e+01 )
      ( 6.241e+03 , 1.619e+01 )
      ( 6.858e+03 , 1.777e+01 )
      ( 7.536e+03 , 1.951e+01 )
      ( 8.282e+03 , 2.142e+01 )
      ( 9.100e+03 , 2.353e+01 )
      ( 1.000e+04 , 2.584e+01 )
    };
    \addplot[dashed, color={rgb,255:red,0;green,255;blue,255}, mark=none, forget plot] coordinates {
      ( 1.000e+00 , 1.382e-01 )
      ( 1.097e+00 , 1.409e-01 )
      ( 1.203e+00 , 1.440e-01 )
      ( 1.319e+00 , 1.473e-01 )
      ( 1.447e+00 , 1.510e-01 )
      ( 1.587e+00 , 1.550e-01 )
      ( 1.741e+00 , 1.595e-01 )
      ( 1.909e+00 , 1.643e-01 )
      ( 2.094e+00 , 1.696e-01 )
      ( 2.297e+00 , 1.755e-01 )
      ( 2.519e+00 , 1.819e-01 )
      ( 2.763e+00 , 1.889e-01 )
      ( 3.031e+00 , 1.966e-01 )
      ( 3.324e+00 , 2.050e-01 )
      ( 3.646e+00 , 2.143e-01 )
      ( 3.999e+00 , 2.244e-01 )
      ( 4.386e+00 , 2.356e-01 )
      ( 4.810e+00 , 2.478e-01 )
      ( 5.276e+00 , 2.612e-01 )
      ( 5.786e+00 , 2.758e-01 )
      ( 6.347e+00 , 2.920e-01 )
      ( 6.961e+00 , 3.096e-01 )
      ( 7.635e+00 , 3.290e-01 )
      ( 8.374e+00 , 3.503e-01 )
      ( 9.184e+00 , 3.736e-01 )
      ( 1.007e+01 , 3.992e-01 )
    };
    \addplot[solid, color=green, mark=none, forget plot] coordinates {
      ( 1.007e+01 , 3.992e-01 )
      ( 1.107e+01 , 4.031e-01 )
      ( 1.217e+01 , 4.074e-01 )
      ( 1.337e+01 , 4.121e-01 )
      ( 1.469e+01 , 4.173e-01 )
      ( 1.615e+01 , 4.230e-01 )
      ( 1.774e+01 , 4.292e-01 )
      ( 1.950e+01 , 4.361e-01 )
      ( 2.143e+01 , 4.437e-01 )
      ( 2.355e+01 , 4.520e-01 )
      ( 2.588e+01 , 4.611e-01 )
      ( 2.844e+01 , 4.712e-01 )
      ( 3.125e+01 , 4.822e-01 )
      ( 3.434e+01 , 4.943e-01 )
      ( 3.774e+01 , 5.076e-01 )
      ( 4.147e+01 , 5.223e-01 )
      ( 4.558e+01 , 5.383e-01 )
      ( 5.009e+01 , 5.560e-01 )
      ( 5.504e+01 , 5.754e-01 )
      ( 6.049e+01 , 5.968e-01 )
      ( 6.647e+01 , 6.202e-01 )
      ( 7.305e+01 , 6.460e-01 )
      ( 8.028e+01 , 6.743e-01 )
      ( 8.822e+01 , 7.055e-01 )
      ( 9.695e+01 , 7.397e-01 )
      ( 1.065e+02 , 7.773e-01 )
      ( 1.171e+02 , 8.186e-01 )
      ( 1.287e+02 , 8.640e-01 )
      ( 1.414e+02 , 9.139e-01 )
      ( 1.554e+02 , 9.688e-01 )
      ( 1.708e+02 , 1.029e+00 )
      ( 1.877e+02 , 1.095e+00 )
      ( 2.062e+02 , 1.168e+00 )
      ( 2.266e+02 , 1.248e+00 )
      ( 2.491e+02 , 1.336e+00 )
      ( 2.737e+02 , 1.432e+00 )
      ( 3.008e+02 , 1.539e+00 )
      ( 3.305e+02 , 1.655e+00 )
      ( 3.632e+02 , 1.783e+00 )
      ( 3.992e+02 , 1.924e+00 )
      ( 4.387e+02 , 2.079e+00 )
      ( 4.821e+02 , 2.249e+00 )
      ( 5.298e+02 , 2.436e+00 )
      ( 5.822e+02 , 2.642e+00 )
      ( 6.398e+02 , 2.867e+00 )
      ( 7.031e+02 , 3.116e+00 )
    };
    \addplot[solid, color=green, mark=none, forget plot] coordinates {
      ( 7.031e+02 , 3.116e+00 )
      ( 7.730e+02 , 3.371e+00 )
      ( 8.499e+02 , 3.651e+00 )
      ( 9.344e+02 , 3.959e+00 )
      ( 1.027e+03 , 4.298e+00 )
      ( 1.130e+03 , 4.671e+00 )
      ( 1.242e+03 , 5.081e+00 )
      ( 1.365e+03 , 5.531e+00 )
      ( 1.501e+03 , 6.027e+00 )
      ( 1.651e+03 , 6.571e+00 )
      ( 1.815e+03 , 7.170e+00 )
      ( 1.995e+03 , 7.828e+00 )
      ( 2.194e+03 , 8.552e+00 )
      ( 2.412e+03 , 9.348e+00 )
      ( 2.652e+03 , 1.022e+01 )
      ( 2.915e+03 , 1.118e+01 )
      ( 3.205e+03 , 1.224e+01 )
      ( 3.524e+03 , 1.340e+01 )
      ( 3.875e+03 , 1.468e+01 )
      ( 4.260e+03 , 1.609e+01 )
      ( 4.684e+03 , 1.763e+01 )
      ( 5.149e+03 , 1.933e+01 )
      ( 5.662e+03 , 2.120e+01 )
      ( 6.225e+03 , 2.325e+01 )
      ( 6.844e+03 , 2.551e+01 )
      ( 7.524e+03 , 2.800e+01 )
      ( 8.273e+03 , 3.073e+01 )
      ( 9.095e+03 , 3.373e+01 )
      ( 1.000e+04 , 3.703e+01 )
    };
    \addplot[dashed, color={rgb,255:red,0;green,255;blue,255}, mark=none, forget plot] coordinates {
      ( 1.000e+00 , 4.478e-01 )
      ( 1.097e+00 , 4.563e-01 )
      ( 1.204e+00 , 4.655e-01 )
      ( 1.321e+00 , 4.757e-01 )
      ( 1.450e+00 , 4.868e-01 )
      ( 1.591e+00 , 4.991e-01 )
      ( 1.746e+00 , 5.125e-01 )
      ( 1.916e+00 , 5.272e-01 )
      ( 2.102e+00 , 5.434e-01 )
      ( 2.306e+00 , 5.611e-01 )
      ( 2.531e+00 , 5.806e-01 )
      ( 2.777e+00 , 6.019e-01 )
      ( 3.047e+00 , 6.254e-01 )
      ( 3.344e+00 , 6.511e-01 )
      ( 3.669e+00 , 6.793e-01 )
      ( 4.026e+00 , 7.103e-01 )
      ( 4.418e+00 , 7.442e-01 )
      ( 4.848e+00 , 7.815e-01 )
      ( 5.320e+00 , 8.224e-01 )
      ( 5.838e+00 , 8.673e-01 )
      ( 6.406e+00 , 9.166e-01 )
      ( 7.029e+00 , 9.706e-01 )
      ( 7.713e+00 , 1.030e+00 )
      ( 8.463e+00 , 1.095e+00 )
      ( 9.287e+00 , 1.166e+00 )
    };
    \addplot[solid, color=green, mark=none, forget plot] coordinates {
      ( 9.287e+00 , 1.166e+00 )
      ( 1.021e+01 , 1.173e+00 )
      ( 1.122e+01 , 1.180e+00 )
      ( 1.233e+01 , 1.188e+00 )
      ( 1.354e+01 , 1.197e+00 )
      ( 1.488e+01 , 1.206e+00 )
      ( 1.636e+01 , 1.216e+00 )
      ( 1.798e+01 , 1.228e+00 )
      ( 1.975e+01 , 1.240e+00 )
      ( 2.171e+01 , 1.254e+00 )
      ( 2.386e+01 , 1.269e+00 )
      ( 2.622e+01 , 1.286e+00 )
      ( 2.881e+01 , 1.304e+00 )
      ( 3.166e+01 , 1.325e+00 )
      ( 3.479e+01 , 1.347e+00 )
      ( 3.824e+01 , 1.371e+00 )
      ( 4.202e+01 , 1.398e+00 )
      ( 4.618e+01 , 1.427e+00 )
      ( 5.075e+01 , 1.459e+00 )
      ( 5.577e+01 , 1.495e+00 )
      ( 6.129e+01 , 1.534e+00 )
      ( 6.735e+01 , 1.577e+00 )
      ( 7.401e+01 , 1.624e+00 )
      ( 8.134e+01 , 1.676e+00 )
      ( 8.938e+01 , 1.732e+00 )
      ( 9.823e+01 , 1.795e+00 )
      ( 1.079e+02 , 1.864e+00 )
      ( 1.186e+02 , 1.939e+00 )
      ( 1.304e+02 , 2.022e+00 )
      ( 1.433e+02 , 2.113e+00 )
      ( 1.574e+02 , 2.213e+00 )
      ( 1.730e+02 , 2.324e+00 )
      ( 1.901e+02 , 2.444e+00 )
      ( 2.089e+02 , 2.577e+00 )
      ( 2.296e+02 , 2.724e+00 )
      ( 2.523e+02 , 2.884e+00 )
      ( 2.773e+02 , 3.061e+00 )
      ( 3.047e+02 , 3.254e+00 )
      ( 3.349e+02 , 3.468e+00 )
      ( 3.680e+02 , 3.702e+00 )
      ( 4.044e+02 , 3.959e+00 )
      ( 4.445e+02 , 4.242e+00 )
      ( 4.884e+02 , 4.553e+00 )
      ( 5.368e+02 , 4.894e+00 )
      ( 5.899e+02 , 5.269e+00 )
      ( 6.482e+02 , 5.682e+00 )
      ( 7.124e+02 , 6.135e+00 )
      ( 7.829e+02 , 6.633e+00 )
      ( 8.603e+02 , 7.181e+00 )
      ( 9.454e+02 , 7.782e+00 )
      ( 1.039e+03 , 8.443e+00 )
      ( 1.142e+03 , 9.170e+00 )
      ( 1.255e+03 , 9.968e+00 )
      ( 1.379e+03 , 1.085e+01 )
      ( 1.515e+03 , 1.181e+01 )
      ( 1.665e+03 , 1.287e+01 )
      ( 1.830e+03 , 1.403e+01 )
      ( 2.011e+03 , 1.531e+01 )
      ( 2.210e+03 , 1.672e+01 )
      ( 2.429e+03 , 1.826e+01 )
      ( 2.669e+03 , 1.996e+01 )
      ( 2.933e+03 , 2.183e+01 )
      ( 3.223e+03 , 2.388e+01 )
      ( 3.542e+03 , 2.613e+01 )
      ( 3.893e+03 , 2.861e+01 )
      ( 4.278e+03 , 3.133e+01 )
      ( 4.701e+03 , 3.432e+01 )
      ( 5.166e+03 , 3.761e+01 )
      ( 5.677e+03 , 4.122e+01 )
      ( 6.239e+03 , 4.519e+01 )
      ( 6.856e+03 , 4.956e+01 )
      ( 7.535e+03 , 5.435e+01 )
      ( 8.280e+03 , 5.962e+01 )
      ( 9.100e+03 , 6.541e+01 )
      ( 1.000e+04 , 7.177e+01 )
    };
    \addplot[dashed, color={rgb,255:red,0;green,255;blue,255}, mark=none, forget plot] coordinates {
      ( 1.000e+00 , 1.089e+00 )
      ( 1.098e+00 , 1.111e+00 )
      ( 1.206e+00 , 1.136e+00 )
      ( 1.325e+00 , 1.163e+00 )
      ( 1.455e+00 , 1.193e+00 )
      ( 1.598e+00 , 1.225e+00 )
      ( 1.756e+00 , 1.261e+00 )
      ( 1.928e+00 , 1.300e+00 )
      ( 2.118e+00 , 1.343e+00 )
      ( 2.326e+00 , 1.390e+00 )
      ( 2.555e+00 , 1.442e+00 )
      ( 2.806e+00 , 1.499e+00 )
      ( 3.082e+00 , 1.562e+00 )
      ( 3.385e+00 , 1.631e+00 )
      ( 3.718e+00 , 1.707e+00 )
      ( 4.084e+00 , 1.790e+00 )
      ( 4.485e+00 , 1.881e+00 )
      ( 4.926e+00 , 1.981e+00 )
      ( 5.411e+00 , 2.091e+00 )
      ( 5.943e+00 , 2.212e+00 )
      ( 6.527e+00 , 2.345e+00 )
      ( 7.169e+00 , 2.491e+00 )
    };
    \addplot[solid, color=green, mark=none, forget plot] coordinates {
      ( 7.169e+00 , 2.491e+00 )
      ( 7.879e+00 , 2.500e+00 )
      ( 8.659e+00 , 2.510e+00 )
      ( 9.516e+00 , 2.522e+00 )
      ( 1.046e+01 , 2.534e+00 )
      ( 1.149e+01 , 2.548e+00 )
      ( 1.263e+01 , 2.563e+00 )
      ( 1.388e+01 , 2.579e+00 )
      ( 1.526e+01 , 2.598e+00 )
      ( 1.677e+01 , 2.618e+00 )
      ( 1.842e+01 , 2.640e+00 )
      ( 2.025e+01 , 2.664e+00 )
      ( 2.225e+01 , 2.690e+00 )
      ( 2.446e+01 , 2.720e+00 )
      ( 2.688e+01 , 2.752e+00 )
      ( 2.954e+01 , 2.787e+00 )
      ( 3.246e+01 , 2.826e+00 )
      ( 3.567e+01 , 2.868e+00 )
      ( 3.921e+01 , 2.915e+00 )
      ( 4.309e+01 , 2.966e+00 )
      ( 4.735e+01 , 3.023e+00 )
      ( 5.204e+01 , 3.085e+00 )
      ( 5.719e+01 , 3.153e+00 )
      ( 6.285e+01 , 3.228e+00 )
      ( 6.907e+01 , 3.311e+00 )
      ( 7.591e+01 , 3.401e+00 )
      ( 8.342e+01 , 3.501e+00 )
      ( 9.168e+01 , 3.610e+00 )
      ( 1.008e+02 , 3.731e+00 )
      ( 1.107e+02 , 3.863e+00 )
      ( 1.217e+02 , 4.008e+00 )
      ( 1.337e+02 , 4.168e+00 )
      ( 1.470e+02 , 4.343e+00 )
      ( 1.615e+02 , 4.536e+00 )
      ( 1.775e+02 , 4.748e+00 )
      ( 1.951e+02 , 4.980e+00 )
      ( 2.144e+02 , 5.236e+00 )
      ( 2.356e+02 , 5.517e+00 )
      ( 2.589e+02 , 5.826e+00 )
      ( 2.846e+02 , 6.166e+00 )
      ( 3.127e+02 , 6.539e+00 )
      ( 3.437e+02 , 6.950e+00 )
      ( 3.777e+02 , 7.400e+00 )
      ( 4.151e+02 , 7.896e+00 )
      ( 4.562e+02 , 8.440e+00 )
      ( 5.014e+02 , 9.039e+00 )
      ( 5.510e+02 , 9.696e+00 )
      ( 6.055e+02 , 1.042e+01 )
      ( 6.655e+02 , 1.121e+01 )
      ( 7.313e+02 , 1.209e+01 )
      ( 8.037e+02 , 1.304e+01 )
      ( 8.833e+02 , 1.410e+01 )
      ( 9.707e+02 , 1.526e+01 )
      ( 1.067e+03 , 1.653e+01 )
      ( 1.172e+03 , 1.793e+01 )
      ( 1.288e+03 , 1.947e+01 )
      ( 1.416e+03 , 2.116e+01 )
      ( 1.556e+03 , 2.301e+01 )
      ( 1.710e+03 , 2.506e+01 )
      ( 1.879e+03 , 2.730e+01 )
      ( 2.066e+03 , 2.976e+01 )
      ( 2.270e+03 , 3.247e+01 )
      ( 2.495e+03 , 3.545e+01 )
      ( 2.742e+03 , 3.872e+01 )
      ( 3.013e+03 , 4.232e+01 )
      ( 3.311e+03 , 4.627e+01 )
      ( 3.639e+03 , 5.061e+01 )
      ( 3.999e+03 , 5.539e+01 )
      ( 4.395e+03 , 6.063e+01 )
      ( 4.830e+03 , 6.640e+01 )
    };
    \addplot[solid, color=green, mark=none, forget plot] coordinates {
      ( 4.830e+03 , 6.640e+01 )
      ( 5.290e+03 , 7.222e+01 )
      ( 5.794e+03 , 7.860e+01 )
      ( 6.346e+03 , 8.559e+01 )
      ( 6.950e+03 , 9.325e+01 )
      ( 7.612e+03 , 1.016e+02 )
      ( 8.337e+03 , 1.108e+02 )
      ( 9.131e+03 , 1.209e+02 )
      ( 1.000e+04 , 1.319e+02 )
    };
    \addplot[dashed, color={rgb,255:red,0;green,255;blue,255}, mark=none, forget plot] coordinates {
      ( 1.000e+00 , 3.217e+00 )
      ( 1.096e+00 , 3.278e+00 )
      ( 1.202e+00 , 3.345e+00 )
      ( 1.318e+00 , 3.419e+00 )
      ( 1.445e+00 , 3.500e+00 )
      ( 1.584e+00 , 3.588e+00 )
      ( 1.736e+00 , 3.685e+00 )
      ( 1.904e+00 , 3.791e+00 )
      ( 2.087e+00 , 3.908e+00 )
      ( 2.288e+00 , 4.035e+00 )
      ( 2.508e+00 , 4.175e+00 )
      ( 2.750e+00 , 4.329e+00 )
      ( 3.015e+00 , 4.497e+00 )
      ( 3.305e+00 , 4.682e+00 )
      ( 3.624e+00 , 4.884e+00 )
      ( 3.973e+00 , 5.106e+00 )
      ( 4.355e+00 , 5.349e+00 )
      ( 4.775e+00 , 5.616e+00 )
      ( 5.235e+00 , 5.908e+00 )
      ( 5.739e+00 , 6.228e+00 )
      ( 6.292e+00 , 6.579e+00 )
    };
    \addplot[solid, color=green, mark=none, forget plot] coordinates {
      ( 6.292e+00 , 6.579e+00 )
      ( 6.917e+00 , 6.596e+00 )
      ( 7.605e+00 , 6.614e+00 )
      ( 8.360e+00 , 6.634e+00 )
      ( 9.191e+00 , 6.655e+00 )
      ( 1.010e+01 , 6.679e+00 )
      ( 1.111e+01 , 6.705e+00 )
      ( 1.221e+01 , 6.734e+00 )
      ( 1.343e+01 , 6.766e+00 )
      ( 1.476e+01 , 6.801e+00 )
      ( 1.623e+01 , 6.839e+00 )
      ( 1.784e+01 , 6.882e+00 )
      ( 1.961e+01 , 6.928e+00 )
      ( 2.156e+01 , 6.979e+00 )
      ( 2.371e+01 , 7.035e+00 )
      ( 2.606e+01 , 7.097e+00 )
      ( 2.865e+01 , 7.165e+00 )
      ( 3.150e+01 , 7.239e+00 )
      ( 3.463e+01 , 7.321e+00 )
      ( 3.807e+01 , 7.411e+00 )
      ( 4.186e+01 , 7.510e+00 )
      ( 4.602e+01 , 7.619e+00 )
      ( 5.059e+01 , 7.739e+00 )
      ( 5.562e+01 , 7.870e+00 )
      ( 6.115e+01 , 8.015e+00 )
      ( 6.723e+01 , 8.174e+00 )
      ( 7.391e+01 , 8.349e+00 )
      ( 8.125e+01 , 8.541e+00 )
      ( 8.933e+01 , 8.753e+00 )
      ( 9.821e+01 , 8.985e+00 )
      ( 1.080e+02 , 9.240e+00 )
      ( 1.187e+02 , 9.521e+00 )
      ( 1.305e+02 , 9.830e+00 )
      ( 1.435e+02 , 1.017e+01 )
      ( 1.577e+02 , 1.054e+01 )
      ( 1.734e+02 , 1.095e+01 )
      ( 1.906e+02 , 1.140e+01 )
      ( 2.096e+02 , 1.190e+01 )
      ( 2.304e+02 , 1.244e+01 )
      ( 2.533e+02 , 1.304e+01 )
      ( 2.785e+02 , 1.370e+01 )
      ( 3.062e+02 , 1.443e+01 )
      ( 3.366e+02 , 1.522e+01 )
      ( 3.700e+02 , 1.610e+01 )
      ( 4.068e+02 , 1.706e+01 )
      ( 4.472e+02 , 1.812e+01 )
      ( 4.917e+02 , 1.928e+01 )
      ( 5.406e+02 , 2.056e+01 )
      ( 5.943e+02 , 2.197e+01 )
      ( 6.534e+02 , 2.351e+01 )
      ( 7.183e+02 , 2.521e+01 )
      ( 7.897e+02 , 2.708e+01 )
      ( 8.682e+02 , 2.914e+01 )
      ( 9.544e+02 , 3.139e+01 )
      ( 1.049e+03 , 3.388e+01 )
      ( 1.154e+03 , 3.661e+01 )
      ( 1.268e+03 , 3.961e+01 )
      ( 1.394e+03 , 4.290e+01 )
      ( 1.533e+03 , 4.653e+01 )
      ( 1.685e+03 , 5.052e+01 )
      ( 1.853e+03 , 5.490e+01 )
      ( 2.037e+03 , 5.972e+01 )
      ( 2.239e+03 , 6.502e+01 )
      ( 2.462e+03 , 7.084e+01 )
      ( 2.706e+03 , 7.725e+01 )
      ( 2.975e+03 , 8.429e+01 )
      ( 3.271e+03 , 9.203e+01 )
      ( 3.596e+03 , 1.005e+02 )
      ( 3.954e+03 , 1.099e+02 )
      ( 4.347e+03 , 1.202e+02 )
      ( 4.779e+03 , 1.315e+02 )
      ( 5.254e+03 , 1.439e+02 )
      ( 5.776e+03 , 1.576e+02 )
      ( 6.350e+03 , 1.726e+02 )
      ( 6.981e+03 , 1.891e+02 )
      ( 7.675e+03 , 2.073e+02 )
    };
    \addplot[solid, color=green, mark=none, forget plot] coordinates {
      ( 7.675e+03 , 2.073e+02 )
      ( 8.383e+03 , 2.246e+02 )
      ( 9.156e+03 , 2.436e+02 )
      ( 1.000e+04 , 2.643e+02 )
    };
    \addplot[dashed, color={rgb,255:red,0;green,255;blue,255}, mark=none, forget plot] coordinates {
      ( 1.000e+00 , 9.636e+00 )
      ( 1.097e+00 , 9.793e+00 )
      ( 1.204e+00 , 9.966e+00 )
      ( 1.320e+00 , 1.015e+01 )
      ( 1.449e+00 , 1.036e+01 )
      ( 1.589e+00 , 1.059e+01 )
      ( 1.744e+00 , 1.084e+01 )
      ( 1.913e+00 , 1.111e+01 )
      ( 2.099e+00 , 1.141e+01 )
      ( 2.302e+00 , 1.174e+01 )
      ( 2.526e+00 , 1.210e+01 )
      ( 2.771e+00 , 1.250e+01 )
      ( 3.040e+00 , 1.293e+01 )
      ( 3.335e+00 , 1.341e+01 )
      ( 3.659e+00 , 1.393e+01 )
      ( 4.014e+00 , 1.451e+01 )
      ( 4.404e+00 , 1.514e+01 )
      ( 4.831e+00 , 1.583e+01 )
      ( 5.301e+00 , 1.659e+01 )
      ( 5.815e+00 , 1.742e+01 )
      ( 6.380e+00 , 1.833e+01 )
      ( 6.999e+00 , 1.933e+01 )
      ( 7.679e+00 , 2.043e+01 )
    };
    \addplot[solid, color=green, mark=none, forget plot] coordinates {
      ( 7.679e+00 , 2.043e+01 )
      ( 8.443e+00 , 2.048e+01 )
      ( 9.284e+00 , 2.054e+01 )
      ( 1.021e+01 , 2.060e+01 )
      ( 1.122e+01 , 2.067e+01 )
      ( 1.234e+01 , 2.074e+01 )
      ( 1.357e+01 , 2.083e+01 )
      ( 1.492e+01 , 2.092e+01 )
      ( 1.641e+01 , 2.102e+01 )
      ( 1.804e+01 , 2.113e+01 )
      ( 1.984e+01 , 2.125e+01 )
      ( 2.181e+01 , 2.138e+01 )
      ( 2.398e+01 , 2.152e+01 )
      ( 2.637e+01 , 2.168e+01 )
      ( 2.899e+01 , 2.186e+01 )
      ( 3.188e+01 , 2.205e+01 )
      ( 3.506e+01 , 2.226e+01 )
      ( 3.855e+01 , 2.250e+01 )
      ( 4.238e+01 , 2.276e+01 )
      ( 4.660e+01 , 2.304e+01 )
      ( 5.124e+01 , 2.335e+01 )
      ( 5.634e+01 , 2.369e+01 )
      ( 6.195e+01 , 2.407e+01 )
      ( 6.812e+01 , 2.448e+01 )
      ( 7.490e+01 , 2.493e+01 )
      ( 8.236e+01 , 2.543e+01 )
      ( 9.056e+01 , 2.598e+01 )
      ( 9.957e+01 , 2.658e+01 )
      ( 1.095e+02 , 2.725e+01 )
      ( 1.204e+02 , 2.798e+01 )
      ( 1.324e+02 , 2.878e+01 )
      ( 1.456e+02 , 2.966e+01 )
      ( 1.600e+02 , 3.063e+01 )
      ( 1.760e+02 , 3.170e+01 )
      ( 1.935e+02 , 3.287e+01 )
      ( 2.128e+02 , 3.416e+01 )
      ( 2.339e+02 , 3.558e+01 )
      ( 2.572e+02 , 3.714e+01 )
      ( 2.828e+02 , 3.886e+01 )
      ( 3.110e+02 , 4.074e+01 )
      ( 3.420e+02 , 4.281e+01 )
      ( 3.760e+02 , 4.509e+01 )
      ( 4.134e+02 , 4.760e+01 )
      ( 4.546e+02 , 5.036e+01 )
      ( 4.999e+02 , 5.339e+01 )
      ( 5.496e+02 , 5.672e+01 )
      ( 6.043e+02 , 6.038e+01 )
      ( 6.645e+02 , 6.441e+01 )
      ( 7.307e+02 , 6.884e+01 )
      ( 8.034e+02 , 7.371e+01 )
      ( 8.834e+02 , 7.907e+01 )
      ( 9.713e+02 , 8.496e+01 )
      ( 1.068e+03 , 9.143e+01 )
      ( 1.174e+03 , 9.855e+01 )
      ( 1.291e+03 , 1.064e+02 )
      ( 1.420e+03 , 1.150e+02 )
      ( 1.561e+03 , 1.245e+02 )
      ( 1.717e+03 , 1.349e+02 )
      ( 1.888e+03 , 1.463e+02 )
      ( 2.075e+03 , 1.589e+02 )
      ( 2.282e+03 , 1.727e+02 )
      ( 2.509e+03 , 1.879e+02 )
      ( 2.759e+03 , 2.047e+02 )
      ( 3.034e+03 , 2.230e+02 )
      ( 3.336e+03 , 2.433e+02 )
      ( 3.668e+03 , 2.655e+02 )
    };
    \addplot[solid, color=green, mark=none, forget plot] coordinates {
      ( 3.668e+03 , 2.655e+02 )
      ( 4.018e+03 , 2.847e+02 )
      ( 4.402e+03 , 3.056e+02 )
      ( 4.822e+03 , 3.286e+02 )
      ( 5.282e+03 , 3.538e+02 )
      ( 5.786e+03 , 3.814e+02 )
      ( 6.339e+03 , 4.116e+02 )
      ( 6.944e+03 , 4.447e+02 )
      ( 7.607e+03 , 4.810e+02 )
      ( 8.333e+03 , 5.207e+02 )
      ( 9.129e+03 , 5.642e+02 )
      ( 1.000e+04 , 6.119e+02 )
    };
    \addplot[dashed, color={rgb,255:red,0;green,255;blue,255}, mark=none, forget plot] coordinates {
      ( 1.000e+00 , 3.499e+01 )
      ( 1.096e+00 , 3.543e+01 )
      ( 1.202e+00 , 3.590e+01 )
      ( 1.317e+00 , 3.643e+01 )
      ( 1.444e+00 , 3.701e+01 )
      ( 1.583e+00 , 3.764e+01 )
      ( 1.735e+00 , 3.833e+01 )
      ( 1.902e+00 , 3.909e+01 )
      ( 2.085e+00 , 3.992e+01 )
      ( 2.285e+00 , 4.083e+01 )
      ( 2.505e+00 , 4.183e+01 )
      ( 2.746e+00 , 4.293e+01 )
      ( 3.010e+00 , 4.413e+01 )
      ( 3.300e+00 , 4.545e+01 )
      ( 3.617e+00 , 4.689e+01 )
      ( 3.965e+00 , 4.848e+01 )
      ( 4.346e+00 , 5.021e+01 )
      ( 4.764e+00 , 5.211e+01 )
      ( 5.223e+00 , 5.420e+01 )
      ( 5.725e+00 , 5.648e+01 )
      ( 6.276e+00 , 5.899e+01 )
      ( 6.879e+00 , 6.173e+01 )
      ( 7.541e+00 , 6.474e+01 )
      ( 8.266e+00 , 6.804e+01 )
      ( 9.061e+00 , 7.166e+01 )
      ( 9.933e+00 , 7.563e+01 )
      ( 1.089e+01 , 7.997e+01 )
    };
    \addplot[solid, color=green, mark=none, forget plot] coordinates {
      ( 1.089e+01 , 7.997e+01 )
      ( 1.197e+01 , 8.011e+01 )
      ( 1.316e+01 , 8.026e+01 )
      ( 1.447e+01 , 8.043e+01 )
      ( 1.591e+01 , 8.061e+01 )
      ( 1.749e+01 , 8.081e+01 )
      ( 1.923e+01 , 8.103e+01 )
      ( 2.114e+01 , 8.127e+01 )
      ( 2.324e+01 , 8.154e+01 )
      ( 2.555e+01 , 8.183e+01 )
      ( 2.809e+01 , 8.215e+01 )
      ( 3.088e+01 , 8.251e+01 )
      ( 3.395e+01 , 8.290e+01 )
      ( 3.732e+01 , 8.333e+01 )
      ( 4.103e+01 , 8.380e+01 )
      ( 4.511e+01 , 8.431e+01 )
      ( 4.959e+01 , 8.488e+01 )
      ( 5.452e+01 , 8.551e+01 )
      ( 5.994e+01 , 8.620e+01 )
      ( 6.590e+01 , 8.695e+01 )
      ( 7.245e+01 , 8.778e+01 )
      ( 7.965e+01 , 8.870e+01 )
      ( 8.756e+01 , 8.970e+01 )
      ( 9.627e+01 , 9.080e+01 )
      ( 1.058e+02 , 9.202e+01 )
      ( 1.164e+02 , 9.335e+01 )
      ( 1.279e+02 , 9.482e+01 )
      ( 1.406e+02 , 9.643e+01 )
      ( 1.546e+02 , 9.821e+01 )
      ( 1.700e+02 , 1.002e+02 )
      ( 1.869e+02 , 1.023e+02 )
      ( 2.055e+02 , 1.047e+02 )
      ( 2.259e+02 , 1.072e+02 )
      ( 2.483e+02 , 1.101e+02 )
      ( 2.730e+02 , 1.132e+02 )
      ( 3.001e+02 , 1.167e+02 )
      ( 3.300e+02 , 1.205e+02 )
      ( 3.628e+02 , 1.246e+02 )
      ( 3.988e+02 , 1.292e+02 )
      ( 4.385e+02 , 1.342e+02 )
      ( 4.820e+02 , 1.397e+02 )
      ( 5.300e+02 , 1.458e+02 )
      ( 5.826e+02 , 1.525e+02 )
      ( 6.405e+02 , 1.599e+02 )
      ( 7.042e+02 , 1.679e+02 )
      ( 7.742e+02 , 1.768e+02 )
      ( 8.512e+02 , 1.866e+02 )
      ( 9.358e+02 , 1.973e+02 )
      ( 1.029e+03 , 2.091e+02 )
      ( 1.131e+03 , 2.221e+02 )
      ( 1.243e+03 , 2.363e+02 )
      ( 1.367e+03 , 2.520e+02 )
      ( 1.503e+03 , 2.693e+02 )
      ( 1.652e+03 , 2.882e+02 )
      ( 1.817e+03 , 3.090e+02 )
      ( 1.997e+03 , 3.320e+02 )
      ( 2.196e+03 , 3.571e+02 )
      ( 2.414e+03 , 3.848e+02 )
      ( 2.654e+03 , 4.153e+02 )
      ( 2.917e+03 , 4.487e+02 )
      ( 3.207e+03 , 4.855e+02 )
      ( 3.526e+03 , 5.260e+02 )
      ( 3.877e+03 , 5.704e+02 )
      ( 4.262e+03 , 6.193e+02 )
      ( 4.686e+03 , 6.731e+02 )
      ( 5.151e+03 , 7.321e+02 )
      ( 5.663e+03 , 7.971e+02 )
      ( 6.226e+03 , 8.685e+02 )
      ( 6.845e+03 , 9.470e+02 )
      ( 7.526e+03 , 1.033e+03 )
      ( 8.274e+03 , 1.128e+03 )
      ( 9.096e+03 , 1.233e+03 )
      ( 1.000e+04 , 1.347e+03 )
    };
    \addplot[dashed, color={rgb,255:red,0;green,255;blue,255}, mark=none, forget plot] coordinates {
      ( 1.000e+00 , 1.314e+02 )
      ( 1.100e+00 , 1.325e+02 )
      ( 1.210e+00 , 1.338e+02 )
      ( 1.331e+00 , 1.352e+02 )
      ( 1.464e+00 , 1.368e+02 )
      ( 1.610e+00 , 1.385e+02 )
      ( 1.771e+00 , 1.403e+02 )
      ( 1.948e+00 , 1.424e+02 )
      ( 2.143e+00 , 1.447e+02 )
      ( 2.357e+00 , 1.471e+02 )
      ( 2.593e+00 , 1.499e+02 )
      ( 2.852e+00 , 1.529e+02 )
      ( 3.138e+00 , 1.562e+02 )
      ( 3.451e+00 , 1.599e+02 )
      ( 3.796e+00 , 1.639e+02 )
      ( 4.176e+00 , 1.683e+02 )
      ( 4.593e+00 , 1.732e+02 )
      ( 5.052e+00 , 1.785e+02 )
      ( 5.558e+00 , 1.844e+02 )
      ( 6.113e+00 , 1.908e+02 )
      ( 6.724e+00 , 1.979e+02 )
      ( 7.397e+00 , 2.058e+02 )
      ( 8.136e+00 , 2.144e+02 )
      ( 8.949e+00 , 2.238e+02 )
      ( 9.844e+00 , 2.342e+02 )
      ( 1.083e+01 , 2.457e+02 )
      ( 1.191e+01 , 2.583e+02 )
      ( 1.310e+01 , 2.721e+02 )
      ( 1.441e+01 , 2.874e+02 )
    };
    \addplot[solid, color=green, mark=none, forget plot] coordinates {
      ( 1.441e+01 , 2.874e+02 )
      ( 1.584e+01 , 2.877e+02 )
      ( 1.742e+01 , 2.881e+02 )
      ( 1.915e+01 , 2.886e+02 )
      ( 2.106e+01 , 2.891e+02 )
      ( 2.315e+01 , 2.896e+02 )
      ( 2.546e+01 , 2.902e+02 )
      ( 2.799e+01 , 2.908e+02 )
      ( 3.077e+01 , 2.915e+02 )
      ( 3.383e+01 , 2.923e+02 )
      ( 3.720e+01 , 2.932e+02 )
      ( 4.089e+01 , 2.941e+02 )
      ( 4.496e+01 , 2.951e+02 )
      ( 4.943e+01 , 2.963e+02 )
      ( 5.435e+01 , 2.975e+02 )
      ( 5.976e+01 , 2.989e+02 )
      ( 6.570e+01 , 3.004e+02 )
      ( 7.223e+01 , 3.021e+02 )
      ( 7.942e+01 , 3.039e+02 )
      ( 8.732e+01 , 3.059e+02 )
      ( 9.600e+01 , 3.081e+02 )
      ( 1.055e+02 , 3.106e+02 )
      ( 1.160e+02 , 3.133e+02 )
      ( 1.276e+02 , 3.162e+02 )
      ( 1.403e+02 , 3.194e+02 )
      ( 1.542e+02 , 3.230e+02 )
      ( 1.696e+02 , 3.269e+02 )
      ( 1.864e+02 , 3.312e+02 )
      ( 2.050e+02 , 3.359e+02 )
      ( 2.254e+02 , 3.411e+02 )
      ( 2.478e+02 , 3.468e+02 )
      ( 2.724e+02 , 3.531e+02 )
      ( 2.995e+02 , 3.600e+02 )
      ( 3.293e+02 , 3.676e+02 )
      ( 3.620e+02 , 3.759e+02 )
      ( 3.981e+02 , 3.851e+02 )
      ( 4.376e+02 , 3.952e+02 )
      ( 4.812e+02 , 4.063e+02 )
      ( 5.290e+02 , 4.185e+02 )
      ( 5.816e+02 , 4.319e+02 )
      ( 6.395e+02 , 4.466e+02 )
      ( 7.031e+02 , 4.628e+02 )
      ( 7.730e+02 , 4.806e+02 )
      ( 8.499e+02 , 5.002e+02 )
      ( 9.344e+02 , 5.218e+02 )
      ( 1.027e+03 , 5.454e+02 )
      ( 1.130e+03 , 5.715e+02 )
      ( 1.242e+03 , 6.001e+02 )
      ( 1.365e+03 , 6.316e+02 )
      ( 1.501e+03 , 6.662e+02 )
      ( 1.651e+03 , 7.042e+02 )
      ( 1.815e+03 , 7.460e+02 )
      ( 1.995e+03 , 7.920e+02 )
      ( 2.194e+03 , 8.426e+02 )
      ( 2.412e+03 , 8.981e+02 )
      ( 2.652e+03 , 9.593e+02 )
      ( 2.915e+03 , 1.026e+03 )
      ( 3.205e+03 , 1.100e+03 )
      ( 3.524e+03 , 1.182e+03 )
      ( 3.875e+03 , 1.271e+03 )
      ( 4.260e+03 , 1.369e+03 )
      ( 4.684e+03 , 1.477e+03 )
      ( 5.149e+03 , 1.596e+03 )
      ( 5.661e+03 , 1.726e+03 )
      ( 6.225e+03 , 1.870e+03 )
      ( 6.844e+03 , 2.027e+03 )
      ( 7.524e+03 , 2.201e+03 )
      ( 8.273e+03 , 2.391e+03 )
      ( 9.095e+03 , 2.601e+03 )
      ( 1.000e+04 , 2.831e+03 )
    };
    \node[anchor=west, inner sep=1pt, fill=white, fill opacity=0.7, text opacity=1, font=\tiny] at (axis cs: 1.100e+04,8.654e+00) {64 DOFs};
    \node[anchor=west, inner sep=1pt, fill=white, fill opacity=0.7, text opacity=1, font=\tiny] at (axis cs: 1.100e+04,1.409e+01) {125 DOFs};
    \node[anchor=west, inner sep=1pt, fill=white, fill opacity=0.7, text opacity=1, font=\tiny] at (axis cs: 1.100e+04,1.694e+01) {216 DOFs};
    \node[anchor=west, inner sep=1pt, fill=white, fill opacity=0.7, text opacity=1, font=\tiny] at (axis cs: 1.100e+04,1.979e+01) {343 DOFs};
    \node[anchor=west, inner sep=1pt, fill=white, fill opacity=0.7, text opacity=1, font=\tiny] at (axis cs: 1.100e+04,2.584e+01) {729 DOFs};
    \node[anchor=west, inner sep=1pt, fill=white, fill opacity=0.7, text opacity=1, font=\tiny] at (axis cs: 1.100e+04,3.703e+01) {1331 DOFs};
    \node[anchor=west, inner sep=1pt, fill=white, fill opacity=0.7, text opacity=1, font=\tiny] at (axis cs: 1.100e+04,7.177e+01) {2744 DOFs};
    \node[anchor=west, inner sep=1pt, fill=white, fill opacity=0.7, text opacity=1, font=\tiny] at (axis cs: 1.100e+04,1.319e+02) {4913 DOFs};
    \node[anchor=west, inner sep=1pt, fill=white, fill opacity=0.7, text opacity=1, font=\tiny] at (axis cs: 1.100e+04,2.643e+02) {9261 DOFs};
    \node[anchor=west, inner sep=1pt, fill=white, fill opacity=0.7, text opacity=1, font=\tiny] at (axis cs: 1.100e+04,6.119e+02) {17576 DOFs};
    \node[anchor=west, inner sep=1pt, fill=white, fill opacity=0.7, text opacity=1, font=\tiny] at (axis cs: 1.100e+04,1.347e+03) {35937 DOFs};
    \node[anchor=west, inner sep=1pt, fill=white, fill opacity=0.7, text opacity=1, font=\tiny] at (axis cs: 1.100e+04,2.831e+03) {68921 DOFs};

   \addplot[solid, color=black, mark=+, forget plot] coordinates {(120, 320)};
   \addplot[solid, color=black, mark=+, forget plot] coordinates {(72, 90)};
   \addplot[solid, color=black, mark=+, forget plot] coordinates {(38, 22)};
   \addplot[solid, color=black, mark=+, forget plot] coordinates {(22, 7)};
   \addplot[solid, color=black, mark=+, forget plot] coordinates {(17, 2.7)};
   \addplot[solid, color=black, mark=+, forget plot] coordinates {(12, 1.15)};
   \addplot[solid, color=black, mark=+, forget plot] coordinates {(8, 0.35)};
   \addplot[solid, color=black, mark=+, forget plot] coordinates {(9, 0.18)};
   \addplot[solid, color=black, mark=+, forget plot] coordinates {(13, 0.095)};
   \addplot[solid, color=black, mark=+, forget plot] coordinates {(22, 0.085)};
    
    \addlegendentry{impl\_mkl}
    \addlegendentry{impl\_cholmod}
    \addlegendentry{expl\_hybrid}
    \addlegendentry{expl\_gpu\_opt}
  \end{loglogaxis}
\end{tikzpicture}

%% file: main.bbl
\begin{thebibliography}{10}

\bibitem{FETI}
Charbel Farhat and Francois-Xavier Roux.
\newblock A method of finite element tearing and interconnecting and its parallel solution algorithm.
\newblock {\em International Journal for Numerical Methods in Engineering}, 32(6):1205--1227, 1991.

\bibitem{BDDC}
Clark~R. Dohrmann.
\newblock A preconditioner for substructuring based on constrained energy minimization.
\newblock {\em SIAM Journal on Scientific Computing}, 25(1):246--258, 2003.

\bibitem{espreso-pasc}
Lubom\'{\i}r \v{R}\'{\i}ha, Tom\'{a}\v{s} Brzobohat\'{y}, Alexandros Markopoulos, Ond\v{r}ej Meca, and Tom\'{a}\v{s} Kozubek.
\newblock {Massively Parallel Hybrid Total {FETI (HTFETI)} Solver}.
\newblock In {\em Proceedings of the Platform for Advanced Scientific Computing Conference}, PASC '16, New York, NY, USA, 2016. Association for Computing Machinery.

\bibitem{BDDC_SCALE}
Santiago Badia, Alberto~F. Mart\'{\i}n, and Javier Principe.
\newblock Multilevel balancing domain decomposition at extreme scales.
\newblock {\em SIAM J. Sci. Comput.}, 38(1):C22–C52, January 2016.

\bibitem{ESPRESO-SC}
Lubom{\'i}r {\v{R}}{\'i}ha, Tom{\'a}{\v{s}} Brzobohat{\'y}, Alexandros Markopoulos, Tom{\'a}{\v{s}} Kozubek, Ond{\v{r}}ej Meca, Olaf Schenk, and Wim Vanroose.
\newblock Efficient implementation of total {FETI} solver for graphic processing units using schur complement.
\newblock In {\em High Performance Computing in Science and Engineering}, pages 85--100, 2016.

\bibitem{BDDS_ACC}
Jakub Šístek and Tomáš Oberhuber.
\newblock Acceleration of a parallel {BDDC} solver by using graphics processing units on subdomains.
\newblock {\em The International Journal of High Performance Computing Applications}, 37(2):151--164, 2023.

\bibitem{FETI_PHI}
Michal Merta, Lubomir Riha, Ondrej Meca, Alexandros Markopoulos, Tomas Brzobohaty, Tomas Kozubek, and Vit Vondrak.
\newblock Intel xeon phi acceleration of hybrid total {FETI} solver.
\newblock {\em Advances in Engineering Software}, 112:124--135, 2017.

\bibitem{pardiso}
Cosmin~G. Petra, Olaf Schenk, Miles Lubin, and Klaus G\"{a}ertner.
\newblock An augmented incomplete factorization approach for computing the schur complement in stochastic optimization.
\newblock {\em SIAM Journal on Scientific Computing}, 36(2):C139--C162, 2014.

\bibitem{PDSEC}
Jakub Homola, Radim Vav{\v{r}}{\'\i}k, Ond{\v{r}}ej Meca, Tom{\'a}{\v{s}} Brzobohat{\`y}, and Lubom{\'\i}r {\v{R}}{\'\i}ha.
\newblock {Assembly of FETI dual operator using CUDA}.
\newblock {\em arXiv preprint arXiv:2502.08382}, 2025.

\bibitem{PCPG}
Chandrika Kamath.
\newblock The {FETI} level 1 method : Theory and implementation.
\newblock 2000.

\bibitem{brzyngeninverse}
T.~Brzobohatý, Z.~Dostál, T.~Kozubek, P.~Kovář, and A.~Markopoulos.
\newblock Cholesky decomposition with fixing nodes to stable computation of a generalized inverse of the stiffness matrix of a floating structure.
\newblock {\em International Journal for Numerical Methods in Engineering}, 88(5):493--509, 2011.

\bibitem{ssolvers}
{Netlib}.
\newblock Direct solvers for sparse matrices, 2025.

\bibitem{Csparse}
Timothy~A. Davis.
\newblock {\em Direct Methods for Sparse Linear Systems}.
\newblock Society for Industrial and Applied Mathematics, 2006.

\bibitem{FETI_radim}
Radim Vavrík and Lubomír Ríha.
\newblock {Acceleration Techniques for FETI Solvers for GPU Accelerators}.
\newblock In {\em 2018 International Conference on High Performance Computing \& Simulation (HPCS)}, pages 546--553, 2018.

\bibitem{cholmod}
Yanqing Chen, Timothy~A. Davis, William~W. Hager, and Sivasankaran Rajamanickam.
\newblock Algorithm 887: Cholmod, supernodal sparse cholesky factorization and update/downdate.
\newblock {\em ACM Trans. Math. Softw.}, 35(3), October 2008.

\bibitem{karolinadocs}
{IT4Innovations Karolina Documentation}.

\bibitem{espresogithub}
{ESPRESO github repository}.

\bibitem{intelmkl}
{Intel Corporation}.
\newblock {Intel oneAPI Math Kernel Library}.

\end{thebibliography}
